\documentclass[12pt]{article}

\usepackage{amssymb}

\def\be{\begin{equation}}
\def\ee{\end{equation}}
\def\cN{{\cal N}}
\def\dfrac#1#2{{\displaystyle\frac{#1}{#2}}}
\def\tfrac#1#2{{\textstyle\frac{#1}{#2}}}

\def\tsum{{\textstyle\sum}}
\def\tr{{\rm tr}}
\def\Tr{{\rm Tr}}
\def\cs{\,{\stackrel{\star}{,}\,}}
\renewcommand\Box{\square}

\def\theequation{\arabic{section}.\arabic{equation}}

\makeatletter
\renewcommand\section{\setcounter{equation}0%
\@startsection {section}{1}{\z@}%
                                   {-3.5ex \@plus -1ex \@minus -.2ex}%
                                   {2.3ex \@plus.2ex}%
                                   {\normalfont\Large\bfseries}}
\makeatother

\begin{document}
\begin{titlepage}
\begin{flushright}
\it
To the memory of Julius Wess
\end{flushright}
\vspace{0.7cm}

\begin{center}
{\Large\bf Gauge theory in deformed ${\cal N}$=(1,1) superspace}
\vspace{1cm}

{\large\bf
I.L. Buchbinder$\,{}^{a,b}$, \ E.A. Ivanov$\,{}^c$, \ O. Lechtenfeld$\,{}^d$, \\[6pt]
I.B. Samsonov$\,{}^{d,e}$, \ B.M. Zupnik$\,{}^c$}\\
\vspace{0.8cm}
{\it
a) Dept.\ of Chemistry and Physics, University of North
Carolina, NC 28372 Pembroke, USA\hfill\phantom{.}
\\
{\tt joseph.buchbinder@uncp.edu}\\[8pt]
b) Dept.\ of Theoretical Physics, Tomsk State Pedagogical University,
634041 Tomsk, Russia \footnote{Permanent address} \\
{\tt joseph@tspu.edu.ru}\\[8pt]
c) Bogoliubov Laboratory of Theoretical Physics, Joint Institute
for Nuclear Research,\hfill\phantom{.}\\
 141980 Dubna, Moscow Region, Russia \\
{\tt eivanov, zupnik@theor.jinr.ru}\\[8pt]
d) Institut f\"ur Theoretische Physik, Leibniz Universit\"at Hannover,
30167 Hannover, Germany\\
{\tt lechtenf, samsonov@itp.uni-hannover.de}\\[8pt]
e) Laboratory of Mathematical Physics, Tomsk Polytechnic University,
634050 Tomsk, Russia\\
{\tt samsonov@mph.phtd.tpu.edu.ru}}
\end{center}
\vspace{0.5cm}

\begin{abstract}
\noindent
We review the non-anticommutative $Q$-deformations of
${\cal N}{=}(1,1)$ supersymmetric theories in four-dimensional
Euclidean harmonic superspace. These deformations preserve
chirality and harmonic Grassmann analyticity. The associated field
theories arise as a low-energy limit of string theory in specific
backgrounds and generalize the Moyal-deformed supersymmetric field
theories. A characteristic feature of the $Q$-deformed theories is
the half-breaking of supersymmetry in the chiral sector of the
Euclidean superspace. Our main focus is on the chiral singlet
$Q$-deformation, which is distinguished by preserving the SO(4)
$\sim$ $Spin(4)$ ``Lorentz'' symmetry and the SU(2) $R$-symmetry. We
present the superfield and component structures of the deformed
${\cal N}{=}(1,0)$ supersymmetric gauge theory as well as of
hypermultiplets coupled to a gauge superfield: invariant actions,
deformed transformation rules, and so on. We discuss quantum
aspects of these models and prove their renormalizability in the
abelian case. For the charged hypermultiplet in an abelian gauge
superfield background we construct the deformed holomorphic
effective action.
\end{abstract}
\end{titlepage}
\tableofcontents
\newpage
\section{Introduction}
By now, the concept of supersymmetry has been organically incorporated
into modern high-energy theoretical physics. Originally, it
was introduced at the mathematical level as a possible kind of
new symmetry which extends the standard space-time symmetries
by spinorial generators and relates
bosons and fermions. Since then, the consequences of the
supersymmetry hypothesis for particle physics has proved so
fruitful that today it is hardly possible to doubt its validity.
At present, the quest for supersymmetric
partners of the known elementary particles is one of the main
occupations of the forthcoming LHC experiments
\footnote{The phenomenological aspects of supersymmetry are
discussed in detail e.g. in \cite{fen}.}.

Let us mention the most impressive achievements of supersymmetry.
First of all, it yields a unified setup for
describing bosons and fermions. In the Standard Model, it suggests
a natural solution of the hierarchy problem. In grand
unification models, it predicts the single-point meeting
of the three basic running couplings (see e.g.
\cite{Hoker}) and solves the problem of the proton lifetime. Finally,
the most popular candidate for unifying gravity with quantum
physics, String Theory, is to large extent based on
the concept of supersymmetry. Supersymmetric theories in various
dimensions originate from the low-energy limit of string theory
with an appropriate choice of background manifold. New applications
of supersymmetry regularly appear in various areas. The present review
is devoted to a recent such development.

We will be concerned only with four-dimensional supersymmetric
theories. The algebra of Poincar\'e supersymmetry in $4D$
Minkowski space is characterized by the number ${\cal N}$ of
fermionic spinorial generators. ${\cal N}{=}1$ supersymmetry is
referred to as {\it simple}, featuring only two
two-component spinorial generators $Q_\alpha$ and $\bar
Q_{\dot\alpha}$. The spinorial generators of {\it extended}
supersymmetry (with ${\cal N}{>}1$ ) carry an index $k$ of
the fundamental representation of the $R$-symmetry group SU$({\cal
N})$. To date, the ${\cal N}{=}1$ supersymmetric theories have
been studied most thoroughly,  both at the classical and at the quantum level,
due to the existence of well established superfield techniques
(see e.g. \cite{BK,GGRS}). Furthermore, only ${\cal N}{=}1$
theories are really interesting for
phenomenological applications. On the other hand, theories with
extended supersymmetry exhibit quite remarkable and unique
properties. For instance, ${\cal N}{=}2$ supersymmetry imposes so
severe constraints on the quantum dynamics
that it becomes possible to find exact expressions (and values)
for some important quantities. It is known that
${\cal N}{=}2$ supersymmetric theories are one-loop exact due to so-called
``non-renormalization'' theorems. Moreover, the low-energy quantum
effective action in ${\cal N}{=}2$ supersymmetric gauge theory
can be exactly evaluated non-perturbatively (so-called
Seiberg-Witten theory \cite{SW}). However, among the supersymmetric field
theories, the unique place belongs to the ${\cal N}{=}4$ supergauge model.
It possesses the maximal number of supersymmetries
admitting spins not higher than one. The restrictions of ${\cal N}{=}4$
supersymmetry on the quantum structure of this theory turn out to
be so strong that they ensure ultraviolet finiteness of this
theory (i.e. it contains no quantum divergences at all). Also
${\cal N}{=}4$ supergauge theory is most intimately related to
superstring theory, e.g. via the renowned AdS/CFT correspondence
(see reviews \cite{ADS}).

Since there is no experimental evidence for supersymmetry at the
energies achievable by now, we must assume it to be broken,
leading to the problem of appropriate theoretical mechanisms for such
breaking.
One possibility is the so-called soft breaking of
supersymmetry. It is used in supersymmetric gauge theories and
adds to the action certain mass terms which preserve gauge invariance
but break supersymmetry.
If supersymmetry is spontaneously broken, auxiliary fields develop
non-vanishing vacuum values and spinorial Goldstone fields (goldstini) appear.
Standard methods of supersymmetry breaking can ruin the remarkable quantum
properties of supersymmetric theories or, at least, limit the
range of their applicability. Therefore, the search for and study of
alternative supersymmetry breaking schemes are of clear importance.

A new mechanism for breaking space-time
symmetries in quantum field theory arises from the hypothesis of
noncommutativity of the space-time coordinates,
\be
[x^m, x^n] = {\rm i}\vartheta^{mn} = const.
\label{1}
\ee
Here, the constants $\vartheta^{mn}$ are the
parameters of the deformation of the commutative algebra of
functions given on standard Minkowski space with
coordinates $x^m$. In the noncommutative field theory based on the
relation (\ref{1}) \cite{NC-reviews,SW1}, Lorentz invariance is
broken but translation invariance is still alive. On general
findings, noncommutativity is implemented by inserting the
so-called $\star$-product everywhere.
In the
case of deformation (\ref{1}) the $\star$-multiplication on fields
is realized with the help of the pseudo-differential
operator $P$ (the Poisson structure operator):
\be
\phi(x)\star
\psi(x) = \phi\, e^{P}\psi\quad \mbox{where} \quad P =
\frac{\rm i}2\overleftarrow{\partial_m}
\vartheta^{mn}\overrightarrow{\partial_n}.
\label{2}
\ee
For
constructing the classical action of noncommutative theories it
suffices to replace the standard multiplication of the fields in
the undeformed Lagrangian by the  $\star$-multiplication
(\ref{2}). In this approach, the free part of the action
preserves Lorentz invariance, while the breaking of Lorentz invariance
due to the deformation
comes out only in the interactions.

The relation (\ref{1}) can be employed also
to deform a superspace. However, the noncommutativity of
bosonic coordinates alone does not trigger any breaking of
supersymmetry. Formally, one can deform the
algebra of both even and odd coordinates in superspace (see e.g.
\cite{NAC}). For Minkowski signature, however, the deformation of
the fermionic superspace coordinates is not very well elaborated
since it is very
difficult to simultaneously maintain reality,
$\star$-product associativity, and the preservation of chiral
supersymmetry representations in the noncommutative theory
(some attempts to overcome this problem were recently undertaken in \cite{ISmi} and \cite{Dimitrievic}).

In the Euclidean version of ${\cal N}{=}1$ superspace, in contrast, the
Grassmann-odd coordinates $\theta^\alpha$ and
$\bar\theta^{\dot\alpha}$ are not related by complex
conjugation. We speak of Euclidean ${\cal N}{=}(n/2, n/2)$ supersymmetry
denoting the number of left-chiral and right-chiral
(antichiral) spinorial generators in the superalgebra.
In ${\cal N}{=}(1/2, 1/2)$ supersymmetric theories
formulated in Euclidean superspace it is therefore consistent
to deform the left-chiral fermionic coordinates \cite{Seiberg},
\be
\{\theta^\alpha, \theta^\beta\} = C^{\alpha\beta} = const
\quad \mbox{while}\quad
\{\bar\theta^{\dot\alpha},\bar\theta^{\dot\beta} \}=
\{\theta^\alpha,\bar\theta^{\dot\beta} \}=0\,,
\label{3}
\ee
thereby replacing a Grassmann algebra with a Clifford algebra.
The parameters $C^{\alpha\beta}$ deform the algebra of functions
on ${\cal N}{=}(1/2, 1/2)$ superspace. The remaining (anti)commutativity
relations in the chiral basis are not altered, in order to preserve
chirality. If such a non-anticommutative deformation is introduced exclusively in the left-chiral sector of
the superspace, the original ${\cal N}{=}(1/2, 1/2)$  Euclidean
supersymmetry gets broken to  ${\cal N}{=}(1/2, 0)$. It is obvious that the opportunity
of such a half-breaking of supersymmetry exists only in Euclidean
superspace.

We point out that the existence of
non-anticommutative deformations preserving chirality derives from
superstring theory \cite{Seiberg,Strings,Ooguri}.
Since the spectrum of IIB supergravity contains the four-form
potential, the ${\cal N}{=}(1,1)$ superstring provides a self-dual
five-form field-strength background, which in first approximation
is assumed to be constant.
After a compactification to the orbifold
$\mathbb{C}^3/(\mathbb{Z}_2\times\mathbb{Z}_2)$ one obtains a
four-dimensional ${\cal N}{=}(1, 1)$ superstring in the
background of a constant selfdual graviphoton field strength
$F^{\alpha\beta}$ (with $F^{\dot\alpha\dot\beta} =0$, i.e. one
considers Euclidean space). It turns out that the correlation
functions
$\langle\theta^{\alpha}(\tau)\theta^{\beta}(\tau')\rangle$
are proportional to the
constant field $F^{\alpha\beta}$, whereas those involving the conjugate variables
$\bar\theta^{\dot\alpha}$ are trivial. In the effective low-energy
field theory, such string variables
become fermionic coordinates of a superspace
with precisely the non-trivial anticommutation relations (\ref{3}).
String models in the background of a constant self-dual gauge field
can have interesting
phenomenological properties. For instance, it was shown in
\cite{Ooguri} that the gluon potential in ${\cal N}{=}(1/2,
1/2)$ supersymmetric theories can be modified by a
non-anticommutative deformation such as to acquire
a nontrivial vacuum expectation value, which may by related to
quark confinement.

Field theories defined in Euclidean superspaces with deformed
anticommutation relations of the type (\ref{3}) are referred to as
${\cal N}{=}1/2$ (or ${\cal N}{=}(1/2, 0)$) non-anticommutative
theories. These theories possess a number of
attractive properties. For example, it was established in
\cite{Britto,Romagnoni,Penati,Lunin,Jack} that the ${\cal N}{=}1/2$ supersymmetric
Wess-Zumino model and the ${\cal N}{=}1/2$ supersymmetric gauge theory
inherit the renormalizability of their undeformed
prototypes. In the Lagrangians of these models,
non-anticommutative deformations (\ref{3}) give rise to additional
terms polynomial in the deformation parameters
$C^{\alpha\beta}$. These terms can be
treated as new interaction vertices, the powers of
$C^{\alpha\beta}$ playing the role of coupling constants with negative mass dimension.
According to the standard lore of quantum field theory, vertices should
give rise to non-renormalizable divergences. However, the extra terms brought
into the action by the non-anticommutative deformations appear in
a non-symmetric way (they are not accompanied by
similar terms with $C^{\dot\alpha\dot\beta}$), and the
renormalization of such theories requires special analysis.
For instance, in \cite{Britto,Romagnoni} it was found that,
a single new term was generated at quantum level, and the non-anticommutative
Wess-Zumino model is multiplicatively renormalizable. For the ${\cal N}{=}1/2$
super Yang-Mills model it was shown \cite{Jack} that all new
divergences owed to the non-anticommutative
deformation can be eliminated by a shift of one spinor field. As
a result of these studies, all considered  ${\cal N}{=}1/2$
theories were found to be renormalizable and, hence, may be of
phenomenological interest (after
performing a Wick rotation to the Minkowski signature). Furthermore,
the effective action of the ${\cal N}{=}1/2$
supersymmetric Wess-Zumino model and the Yang-Mills theory was
studied in \cite{B}.

Let us turn to the extended supersymmetry and its non-anticommutative
deformation. We consider Euclidean ${\cal N}{=}(1,
1)$ superspace with Grassmann coordinates $\theta_i^\alpha,
\bar\theta^{\dot\alpha j}$ with $i, j = 1,2$ and $\alpha =1, 2$ and $\dot\alpha
= \dot{1}, \dot{2}\,$. The left-chiral deformation
(\ref{3}) generalizes to \cite{singlet,singlet1}
\be
\{\hat\theta^\alpha_i, \hat\theta^\beta_j \} = C^{\alpha\beta}_{ij}=const,
\label{4}
\ee
with all other (anti)commutation relations between chiral coordinates
of the ${\cal N}{=}(1, 1)$ superspace remaining undeformed. The constant tensor
$C^{\alpha\beta}_{ij}$ decomposes into irreducible pieces,
\be
C^{\alpha\beta}_{ij} =
C^{(\alpha\beta)}_{(ij)} + \varepsilon^{\alpha\beta}
\varepsilon_{ij} I.
\ee
Putting all components but $C^{\alpha\beta}_{11}$ to zero, we recover
the deformation (\ref{3}). Various types of such
deformations were studied in \cite{CILQ,Sasaki,Araki1}. Of particular
interest is the pure-trace deformation
$C^{\alpha\beta}_{ij} = \varepsilon^{\alpha\beta} \varepsilon_{ij}
I\,$. This type was named non-anticommutative
chiral singlet deformation \cite{singlet,singlet1}. Since the chiral singlet
deformation is fully specified by a single parameter $I$ which
carries no indices, it does not break Euclidean
SO(4) invariance or SU(2) $R$-symmetry. However, it breaks ${\cal N}{=}(1, 1)$
supersymmetry down to ${\cal N}{=}(1, 0)\,$. Apart from these special properties,
chiral singlet deformations can be given a stringy interpretation
\cite{FILSZ}. Unlike the ${\cal N}{=}(1/2, 0)$ case, for
deriving chiral singlet deformations one must
consider the ${\cal N}{=}4$ superstring in the background of a constant
axion field strength compactified on the orbifold
$\mathbb{C}\times\mathbb{C}^2/\mathbb{Z}_2$.
The stringy origin of non-singlet deformations of ${\cal N}{=}(1, 1)$
supersymmetry was discussed in \cite{Ito}.

Like for the bosonic deformation (\ref{1}), the
relations (\ref{3}) are also implemented in terms of an
appropriate $\star$-product, which now operates on superfunctions
of the coordinates of the undeformed ${\cal N}{=}(1/2, 1/2)$
superspace $z = (x^m, \theta^\alpha, \bar\theta^{\dot\alpha})$:
\be
A(z)\star B (z) = A\, e^{P_C} B\quad
\mbox{with} \quad
P_C = -\frac12\overleftarrow{Q_\alpha}
C^{\alpha\beta}\overrightarrow{Q_\beta},   \label{5}
\ee
where $Q_\alpha$ are the left-chiral supercharges. In the chiral basis
the generators $Q_\alpha$ coincide with the partial derivatives
with respect to the left Grassmann coordinates, $Q_\alpha =
\partial_\alpha\,$.
Non-anticommutative models with simple supersymmetry are
obtained from the corresponding undeformed models via insertion of
the $\star$-multiplication (\ref{5}) everywhere inside the corresponding
superfield Lagrangians \cite{Seiberg}. The criterion of preserving
some symmetry of the ``classical'' (undeformed) action in the non-anticommutative case
is the commuting of the symmetry generator with the Poisson
operator $P_C$ in (\ref{5}).

The expression (\ref{5}) for the $\star$-product can easily be
generalized to the extended supersymmetry (\ref{4})
\cite{singlet,singlet1}:
\be
A(z)\star B(z) = A\,e^{P_C}B
\quad\mbox{with} \quad P_C = -\overleftarrow{Q}{}^i_\alpha C^{\alpha\beta}_{ij}
\overrightarrow{Q}{}^j_\beta.
\label{6}
\ee
Here, the ${\cal N}{=}(1, 0)$ supersymmetry generators $Q^i_\alpha$
can be chosen in the chiral basis, $Q^i_\alpha =
\partial/\partial \theta^\alpha_i\equiv \partial^i_\alpha$,
 and $A(z)$ and $B(z)$ are
arbitrary superfunctions on the extended superspace $z = (x^m,
\theta^\alpha_i, \bar\theta^{\dot\alpha j})\,$. The most appropriate
superfield formulation of models with  ${\cal N}{=}(1, 1)$
supersymmetry is provided by the harmonic superspace approach,
which has been worked out in detail for ${\cal N}{=}2$
supersymmetric theories in Minkowski space \cite{HSS,Book}. This
approach allows one to write down superfield actions for
non-anticommutative models in manifestly
${\cal N}{=}(1, 0)$ supersymmetric form, and it also ensures the preservation of
supersymmetry at all stages of the quantum calculations.
Non-anticommutative deformations of the type (\ref{4}) for
harmonic superspace were introduced in \cite{singlet,singlet1}, while
non-anticommutative ${\cal N}{=}(1,0)$ models of
hypermultiplets and gauge superfields were introduced and studied
in \cite{singlet,singlet1,Araki1,FILSZ,ILZ,IZ,Araki}. In these papers, the component
structure of the corresponding classical deformed actions has been
established.

The Poisson operators $P_C$ generating the non-anticommutative
deformations (\ref{5}) and (\ref{6}) are composed from the
supercharges of the unbroken ${\cal N}{=}(1/2, 0)$ or ${\cal N}{=}(1,
0)$ supersymmetries, respectively, hence such deformations are called
$Q$-deformations. By definition, the operators $P_C$ do not
commute with the ${\cal
N}{=}(0,1/2)$ or ${\cal N}{=}(0, 1)$ supercharges, or generally with
the generators of bosonic symmetries
realized on the supercharges $Q_\alpha$ or $Q_\alpha^i$. On the other hand,
the operators $P_C$ commute with the covariant spinor
derivatives $D_\alpha, \bar D_{\dot\alpha}$ or $D_\alpha^k, \bar
D_{k \dot\alpha}$ defined in the corresponding superspaces.
Therefore, $Q$-deformations preserve superfield constraints
involving these spinor derivatives, in particular the conditions
of chirality, antichirality and Grassmann harmonic analyticity. An
alternative possibility is the non-anticommutative $D$-deformation
\cite{NAC}, defined by a the Poisson operator bilinear in the
covariant spinor derivatives. Such deformations preserve the
entire supersymmetry but break chirality, which makes it difficult to
construct $D$-deformed interactions of chiral superfields
\footnote{The singlet $D$-deformation of the ${\cal N}{=}(1, 1)$
gauge theory was considered in \cite{singlet,singlet1}. In this model
supersymmetry is preserved, the superfield geometry in the full
superspace is deformed, but the Grassmann-analytic representations
remain undeformed.}. As distinct from the $Q$-deformations,
no stringy interpretation is known for the $D$-deformation.

Non-anticommutative $Q$-deformations (\ref{5}) or (\ref{6})
differ in a crucial aspect from the
bosonic deformations (\ref{2}). Their Poisson operators $P_C$ are built
of mutually anticommuting
operators satisfying the nilpotency property $(Q_\alpha)^3 = 0$
or $(Q^k_\alpha)^5 = 0$, respectively. Therefore, the power expansions of the
exponentials in (\ref{5}) and (\ref{6}) terminate at corresponding orders.
As a result, the ensueing models
contain only a {\it finite} number of local deformation terms in
their Lagrangians. In other words, non-anticommutative theories
are always local, as opposed to Moyal-deformed theories
based on (\ref{2}), which bring an infinite number of new vertices into the
Lagrangian.

Mathematically rigorous treatment of the
$\star$-products for the deformation of both bosonic and Grassmann
coordinates in the framework of noncommutative field theory is
discussed in \cite{Q} using the language of quantum
(super)groups and Hopf algebras. In this interpretation, the
broken space-time symmetries and supersymmetries of the noncommutative
theories are not lost but just deformed. The generators of the
deformed (quantum) symmetries by definition act covariantly on the
$\star$-products of the corresponding fields or superfields, which
guarantees the invariance of the action under the deformed (quantum)
(super)symmetry transformations. In this review we will not deal with the deformed
(quantum) (super)symmetries, since their implications for
non-anticommutatively deformed theories are still obscure.

The renormalizability and other quantum aspects of theories with
non-anticommutative $Q$-deformations of ${\cal N}{=}(1/2, 1/2)$
supersymmetry were considered in detail in
\cite{Britto}-\cite{B}. Up to now, the case of extended quantum
supersymmetry has been studied only for the particular case of chiral singlet
$Q$-deformations (in the harmonic superspace
approach) \cite{BILSZ,BLS}. In particular, it was
found that the non-anticommutative models of the abelian gauge
superfield and the neutral hypermultiplet are renormalizable. These
results were obtained by computing the divergent contributions to
the quantum effective actions. These
divergent contributions do not have the
form of classical interactions, whence one might conclude
that multiplicative renormalizability is jeopardized. Yet,
all the divergent terms in the
effective action can be removed by a simple field
redefinition, viz. by a shift of the
scalar field $\phi$ in the vector gauge multiplet.
Since such a field redefinition does not influence the dynamics of
the theory and it follows that the divergences in the given case are
unphysical. Therefore, the considered theories are not only
renormalizable, but actually finite.
An analogous situation had been observed in  \cite{Jack} while proving
the renormalizability of the ${\cal N}{=}1/2$ supersymmetric gauge
theory. In this case, the divergences are removed by shifting
one of the gaugini belonging to the gauge supermultiplet. We remark
that in the undeformed limit the actions of the abelian gauge superfield and
the neutral hypermultiplet considered in \cite{BILSZ} reduce to
free ones. This implies that all interactions in these deformed
theories are caused by the deformation.

In \cite{BLS} we also studied the quantum structure of the
non-anticommutative charged hypermultiplet model introduced
in \cite{ILZ}. This model is of interest
because in the undeformed limit it remains interacting, becoming the
${\cal N}{=}(1, 1)$
supersymmetric extension of electrodynamics. It is well known that
the low-energy effective action of the latter model is described
by a holomorphic potential which plays an important role in ${\cal
N}{=}2$ Seiberg-Witten theory \cite{SW}. In \cite{BLS}, by
quantum superfield calculations in harmonic superspace, it was
established that this non-anticommutative model is
renormalizable in the standard sense. In addition, finite
contributions to the low-energy effective action were obtained
including the holomorphic potential, which turned out to be
deformed in the naive sense.
Thus, by now, all abelian models with
non-anticommutative chiral singlet $Q$-deformation of ${\cal
N}{=}(1, 1)$ supersymmetry have been proved renormalizable.

The review is organized as follows. In Sect.\ 2 we
introduce general chiral $Q$-deformations of ${\cal N}{=}(1,
1)$ superspace and consider chiral singlet $Q$-deformations
in harmonic superspace. In Sect.\ 3 we present superfield
formulations of the classical actions for the supersymmetric gauge
multiplet and hypermultiplet models with chiral singlet
$Q$-deformation of ${\cal N}{=}(1, 1)$ supersymmetry in harmonic
superspace. In Sect.\ 4 the component structure of these actions in
the abelian case is given. Sect.\ 5 is devoted to proving
renormalizability of the abelian theories of the hypermultiplet
and gauge superfield. In Sect.\ 6 we describe the general structure
of the effective action in the charged hypermultiplet model and
evaluate the leading (holomorphic) contributions to the effective
action. In Sect.\ 6 we also study the component structure of the
new contributions to the low-energy effective action induced by
the non-anticommutativity. In the Conclusions the main results are
summarized and some further directions are
outlined. Two Appendices contain basic relations of Euclidean
${\cal N}{=}(1, 1)$ supersymmetry and Euclidean harmonic superspace.

The review is mostly based on our papers
\cite{singlet,FILSZ,ILZ,BILSZ,BLS}. We shall
keep to the notation used in these works.

\setcounter{equation}{0}
\section{Chiral deformations of ${\cal N}{=}(1,1)$ supersymmetry}
\subsection{Chiral deformations of ${\cal N}{=}(1,1)$ superspace}
The non-anticommutative chiral deformations are possible only in
the Euclidean superspace. Therefore we consider the Euclidean
$\cN{=}(1,1)$ superspace parametrized by the coordinates
$z=(x^m,\theta^\alpha_i, \bar\theta^{\dot\alpha i})$, where $x^m$
are the coordinates of the Euclidean space $\mathbb{R}^4$ and
$\theta^\alpha_i$, $\bar\theta^{\dot\alpha i}$ are Grassmann
coordinates. Here $\alpha,\dot\alpha=1,2$ denote the spinor
indices, $i=1,2$ is the index of the $R$-symmetry group ${\rm
SU}(2)$. Note that the group ${\rm SO}(4)$ of rotations of the
Euclidean space $\mathbb{R}^4$ plays the role similar to the Lorentz
group for the Minkowski space $\mathbb{R}^{3,1}$. The
corresponding universal covering group for ${\rm SO}(4)$ is
$Spin(4)={\rm SU}(2)_L\times {\rm SU}(2)_R$. Therefore the spinors
of different chiralities transform independently with respect to the
subgroups ${\rm SU}(2)_L$, ${\rm SU}(2)_R$
and they are not related to each other by the complex conjugation.
The basic definitions related to the $\cN{=}(1,1)$ superspace
are collected in the Appendix 1.

It is important to realize that there are two different types of
complex conjugation in the $\cN{=}(1,1)$ superspace
\cite{singlet}. The first one, by definition, acts on the
superspace coordinates and superfields as follows \be
\widetilde{\theta_k^\alpha}=\varepsilon^{kj}\varepsilon_{\alpha\beta}\theta_j^\beta,
\qquad \widetilde{\bar\theta^{\dot\alpha k}}=-\varepsilon_{kj}
\varepsilon_{\dot\alpha\dot\beta}\bar\theta^{\dot\beta j},\qquad
\widetilde{x^m}=x^m,\qquad \widetilde{AB}=\tilde B\tilde A.
\label{conj1} \ee Clearly, the conjugation (\ref{conj1}) squares
to the identity on any object and is compatible with both
$Spin(4)$ and  $R$-symmetry ${\rm SU}(2)$ groups of $\cN{=}(1,1)$
superspace, preserving the irreducible representations of these
groups. However, this conjugation is incompatible with the
reduction of $\cN{=}(1,1)$ supersymmetry down to
$\cN{=}(1/2,1/2)\,$ since it is impossible to define the invariant
under (\ref{conj1}) subset of supercharges forming the
$\cN{=}(1/2,1/2)\,$ supersymmetry \footnote{Respectively, in
$\cN{=}(1,1)\,$ superspace with the conjugation (\ref{conj1}) there
are no subspaces closed under the $\cN{=}(1/2,1/2)\,$ supersymmetry .}.

There is an alternative conjugation in $\cN{=}(1,1)$ superspace
denoted by ``$*$'' and defined by the rules \be
(\theta^\alpha_k)^*=\varepsilon_{\alpha\beta}\theta^\beta_k,\qquad
(\bar\theta^{\dot\alpha
k})^*=\varepsilon_{\dot\alpha\dot\beta}\bar\theta^{\dot\beta k},
\qquad (x^m)^*=x^m,\qquad (AB)^*=B^* A^*. \label{conj2} \ee The
conjugation (\ref{conj2}) is compatible with the reduction of the
$\cN{=}(1,1)$ supersymmetry down to $\cN{=}(1/2,1/2)$ since it allows to single
out the invariant $\cN{=}(1/2,1/2)$ subspaces in the $\cN{=}(1,1)$
superspace. It respects also the action of the group $Spin(4)$.
However, this conjugation squares to the identity only on the
bosonic coordinates and fields, while for the spinor fields the
double conjugation yields $-1\,$. Therefore it is natural to refer to
the involution (\ref{conj2}) as a ``pseudoconjugation''.
The action of the $R$-symmetry group ${\rm SU}(2)$ on
$\cN{=}(1,1)$ supercharges and fermionic coordinates of
$\cN{=}(1,1)$ superspace is incompatible with the
pseudoconjugation (\ref{conj2}), while it preserves the ${\rm SL
}(2,R)$ group which plays the role of $R$-symmetry group in this
case. This means that the pseudoconjugation * corresponds to
another real form of $\cN{=}(1,1)$ supersymmetry with a non-compact
group of internal automorphisms. The undeformed real superfield
actions in these two different Euclidean $\cN{=}(1,1)$ superspaces
are related to each other and to $\cN{=}2$ supersymmetric
actions in Minkowski space by the Wick rotations. It should be
pointed out that, when the deformations of supersymmetry (or other
symmetries) are introduced, the actions which are real with respect to one
conjugation can be complex with respect to the other, and vice
versa. In what follows we shall deal with only one type of the
conjugation, that given by eq. (\ref{conj1}).

The chiral deformations of supersymmetry appear most naturally in
the chiral coordinates,
\be
z_L=(x^m_L,\theta^\alpha_i,\bar\theta^{\dot\alpha
k}),\qquad
x^m_L=x^m+{\rm i}
(\sigma^m)_{\alpha\dot\alpha}\theta^\alpha_k\bar\theta^{\dot\alpha
k},
\label{chir}
\ee
where the Euclidean sigma-matrices are given in the Appendix 1,
(\ref{A2}). The supertranslations act on the coordinates $z_L$ as follows
\be
\delta_\epsilon x^m_L=2{\rm i}(\sigma^m)_{\alpha\dot\alpha}
\theta^\alpha_k\bar\epsilon^{\dot\alpha k},
\qquad
\delta_\epsilon\theta^\alpha_k=\epsilon^\alpha_k,
\qquad
\delta_\epsilon\bar\theta^{\dot\alpha k}=\bar\epsilon^{\dot\alpha
k},
\label{susy_}
\ee
where $\epsilon^\alpha_k$, $\bar\epsilon^{\dot\alpha k}$ are
anticommuting parameters. In the chiral coordinates,
the supercharges and covariant spinor derivatives (\ref{A3}) read
\begin{eqnarray}
&&Q^i_\alpha=\partial^i_\alpha,\qquad
\bar Q_{\dot\alpha i}=-\bar\partial_{\dot\alpha i}
+2{\rm i}\theta_i^\alpha (\sigma_m)_{\alpha\dot\alpha}\frac\partial{\partial
x^m_L},
\label{charges}\\&&
D^i_\alpha=\partial^i_\alpha+2{\rm i}\bar\theta^{\dot\alpha i}(\sigma_m)_{\alpha\dot\alpha}
\frac\partial{\partial x^m_L},\qquad
\bar D_{\dot\alpha i}=-\bar\partial_{\dot\alpha i}.
\label{D}
\end{eqnarray}

Consider now the operator $P_C$ defined in the coordinate basis
(\ref{chir}) by the following expression
\be
P_C=-\overleftarrow{\partial}{}^i_\alpha C^{\alpha\beta}_{ij}
\overrightarrow{\partial}{}^j_\beta = -\overleftarrow{Q}{}^i_\alpha C^{\alpha\beta}_{ij}
\overrightarrow{Q}{}^j_\beta.
\label{P}
\ee
It acts on the arbitrary superfields $A$, $B$ according to the rules
\begin{eqnarray}
&&AP_CB
=-(-1)^{p(A)}(\partial^i_\alpha A)C^{\alpha\beta}_{ij}(\partial^j_\beta B),
\label{APB}\\
&&AP_C^2B=(\partial^k_\gamma\partial^i_\alpha A)C^{\alpha\beta}_{ij}
C^{\gamma\rho}_{kl}(\partial_\rho^l\partial^j_\beta B).\nonumber
\end{eqnarray}
Here $C^{\alpha\beta}_{ij}$ are some constants and $p(A)$ is
the Grassmann parity of the superfield $A$. The operator
$P_C$ defines the Moyal-Weyl $\star$-product of superfields
(see (\ref{6})),
\be
A\star B=Ae^{P_C} B=AB+AP_CB+\frac12 AP^2_CB+\frac16AP^3_CB
+\frac1{24}AP^4_CB.
\label{star}
\ee
The operator $P_C$ is nilpotent since
$(\partial^i_\alpha)^5=0\,$. Therefore the $\star$-deformation (\ref{star})
never produces non-localities, in contrast to the deformations of bosonic
coordinates (\ref{1}) (see, e.g., \cite{SW1}).

As the operator (\ref{P}) is built out only of the supercharges,
and they anticommute with the covariant derivatives (\ref{D}),
the product (\ref{star}) preserves both chirality and
antichirality,
\begin{eqnarray}
&&D^i_\alpha(A\star B)=(D^i_\alpha A)\star B+A\star(D^i_\alpha B),\nonumber\\
&&\bar{D}_{i\dot\alpha}(A\star B)=(\bar{D}_{i\dot\alpha} A)\star B
+A\star(\bar{D}_{i\dot\alpha} B).
\end{eqnarray}
What is more important for $\cN{=}(1,1)$ supersymmetric theories,
the $\star$-mul\-ti\-pli\-ca\-ti\-on also respects
the Grassmann harmonic analyticity (see the next
subsection for details). Since all $\cN{=}(1,1)$ supersymmetric Euclidean
theories are well defined  only if the chirality and harmonic analyticity are
preserved (similarly to  $\cN{=}2$ models in Minkowski space),
it is a consistent deformation of these theories  when the standard
multiplication in their classical actions is replaced by the $\star$-product (\ref{star}).

It is natural to demand the multiplication (\ref{star}) to be consistent
with the reality properties. Since in the Euclidean superspaces there are two different conjugations
(\ref{conj1}) and (\ref{conj2})  which
respect either SU(2) or SL$(2,R)$ $R$-symmetry groups,
the preservation of reality puts two different
constraints on the parameters of deformations
$C^{\alpha\beta}_{ij}$:
\begin{eqnarray}
\widetilde{(A\star B)}=\tilde B\star \tilde A\quad&
\Longrightarrow \quad&
\widetilde{C^{\alpha\beta}_{ij}}=C^{ij}_{\alpha\beta},
\label{C2}\\
(A\star B)^*=B^*\star A^*\quad& \Longrightarrow\quad&
(C^{\alpha\beta}_{ij})^*=C_{\alpha\beta ij}
\label{C1}.
\end{eqnarray}
In our further consideration we restrict ourselves to the case of
the conjugation (\ref{C2}).

In general, since the constants $C^{\alpha\beta}_{ij}$ have both
the spinor and $R$-symmetry group indices, the Euclidean
${\rm SO}(4)$ and ${\rm SU}(2)_L$ groups, as well as the $R$-symmetry
group ${\rm SU}(2)$, are broken in the theories with the
deformations induced by the $\star$-product (\ref{star}). Moreover,
the $\cN{=}(1,1)$ supersymmetry is also broken down to
$\cN{=}(1,0)$ since the product (\ref{star}) involves
only the $\cN{=}(1,0)$ supercharges $Q^i_\alpha$ which have
non-vanishing anticommutators with the $\cN{=}(0,1)$ supercharges
$\bar Q_{i\dot\alpha}$.

Taking into account the definition (\ref{P}),
the $\star$-multiplication (\ref{star}) of two superfields can
always be written as
\be
A\star B=A\, B+Q^k_\alpha N^\alpha_k(A,B),
\label{P6}
\ee
where $N^\alpha_k(A,B)$ is some function of the superfields $A$, $B$
and constants $C^{\alpha\beta}_{ij}$. Eq. (\ref{P6})
implies that in the full superspace integral the $\star$-product of
two superfields reduces to the usual product,
\be
\int d^4x_L d^4\theta \,d^4\bar\theta\, A\star B=\int d^4x_L d^4\theta\,
d^4\bar\theta\, A\,B\,.
\label{int1}
\ee
In a similar way one can check that under the superspace integral the $\star$-product of
three superfields obeys the cyclic property
\be
\int d^8z \, A\star B\star C=\int d^8z \,C\star A\star B.
\label{int2}
\ee
There is also an analog of the relation (\ref{int1}) for the
chiral subspace,
\begin{eqnarray}
\int d^4x_L d^4\theta \, A\star B=\int d^4x_L d^4\theta\,
 A\,B\,.
\label{int2_}
\end{eqnarray}
The relation (\ref{int2_}) is formally valid not only for the chiral superfields, but
also for general ones $A$, $B$ (i.e. those given on the full $\cN{=}(1,1)$
superspace). However, this is not case for the
general $\cN{=}(1,1)$ superfields under the antichiral integral,
\be
\int d^4x_R d^4\bar\theta \, A\star
B\neq\int d^4x_R d^4\bar\theta\,
 A\,B\,.
\label{int3}
\ee
Only for the antichiral superfields $\bar\Phi$, $\bar\Lambda$
the equality sign in (\ref{int3}) is restored,
\be
\int d^4x_R d^4\bar\theta \, \bar\Phi\star\bar\Lambda=\int d^4x_R d^4\bar\theta\,
 \bar\Phi\,\bar\Lambda.
\label{int4}
\ee
Note that in the antichiral coordinates one should
use the following expressions for supercharges and covariant
spinor derivatives
\begin{eqnarray}
&&Q^i_\alpha=\partial^i_\alpha-2{\rm i}\bar\theta^{\dot\alpha i}(\sigma_m)_{\alpha\dot\alpha}
\frac\partial{\partial x^m_R},\qquad
\bar Q_{\dot\alpha i}=-\bar\partial_{\dot\alpha i},
\label{Rcharges}\\&&
D^i_\alpha=\partial^i_\alpha,\qquad
\bar D_{\dot\alpha i}=-\bar\partial_{\dot\alpha i}-2{\rm i}\theta_i^\alpha (\sigma_m)_{\alpha\dot\alpha}\frac\partial{\partial
x^m_R},
\label{DR}
\end{eqnarray}
where $x_R^m=x^m-{\rm i} (\sigma^m)_{\alpha\dot\alpha}\theta^\alpha_k\bar\theta^{\dot\alpha k}$.

Now, let us define the $\star$-commutators and anticommutators of
operators and superfields as follows
\be
[A\cs B]=A\star B-B\star A,\qquad
\{A\cs B\}=A\star B+B\star A.
\label{comm}
\ee
It is instructive to find the $\star$-(anti)commutators of the bosonic
and fermionic superspace coordinates,
\begin{eqnarray}
&&\{\theta^\alpha_k\cs\theta^\beta_j \}=2C^{\alpha\beta}_{ij},\qquad
[x^m_L\cs x^n_L]=0,\qquad
[x^m_L\cs\theta^\alpha_k]=0,\nonumber\\
&&[x^m_L\cs\bar\theta^{\dot\alpha k}]=0,\qquad
\{\theta^\alpha_k\cs\bar\theta^{\dot\beta j} \}=0,
\qquad
\{\bar\theta^{\dot\alpha k}\cs\bar\theta^{\dot\beta j} \}=0.
\label{comm1}
\end{eqnarray}
Eqs. (\ref{comm1}) tell us that in the chiral basis the
$\star$-product affects only the anticommutator of left-chiral
coordinates $\theta^\alpha_i$.

The constant tensor $C^{\alpha\beta}_{ij}$ can be decomposed into
the traceless part and trace with respect to the SU$(2)_L$ spinor and
SU$(2)$ $R$-symmetry indices,
\be
C^{\alpha\beta}_{ij}=C^{(\alpha\beta)}_{(ij)}+\varepsilon^{\alpha\beta}
\varepsilon_{ij}I.
\label{C3}
\ee
The Poisson operator (\ref{P}) acquires the most simple form in the
particular case $C^{(\alpha\beta)}_{(ij)}=0\,$:
\be
P_s=-\overleftarrow{Q}{}^i_\alpha
I\varepsilon^{\alpha\beta}\varepsilon_{ij}\overrightarrow{Q}{}^j_\beta
=-\overleftarrow{Q}{}^i_\alpha
I\overrightarrow{Q}{}_i^\alpha=
-\overleftarrow{\partial}{}^i_\alpha
I\overrightarrow{\partial}{}_i^\alpha.
\label{Ps}
\ee
The operator $P_s$ produces the following $\star$-product
\be
A\star B=A e^{P_s} B.
\label{sstar}
\ee
Clearly, the deformation (\ref{sstar}) does not break the
symmetries with respect to the Euclidean rotation group ${\rm SO}(4)$ and the
$R$-symmetry group ${\rm SU}(2)$. However, in
the deformed theories corresponding to the operator $P_s\,$,
${\cal N}{=}(1,1)$ supersymmetry is still broken by half.

The non-anticommutative $Q$-deformation associated with the
$\star$-product (\ref{sstar}) and preserving the maximal number of
symmetries will be  referred to as the chiral singlet deformation.
In what follows we will consider only this type of deformations because
of its uniqueness and relative simplicity.

\subsection{Chiral singlet deformation of ${\cal N}{=}(1,1)$ harmonic superspace}
$\cN{=}2$, $D=4$ harmonic superspace (its Minkowski space version) was
pioneered in \cite{HSS}. The pedagogical introduction to the harmonic superspace
approach can be found in the book \cite{Book}. Here, following ref. \cite{singlet},
we present how the non-anticommutative deformations given by the operator (\ref{Ps})
are realized in  harmonic superspace. The salient features of the Euclidian version of harmonic
superspace are collected in the Appendix 2.

The complex conjugation (\ref{conj1}) can be naturally
extended to the harmonic variables,
\be
\widetilde{u^\pm_k}=u^{\pm k}.
\label{conj3}
\ee
Using (\ref{conj1}) and (\ref{conj3}) one can find the complex conjugation rules
for the harmonic superspace coordinates
$x^m_A$,
$\theta^{\pm\alpha}$, $\bar\theta^{\pm\dot\alpha}$
\begin{eqnarray}
&&\widetilde{x^m_A}=x^m_A,\qquad
\widetilde{\theta^{\pm\alpha}}=\varepsilon_{\alpha\beta}\theta^{\pm\beta},\qquad
\widetilde{\bar\theta^{\pm\dot\alpha}}=\varepsilon_{\dot\alpha\dot\beta}
\bar\theta^{\pm\dot\beta}.
\label{conj4}
\end{eqnarray}
Note that the involution $\widetilde{\phantom{m}}$ is a pseudoconjugation since
it squares to $-1\,$ while acting on the harmonics and harmonic projections of Grassmann coordinates.

Let us now apply to the Poisson operator $P_s$ (\ref{Ps}) of the chiral singlet deformations
with the $\star$-product (\ref{sstar}).
In harmonic superspace,  this operator can be written as
\be
P_s=I(\overleftarrow{Q}{}^{+\alpha}\overrightarrow Q{}^-_\alpha
-\overleftarrow Q{}^{-\alpha}\overrightarrow Q{}^+_\alpha),
\label{Ps1}
\ee
where $Q^\pm_\alpha=Q^i_\alpha u^\pm_i$ are the harmonic
projections of supercharges. In terms of the supercharges $Q^\pm_\alpha$
the $\star$-product (\ref{sstar}) is rewritten as
\be
\star=e^{P_s}=1+P_s+\frac12 P_s^2+\frac16 P_s^3+\frac1{24}P_s^4,
\label{sstar1}
\ee
where
\begin{eqnarray}
\frac12 P_s^2&=&-\frac{I^2}4[(\overleftarrow Q{}^+)^2(\overrightarrow Q{}^-)^2
+(\overleftarrow Q{}^-)^2(\overrightarrow Q{}^+)^2]-I^2
\overleftarrow Q{}^{+\alpha}\overleftarrow Q{}^{-\beta}
\overrightarrow Q{}^-_\alpha \overrightarrow Q{}^+_\beta,\nonumber\\
\frac16(P_s)^3&=&-\frac{I^3}3[
(\overleftarrow Q{}^-)^2\overleftarrow Q{}^{+\alpha}
(\overrightarrow Q{}^+)^2\overrightarrow Q{}^-_\alpha
-(\overleftarrow Q{}^+)^2\overleftarrow Q{}^{-\alpha}
(\overrightarrow Q{}^-)^2\overrightarrow Q{}^+_\alpha],\nonumber\\
\frac1{24}(P_s)^4&=&\frac{I^4}{16}(\overleftarrow Q{}^-)^2
(\overleftarrow Q{}^+)^2(\overrightarrow Q{}^+)^2
(\overrightarrow Q{}^+)^2.
\label{Ps2}
\end{eqnarray}
Note that $P_s$ commutes with the spinor derivatives in the analytic basis (they are
defined in (\ref{deriv})),
\be
[P_s,D^\pm_\alpha]=0,\qquad
[P_s,\bar D^\pm_{\dot\alpha}]=0.
\label{comm_1}
\ee
The property (\ref{comm_1}) shows that the $\star$-product (\ref{sstar1})
preserves the harmonic Grassmann analyticity. In other words, the
$\star$-product of two analytic superfields $\Phi_A$, $\Psi_A$ is
again an analytic superfield,
\be
(D^\pm_\alpha,~\bar D^\pm_{\dot\alpha})(\Phi_A\star\Psi_A)=0.
\ee

Due to the simple relation between the supercharges and
covariant spinor derivatives
\be
Q^\pm_\alpha=D^\pm_\alpha+2{\rm i}\bar\theta^{\pm\dot\alpha}
(\sigma_m)_{\alpha\dot\alpha}\partial_m,
\label{rel1}
\ee
the following relations are valid for an arbitrary analytic superfield
$\Phi_A$
\be
Q^+_\alpha\Phi_A=2{\rm i}\bar\theta^{+\dot\alpha}(\sigma_m)_{\alpha\dot\alpha}
\partial_m\Phi_A,\qquad
(Q^+)^2\Phi_A=4(\bar\theta^+)^2\square\Phi_A.
\label{rel_2}
\ee
The equations (\ref{rel_2}) imply that in the decomposition
of the $\star$-product (\ref{sstar1}) any term involving more than
two $Q^+_\alpha$ supercharges on the analytic superfields vanishes, e.g.,
\be
(Q^+)^2\Phi_A Q^+_\alpha \Psi_A=4{\rm i}(\bar\theta^+)^2\square\Phi_A
\bar\theta^{+\dot\alpha}(\sigma_m)_{\alpha\dot\alpha}\partial_m\Psi_A=0.
\label{rel2}
\ee
As a consequence, the singlet $\star$-product of two analytic
superfields is at most quadratic in the deformation
parameter $I$:
\begin{eqnarray}
&&\Phi_A\star \Psi_A
=\Phi_A\Psi_A+I(-1)^{p(\Phi)}(Q^{+\alpha}\Phi_A Q^-_\alpha\Psi_A-
Q^{-\alpha}\Phi_A Q^+_\alpha\Psi_A)\nonumber\\&&
-\frac{I^2}4[(Q^+)^2\Phi_A(Q^-)^2\Psi_A+(Q^-)^2\Phi_A(Q^+)^2\Psi_A]
-I^2Q^{+\alpha}Q^{-\beta}\Phi_A Q^-_\alpha Q^+_\beta\Psi_A.
\label{star-anal}
\end{eqnarray}
Then it is easy to see that the $\star$-commutator of
analytic superfields is linear in $I$
\begin{eqnarray}
[\Phi_A\cs\Psi_A]&=&\Phi_A P_s \Psi_A-\Psi_A P_s \Phi_A=2\Phi_A P_s \Psi_A
\nonumber\\&
=&2I(Q^{+\alpha}\Phi_A Q^-_\alpha\Psi_A-
Q^{-\alpha}\Phi_A Q^+_\alpha\Psi_A).
\label{comm-anal}
\end{eqnarray}

The operator of chiral singlet deformations
(\ref{Ps}) also commutes with the harmonic derivatives
(\ref{harm-deriv}),
\be
[P_s,D^{++}]=0,\qquad [P_s,D^{--}]=0.
\label{comm-harm}
\ee
As a result, the chiral
singlet deformation does not break the internal symmetry group
${\rm SU}(2)$ represented by the harmonics $u^\pm_i\,$.
It also preserves the Grassmann shortness conditions,
 $D^{\pm\pm}\Phi = 0\,$. This makes it possible to utilize short
multiplets while constructing the actions, like in the
undeformed theories.

The properties of the chiral singlet deformation listed above
(the preservations of left and right chiralities, as well as of the Grassmann
analyticity and Grassmann shortness) indicate that the
the harmonic superspace approach is equally applicable to the
$\cN{=}(1,1)$ non-anticommutative superfield theories, as to the
conventional $\cN{=}2$ supersymmetric ones.

\section{Classical non-anticommutative models in ${\cal N}{=} (1,1)$ harmonic
superspace}
In constructing classical superfield actions of non-anticommutative
theories we follow the simple rule: in order to obtain the action of a non-anticommutative
model one should replace the usual product of superfields in the action of the corresponding
undeformed model by the $\star$-product (\ref{sstar}).

\subsection{Super-Yang-Mills model}
In the harmonic superspace approach \cite{Book} the gauge
multiplet of $\cN{=}2$ or $\cN{=}(1,1)$ supersymmetry is described
by the analytic superfield $V^{++}$ with the harmonic ${\rm U}(1)$ charge $+2$.
In general, this superfield is valued in the Lie algebra of the
gauge group ${\rm U}(n)$, i.e., it can be written as
$V^{++}=V^{++\,M}T^M$, where $T^M$ are the generators of ${\rm
U}(n)$.

Under the deformed ${\rm U}(n)$ gauge group
the gauge superfield is assumed to transform as
\be
\delta_\Lambda V^{++}=D^{++}\Lambda+[V^{++}\cs
\Lambda], \label{gauge-tr}
\ee
where $\Lambda$ is an analytic superfield parameter also taking values in the algebra of the gauge group.
Note that even in the ${\rm U}(1)$ case, i.e. with  only one copy of  $V^{++}$ and $\Lambda$, the transformation rule
(\ref{gauge-tr}) is still non-abelian due to the presence of the $\star$ product
in the second term. In the undeformed limit this ``non-abelian'' piece vanishes.
In the case of ``genuine'' non-abelian $\cN{=}(1,1)$ gauge theory there are two sources of the non-abelian structure,
the standard one surviving in the undeformed limit and the one induced by the $\star$-product.

To construct an action which is invariant under the gauge transformations
(\ref{gauge-tr}) representing the deformed gauge ${\rm U}(n)$ group we follow the same steps as in the non-abelian
$\cN{=}2$ super Yang-Mills theory in harmonic superspace
\cite{HSS,Book}. We introduce the superfield $V^{--}$ as a
solution of the harmonic zero-curvature equation,
\be
D^{++}V^{--}-D^{--}V^{++}+[V^{++}\cs V^{--}]=0.
\label{zero-curv}
\ee
The solution of (\ref{zero-curv}) is given by the
following series \cite{Zu1}
\be
V^{--}(z,u)=\sum_{n=1}^\infty
(-1)^n\int du_1\ldots du_n \frac{V^{++}(z,u_1)\star
V^{++}(z,u_2)\star\ldots\star
V^{++}(z,u_n)}{(u^+u^+_1)(u^+_1u^+_2)\ldots(u^+_n u^+)},
\label{V--}
\ee
where $(u^+_1 u^+_2)^{-1}$ is a harmonic
distribution introduced in \cite{HSS}. Using the superfield
$V^{--}\,$, one can construct the gauge superfield strengths in the standard
manner,
\be
W=-\frac14(\bar D^+)^2 V^{--},\qquad \bar
W=-\frac14(D^+)^2 V^{--}.
\label{W}
\ee
As in the usual $\cN{=}2$
SYM theory, these superfields satisfy the Bianchi identity
$(D^+)^2W=(\bar D^+)^2\bar{W}$. Applying the relations
(\ref{gauge-tr}), (\ref{zero-curv}) and the gauge transformation
rule for the $V^{--}$ prepotential, $\delta_\Lambda
V^{--}=D^{--}\Lambda+[V^{--}\cs \Lambda]\,$, it is easy to show
that the superfield strengths (\ref{W}) transform covariantly
under the gauge group,
\be
\delta_\Lambda W=[W\cs\Lambda],\qquad
\delta_\Lambda \bar W= [\bar W\cs\Lambda].
\label{deltaW}
\ee
Moreover, they are covariantly (anti)chiral
\begin{eqnarray}
\bar D^+_{\dot\alpha}W&=&0,\qquad \bar D^-_{\dot\alpha}W-
[\bar D^+_{\dot\alpha}V^{--}\cs W]=0,\nonumber\\
D^+_\alpha\bar W&=&0,\qquad  D^-_\alpha\bar W-[D^+_\alpha V^{--}\cs \bar
W]=0,
\label{cov-chir}
\end{eqnarray}
and are covariantly independent of harmonics,
\be
D^{++}W+[V^{++}\cs W]=0,\qquad
D^{++}\bar W+[V^{++}\cs \bar W]=0.
\label{cov-harm}
\ee

The equations (\ref{gauge-tr})--(\ref{cov-harm}) have exactly the
same form as in the corresponding undeformed non-abelian $\cN{=}2$
super Yang-Mills theory. Therefore, the classical action of the
non-an\-ti\-com\-mu\-ta\-ti\-ve supersymmetric gauge theory can be
also represented as an integral over the chiral subspace
\be
S_{SYM}=\frac14\tr\int d^4x_L
d^4\theta \, W^2.
\label{SYM}
\ee
Note that, due to the property (\ref{int1}), the $\star$-product of two
superfield strengths in (\ref{SYM}) is reduced to the ordinary product. However, despite
the absence of the $\star$-product in (\ref{SYM}), the
non-anticommutative deformation is still present in this expression through
the superfield strengths (\ref{W}) and the prepotential $V^{--}$ (\ref{V--}).
It is easy to check that the action (\ref{SYM}) is gauge invariant,
\be
\delta_\Lambda
S_{SYM}=\frac14\tr \int d^4x_L d^4\theta \,[W^2\cs \Lambda]=0.
\label{deltaS}
\ee
Here we have applied the equations
(\ref{int1}), (\ref{deltaW}). One can also show that this action
does not depend on the harmonic and Grassmann variables,
\be
D^{++}S_{SYM}=0,\qquad \bar D^\pm_{\dot\alpha}S_{SYM}=0.
\label{DS}
\ee
It should also be noted that the chiral action
(\ref{SYM}) is real in the Euclidean case.

The classical action of non-anticommutative supersymmetric gauge
theory can be expressed as a full superspace integral of the
Lagrangian written in terms of the analytic superfields $V^{++}$,
quite analogously to the action of the usual non-abelian
$\cN{=}2$ gauge theory \cite{Zu1},
\be
S_{SYM}[V^{++}]= \sum_{n=2}^\infty\frac{(-1)^n}{n} \tr \int d^{12}z
du_1\ldots du_n\frac{ V^{++}(z,u_1)\star
V^{++}(z,u_2)\star\ldots\star V^{++}(z,u_n) }{(u^+_1 u^+_2)(u^+_2
u^+_3)\ldots (u^+_n u^+_1)}.
\label{SYM1}
\ee
While passing to the quantum theory, the representation (\ref{SYM1}) for the classical action
proves to be more advantageous.

\subsection{Hypermultiplet model}
The hypermultiplet in harmonic superspace is described either by a
complex analytic superfield $q^+$ with the ${\rm
U}(1)$ charge $+1$ or by a real analytic chargeless superfield  $\omega$.
Both these descriptions are known to be related to each other via some sort of duality \cite{Book}.
Therefore we can confine our consideration to the  $q$-hypermultiplet models.

The free classical action  of the $q^+$ superfield in harmonic
superspace is given by
\be
S_0[q^+]=-\int d\zeta du\,\tilde q^+D^{++}q^+.
\label{Sq0}
\ee
Here $\tilde q^+$ is a superfield conjugated to $q^+$ and $d\zeta du=d^4x_A
d^4\theta^- du$ is the integration measure of the analytic superspace. The
rules of integration in harmonic superspace are given in the
Appendix 2, eq. (\ref{int-def}). Note that the identity
(\ref{int1}) allows us to omit the $\star$-product in the free
superfield actions like (\ref{Sq0}). In other words, the
chiral singlet deformation does not modify the free
actions and affects only the interaction terms. We will show that
 both the hypermultiplet self-interaction and the
interaction of hypermultiplet with a vector multiplet are deformed
due to non-anticommutativity. In some special cases considered below
this interaction disappears when the deformation is turned off.

It is easy to write the quartic interaction term of the $q$-superfields
\cite{singlet},
\be
S_4[q^+]=\int d\zeta du(a\,\tilde q^+\star q^+\star \tilde q^+\star q^+
+b\, q^+\star q^+\star\tilde q^+\star\tilde q^+),
\label{Sq4}
\ee
where $a$, $b$ are coupling constants. Note that two terms in the
action (\ref{Sq4}) differ only by ordering of superfields with
respect to the $\star$-product. In the undeformed limit $I\to0$
both these terms are reduced to the single standard interaction term $(\tilde q^+
q^+)^2$, with the coupling constant $a+b$.

Let us now introduce the interaction of hypermultiplet with the
background gauge superfields.

As is well known, the interaction of matter fields with the gauge
ones is to large extent specified by the choice of the representation of the gauge
group to which matter fields belong. In particular, the fundamental and adjoint representations
are of the main interest in quantum field theory.

Let us start with the fundamental representation. In this case the superfield $V^{++}$ is a matrix which
belongs to the Lie algebra of the gauge group ${\rm U}(n)$ acting on the complex $n$-plet of superfields $q^+$.

Based on the analogy with the ordinary  ${\rm U}(n)$ gauge theory, the model (\ref{Sq0}) can be coupled to the
gauge superfield in the standard way, i.e. just by replacing the flat harmonic derivative
$D^{++}$ with the corresponding covariant one
$\nabla^{++}=D^{++}+V^{++}\star $. As a result, the action of the
non-anticommutative hypermultiplet superfield interacting with the
vector superfield in the fundamental representation of the deformed  ${\rm U}(n)$ group is given by
\be
S_f[q^+,V^{++}]=-\int d\zeta du\, \tilde
q^+\star(D^{++}+V^{++})\star q^+.
\label{Sqf}
\ee
Here $\tilde q^+$ is conjugated to $q^+$.
It is easy to check that the action (\ref{Sqf}) is invariant under
the gauge transformations of vector superfield (\ref{gauge-tr})
supplemented by the following hypermultiplet transformations
\be
\delta_\Lambda \tilde q^+=\tilde q^+\star\Lambda,\qquad
\delta_\Lambda q^+=\Lambda\star q^+.
\label{gauge-tr1}
\ee
We refer to the model (\ref{Sqf}) as a non-anticommutative model
of charged hypermultiplet \cite{ILZ}. It should be emphasized that
the transformation laws (\ref{gauge-tr1}) are essentially non-abelian (they possess a non-zero Lie
bracket) even in the ${\rm U}(1)$ case. They become the standard ${\rm U}(1)$ transformations only in
the undeformed limit, when the $\star$-product turns into the ordinary one.

In the adjoint representation the hypermultiplet superfield is transformed
on pattern of the second term in the transformation law
(\ref{gauge-tr})
\be
\delta_\Lambda \tilde q^+=[\tilde q^+\cs \Lambda],\qquad
\delta_\Lambda q^+=[q^+\cs\Lambda].
\label{gauge-tr2}
\ee
Here $q^+$ is a matrix in the Lie algebra of the gauge group, and it can be
expanded over the generators of the gauge group as $q^+=q^{+ M}T^M\,$.
The corresponding classical action is given by
\be
S_{ad}[q^+,V^{++}]=-\tr\int d\zeta du\,
\tilde q^+\star(D^{++}q^++[V^{++}\cs q^+]).
\label{Sqad}
\ee
It is easy to check that (\ref{Sqad})
is invariant under the gauge transformations (\ref{gauge-tr})
supplemented by (\ref{gauge-tr2}).

We refer to the model with the classical action (\ref{Sqad}) and
deformed gauge group U(1) as the non-an\-ti\-com\-mu\-ta\-ti\-ve model of neutral
hypermultiplet. It is worth noting that in the case of U(1)
gauge group the interaction with the gauge superfield in
(\ref{Sqad}) is only due to the non-anticommutative
deformation. This interaction disappears in the limit $I\to0$ and
the model (\ref{Sqad}) becomes free. This is a new feature
specific only for the non-anticommutative neutral hypermultiplet
model with the U$(1)$ gauge group. The interaction
still survives in the limit $I\to0$  for the non-abelian neutral
hypermultiplet or for the charged hypermultiplet (even with
the U$(1)$ gauge group). In our further consideration we restrict
ourselves only to the models with deformed U$(1)$ gauge group.

It is instructive to rewrite the actions
(\ref{Sqf}), (\ref{Sqad}) in a unified form.
For this purpose we combine the hypermultiplet superfields
$\tilde q^+$, $q^+$ into a single ${\rm SU}(2)$ doublet $q^{+a}$,
\be
q^{+a}=\varepsilon^{ab}q_b^+=(\tilde
q^+,q^+)=\widetilde{q^+_a},\qquad
a=1,2.
\label{dublet}
\ee
The covariant harmonic derivative $\nabla^{++}$ acts on the
doublet $q^{+a}$ in a different way for the adjoint and fundamental
representations of the U(1) gauge group,
\begin{eqnarray}
\mbox{Adj. rep.:}&&\nabla^{++}q^{+a}=D^{++}q^{+a}+
[V^{++}\cs q^{+a}],\label{nabla-ad}\\
\mbox{Fund. rep.:}&&\nabla^{++}q^{+a}=D^{++}q^{+a}
+\frac12[V^{++}\cs q^{+a}]-\frac12(\tau_3)_b^a
\{V^{++}\cs q^{+b} \}.
\label{nabla-f}
\end{eqnarray}
Here $\tau_3={\rm diag}(1,-1)$ is the Pauli matrix.
According to the definition (\ref{nabla-ad}), the expression
$\nabla^{++}q^{+a}$ is covariant with respect to the additional
symmetry group ${\rm SU}(2)_{PG}$ which is called the Pauli-G\"ursey
group \cite{Book}. The matrices of this group act on the index
$a$ of $\nabla^{++}q^{+a}$. Using the new notation, the
actions (\ref{Sqf}), (\ref{Sqad}) can be uniformly written as
\be
S[q^+,V^{++}]=\frac12\int d\zeta du\,q^+_a\nabla^{++}q^{+a}.
\label{Sq-pg}
\ee
In the case of fundamental representation, the symmetry group
${\rm SU}(2)_{PG}$ is broken down to U(1) with the generator
$\tau_3$.

\section{The component structure of ${\cal N}{=} (1,0)$ non-an\-ti\-com\-mu\-ta\-ti\-ve
abelian models}
In the previous section we have shown that in the superfield
Lagrangians the chiral singlet deformation leads to some new
interaction terms induced by the $\star$-product.
It is important that this new interaction is always local owing to the
nilpotency of the operator $P_s$. Here we study these new
interaction terms at the component level. The most important
features of such Lagrangians can be most clearly exhibited on the examples of
abelian models of gauge superfield and hypermultiplet.

\subsection{Gauge superfield model}
\label{SYMsect} The gauge multiplet of $\cN{=}(1,1)$ supersymmetry
consists of two independent real scalar fields $\phi$, $\bar\phi$,
independent Weyl spinors $\Psi^k_\alpha$, $\bar\Psi^{\dot\alpha
k}$ with the internal symmetry group index $k=1,2$ and a triplet
of auxiliary fields ${\cal D}^{(kl)}$. The component structure of
the $\cN{=}(1,0)$ non-anticommutative abelian supergauge model in
terms of these fields was studied in \cite{FILSZ,Araki}.

The classical action of non-anticommutative super Yang-Mills model is given
by (\ref{SYM}). In the abelian case we can omit the trace in (\ref{SYM}),
\be
S_{SYM}=\frac14\int d^4x_L d^4\theta\, W^2.
\label{ab-SYM}
\ee
Note that, according to (\ref{cov-chir}), the superfield $W$ is
covariantly chiral rather than manifestly chiral. Therefore it depends on
the variables $\bar\theta^+_{\dot\alpha}$:
\be
W={\cal A}+\bar\theta^+_{\dot\alpha}\tau^{-\dot\alpha}+
(\bar\theta^+)^2\tau^{-2},
\label{W-expand}
\ee
where $\cal A$, $\tau^{-\dot\alpha}$, $\tau^{-2}$ are some chiral
superfields. Remarkably, among these superfields only
$\cal A$ contributes to the action (\ref{ab-SYM}).
Indeed, the relations (\ref{cov-chir}) and (\ref{cov-harm})
show that the terms involving the superfields $\tau^{-\dot\alpha}$,
$\tau^{-2}$ are always proportional to some $\star$-commutators of
superfields and therefore vanish under the integral over $d^4\theta$.
As a result, the action (\ref{ab-SYM}) acquires the
following form
\be
S_{SYM}=\frac14\int d^4x_L d^4\theta\, {\cal A}^2.
\label{ab-SYM1}
\ee

Let us find the component structure of the superfield $\cal A$.
For this purpose we have to fix the component structure of the gauge
superfield $V^{++}\,$. Using the gauge freedom
(\ref{gauge-tr}) one can eliminate the lowest components of $V^{++}$
by effecting the Wess-Zumino gauge,
\begin{eqnarray}
V^{++}_{WZ}(x^m_A,\theta^{+\alpha},\bar\theta^{+\dot\alpha},u)&=&(\theta^+)^2\bar\phi(x_A)+
(\bar\theta^+)^2\phi(x_A)+2(\theta^+\sigma_m\bar\theta^+)A_m(x_A)
+4(\bar\theta^+)^2\theta^+\Psi^-(x_A)\nonumber\\
&&+4(\theta^+)^2\bar\theta^+\bar\Psi^-(x_A)
+3(\theta^+)^2(\bar\theta^+)^2{\cal D}^{--}(x_A),
\label{Vwz}
\end{eqnarray}
where
\be
\Psi^-_\alpha(x_A)=\Psi^k_\alpha(x_A)u^-_k,\qquad
\bar\Psi^{\dot\alpha-}(x_A)=\bar\Psi^{\dot\alpha k}(x_A)u^-_k,
\qquad {\cal D}^{--}={\cal D}^{kl}(x_A)u^-_k u^-_l.
\label{redif1}
\ee
The residual gauge transformation of the superfield (\ref{Vwz}) reads
\be
\delta_r V^{++}_{WZ}=D^{++}\Lambda_r+[V^{++}\cs\Lambda_r],\qquad
\Lambda_r={\rm i} \lambda(x_A),
\label{residual}
\ee
where $\lambda(x_A)$ is an arbitrary real function. The
transformation (\ref{residual}) amounts to the following gauge
transformations for the component fields
\be
\begin{array}{ll}
\delta\phi=-8IA_m\partial_m \lambda,  &\delta\bar\phi=0,\\
\delta\Psi^k_\alpha=-4I(\sigma_m\bar\Psi^k)_\alpha\partial_m\lambda,\quad
& \delta \bar\Psi^k_{\dot\alpha}=0,\\
\delta A_m=(1+4I\bar\phi)\partial_m\lambda,&
\delta {\cal D}^{kl}=0.
\end{array}
\label{resid-comp}
\ee
As is seen from (\ref{resid-comp}), the gauge transformations of
fields $\phi$, $A_m$, $\Psi_\alpha^k$ are deformed due to the
non-anticommutativity. In the limit $I=0$ we are left with the
standard abelian gauge transformation for the vector potential
$A_m(x)\,$, $\delta A_m(x) = \partial_m\lambda(x)\,$.

The chiral coordinates are best suited for the chiral
singlet deformation since the latter preserves chirality. In what follows
we pass from the analytic coordinates
$\{ x^m_A,\theta^\pm_\alpha,\bar\theta^\pm_{\dot\alpha}\}$ to the
mixed chiral-analytic ones $\{z_C=(x^m_L,\theta^\pm_\alpha)$,
$\bar\theta^{\pm}_{\dot\alpha}\}$ by the rule
\be
x^m_A=x^m_L-2{\rm i}\theta^-\sigma^m\bar\theta^+.
\label{change-var}
\ee
For example, in the chiral-analytic basis the operator of chiral singlet
deformations (\ref{Ps}) is simplified to the form,
\be
P_s=I(\overleftarrow{\partial}{}^\alpha_+\overrightarrow{\partial}_{-\alpha}
-\overleftarrow{\partial}{}^\alpha_-\overrightarrow{\partial}_{+\alpha}),
\label{Ps-chir}
\ee
where $\partial_{\pm\alpha}=\partial/\partial\theta^{\pm\alpha}$.
Let us also rewrite the component structure of the prepotential
(\ref{Vwz}) in these coordinates,
\be
V^{++}_{WZ}(z_C,\bar\theta^+,u)=
v^{++}(z_C,u)+\bar\theta^+_{\dot\alpha}v^{+\dot\alpha}(z_C,u)
+(\bar\theta^+)^2 v(z_C,u)\,.
\label{Vwz-chir}
\ee
Here
\begin{eqnarray}
v^{++}&=&(\theta^+)^2\bar\phi,\nonumber\\
v^{+\dot\alpha}&=&-2\theta^+_\alpha A^{\alpha\dot\alpha}
+4(\theta^+)^2\bar\Psi^{-\dot\alpha}+2{\rm i}(\theta^+)^2
\theta^-_\alpha\partial^{\alpha\dot\alpha}\bar\phi,\nonumber\\
v&=&\phi+4\theta^+\Psi^-+3(\theta^+)^2{\cal D}^{--}
-2{\rm i}(\theta^+\theta^-)\partial_m A_m
-\theta^-\sigma_{mn}\theta^+ F_{mn}\nonumber\\&&
-(\theta^+)^2(\theta^-)^2\square\bar\phi
+4{\rm i}(\theta^+)^2\theta^-\sigma_m\partial_m\bar\Psi^-
\label{v}
\end{eqnarray}
and $F_{mn}=\partial_m A_n-\partial_n A_m$.

Consider now the zero-curvature equation (\ref{zero-curv}),
\be
D^{++}V^{--}-D^{--}V^{++}_{WZ}+[V^{++}_{WZ}\cs V^{--}]=0.
\label{zero-curv-wz}
\ee
Developing the $\star$-product in (\ref{zero-curv-wz}) and applying
(\ref{Ps-chir}), we have
\begin{eqnarray}
&&D^{++}V^{--}-D^{--}V^{++}_{WZ}+2I
(\partial^\alpha_+ V^{++}_{WZ}\partial_{-\alpha}V^{--}
-\partial^\alpha_- V^{++}_{WZ}\partial_{+\alpha}V^{--})\nonumber\\&&
+\frac12I^3[\partial^\alpha_-
(\partial_+)^2 V^{++}_{WZ}\partial_{+\alpha}(\partial_-)^2V^{--}
-\partial_+^\alpha(\partial_-)^2V^{++}_{WZ}\partial_{-\alpha}(\partial_+)^2V^{--}]=0.
\label{zero-curv1}
\end{eqnarray}
We seek for a solution of the equation (\ref{zero-curv1}) as an
expansion over $\bar\theta^\pm_{\dot\alpha}$,
\begin{eqnarray}
V^{--}&=&v^{--}+\bar\theta^+_{\dot\alpha}v^{-3\dot\alpha}+\bar\theta^-_{\dot\alpha}
v^{-\dot\alpha}+(\bar\theta^+)^2v^{-4}+(\bar\theta^-)^2{\cal A}
+(\bar\theta^+\bar\theta^-)\varphi^{--}\nonumber\\
&&+\bar\theta^{+\dot\alpha}\bar\theta^{-\dot\beta}\varphi^{-2}_{(\dot\alpha\dot\beta)}
+(\bar\theta^-)^2\bar\theta^+_{\dot\alpha}\tau^{-\dot\alpha}
+(\bar\theta^+)^2\bar\theta^-_{\dot\alpha}\tau^{-3\dot\alpha}
+(\bar\theta^+)^2(\bar\theta^-)^2\tau^{-2},
\label{V--1}
\end{eqnarray}
where $v^{--}$, $v^{-3\dot\alpha}$, $v^{-4}$, $\cal A$,
$\varphi^{--}$, $\varphi^{-2}_{(\dot\alpha\dot\beta)}$, $\tau^{-\dot\alpha}$,
$\tau^{-3\dot\alpha}$, $\tau^{-2}$ are the superfields depending
only on the chiral-analytic variables $x_L^m,\theta^\pm_\alpha,
 u^\pm_k$. Note that the superfield
$\cal A$ that defines the classical SYM action (\ref{ab-SYM1})
appears as one of the components in the expansion (\ref{V--1}).
Now we substitute the expressions (\ref{V--1}), (\ref{Vwz-chir}) into (\ref{zero-curv1})
and equate to zero the coefficients at the corresponding powers of
$\bar\theta^\pm_{\dot\alpha}\,$. In this way we obtain the following set of equations:
\begin{eqnarray}
{\bf D}^{++}v^{--}
-D^{--}v^{++}&=&0\,,\label{eq1_}\\
{\bf D}^{++}v^{-\dot\alpha}
-v^{+\dot\alpha}&=&0\,,\label{eq2_}\\
{\bf D}^{++} v^{-3\dot\alpha}+
v^{-\dot\alpha}-{\bf D}^{--}v^{+\dot\alpha}&=&0,\label{eq3_}\\
{\bf D}^{++}{\cal A}&=&0\,,
\label{eq4_}\\
{\bf D}^{++}\varphi^{--}+2{\cal A}-2v+\frac12\{v^{+\dot\alpha}\cs v^{-}_{\dot\alpha}
 \}&=&0\,,
\label{eq_5}\\
{\bf D}^{++}v^{-4}-{\bf D}^{--}v+\varphi^{--}
+\frac12\{v^{+\dot\alpha}\cs v^{-3}_{\dot\alpha} \}&=&0\,,
\label{eq6_}\\
{\bf D}^{++}\varphi^{-2}_{(\dot\alpha\dot\beta)}
+\frac12\{v^+_{\dot\alpha}\cs v^-_{\dot\beta} \}
+\frac 12\{v^+_{\dot\beta}\cs v^-_{\dot\alpha} \}&=&0\,,
\label{eq_7}\\
{\bf D}^{++}\tau^{-\dot\alpha}+[v^{+\dot\alpha}\cs {\cal A}]&=&0\,,
\label{eq_8}\\
{\bf D}^{++}\tau^{-3\dot\alpha}-\tau^{-\dot\alpha}
+[v\cs v^{-\dot\alpha}]
-\frac12[v^{+\dot\alpha}\cs \varphi^{--}]
+\frac12[v^+_{\dot\beta}\cs
\varphi^{-2(\dot\alpha\dot\beta)}]&=&0\,,
\label{eq_9}\\
{\bf D}^{++}\tau^{-2}+\frac12\{v^{+\dot\alpha}\cs\tau^-_{\dot\alpha} \}
+[v\cs {\cal A}]&=&0.
\label{eq_10}
\end{eqnarray}
Here we used the notations
\begin{eqnarray}
{\bf D}^{++}&=&D^{++}+[v^{++}\cs \cdot ]
 =u^+_i \frac\partial{\partial u^-_i}
 +L\,\theta^{+\alpha}\partial_{-\alpha}\,,
\label{defin1}\\
{\bf D}^{--}&=&D^{--}+[v^{--}\cs \cdot]
 =u^-_i\frac\partial{\partial u^+_i}
 +\frac1L\theta^{-\alpha}\partial_{+\alpha}\,,
\label{defin2}\\
L&=&1+4I\bar\phi\,.
\label{defin3}
\end{eqnarray}
It is straightforward (though somewhat lengthy) to find the
solutions of (\ref{eq1_})--({\ref{eq_10}),
\begin{eqnarray}
v^{--}(z_C,u)&=&(\theta^-)^2\frac{\bar\phi}L,\label{eq6__}\\
v^{-3\dot\alpha}(z_C,u)&=&2(\theta^-)^2\frac{\bar\Psi^{-\dot\alpha}}{L^2},
\label{eq7_}\\
v^-_{\dot\alpha}(z_C,u)&=&\frac{2}{L}\theta^{-\alpha}A_{\alpha\dot\alpha}
-\frac{2}{L^2}(\theta^-)^2\bar\Psi^+_{\dot\alpha}
+\frac{4}{L}(\theta^+\theta^-)\bar\Psi^-_{\dot\alpha}
+\frac{2{\rm i}}{L}(\theta^-)^2\theta^{+\alpha}\partial_{\alpha\dot\alpha}\bar\phi,\label{eq8_}\\
{\cal A}(z_C,u)&=&\displaystyle[\phi+\frac{4IA_mA_m}{L}
+\frac{16I^3(\partial_m\bar\phi)^2}{L}]
\nonumber\\&&
+2\theta^+[\Psi^-+\frac{4I(\sigma_m\bar\Psi^-)A_m}{L}]
-\frac{2\theta^-}{L}[\Psi^++\frac{4I(\sigma_m\bar\Psi^+)A_m}{L}]
\nonumber\\&&
+(\theta^+)^2[\frac{8I(\bar\Psi^-)^2}{L}+{\cal D}^{--}]
+\frac{(\theta^-)^2}{L^2}[\frac{8I(\bar\Psi^+)^2}{L}+{\cal
D}^{++}]
\nonumber\\&&
-\frac{2(\theta^+\theta^-)}{L}[\frac{8I(\bar\Psi^+\bar\Psi^-)}{L}
+{\cal D}^{+-}]
+(\theta^+\sigma_{mn}\theta^-)(F_{mn}-\frac{8I\partial_{[m}\bar\phi A_{n]}}{
L})\nonumber\\&&\displaystyle
+2{\rm i}(\theta^-)^2\theta^+\sigma_m\partial_m\frac{\bar\Psi^+}{L}
+2{\rm i}(\theta^+)^2L\theta^-\sigma_m\partial_m\frac{\bar\Psi^-}{L}
-(\theta^+)^2(\theta^-)^2\square\bar\phi\,,
\label{A}\\
\varphi^{--}(z_C,u)&=&\frac4L\theta^{-\alpha}\Psi^-_\alpha
+\frac{8I}{L^2}\theta^{-\alpha}A_{\alpha\dot\alpha}\bar\Psi^{-\dot\alpha}
+(\theta^+\theta^-)[\frac4L{\cal D}^{--}+\frac{16I}{L^2}
\bar\Psi^-_{\dot\alpha}\bar\Psi^{-\dot\alpha}]
\nonumber\\&&
-(\theta^-)^2[\frac{2{\rm i}}L\partial_m A_m
+\frac2{L^2}{\cal D}^{+-}+\frac{16I}{L^3}
\bar\Psi^-_{\dot\alpha}\bar\Psi^{+\dot\alpha}]
\nonumber\\&&
-4{\rm i}(\theta^-)^2\theta^{+\alpha}\frac1L
 [\partial_{\alpha\dot\alpha}\bar\Psi^{-\dot\alpha}-\frac{2I}{L}
 \bar\Psi^{-\dot\alpha}\partial_{\alpha\dot\alpha}\bar\phi]\,,
\label{eq9_}\\
v^{-4}(z_C,u)&=&(\theta^-)^2[\frac2{L^2}{\cal D}^{--}+\frac{16I}{L^3}
\bar\Psi^-_{\dot\alpha}\bar\Psi^{-\dot\alpha}]\,,\\
\varphi^{-2}_{(\dot\alpha\dot\beta)}(z_C,u)&=&
-\frac12\{v^-_{\dot\alpha}\cs v^-_{\dot\beta} \}=
8IL^{-2}(\theta^{-\alpha}\bar\Psi^-_{\dot\alpha}A_{\alpha\dot\beta}
+\theta^{-\alpha}\bar\Psi^-_{\dot\beta}A_{\alpha\dot\alpha})
\nonumber\\&&
+4{\rm i} IL^{-2}(\theta^-)^2(\partial^\alpha_{\dot\alpha}\bar\phi A_{\alpha\dot\beta}
+\partial^\alpha_{\dot\beta}\bar\phi A_{\alpha\dot\alpha})
-16IL^{-3}(\theta^-)^2(\bar\Psi^+_{\dot\beta}\bar\Psi^-_{\dot\alpha}
+\bar\Psi^+_{\dot\alpha}\bar\Psi^-_{\dot\beta})\nonumber\\&&
+8{\rm i} IL^{-2}(\theta^-)^2\theta^{+\alpha}(\bar\Psi^-_{\dot\alpha}
\partial_{\alpha\dot\beta}\bar\psi+
\bar\Psi^-_{\dot\beta}\partial_{\alpha\dot\alpha}\bar\phi)\,,
\label{eq10_}\\
\tau^{-\dot\alpha}&=&[{\cal A}\cs v^{-\dot\alpha}]\,,\label{eq10__}\\
\tau^{-3\dot\alpha}&=&-{\bf D}^{--}\tau^{-\dot\alpha}
+\frac12[v^{-\dot\alpha}\cs \varphi^{--}]
-\frac12[v^-_{\dot\beta}\cs
\varphi^{-2(\dot\alpha\dot\beta)}]\,,\label{eq11_}\\
\tau^{-2}&=&-\frac12[\varphi^{--}\cs {\cal A}]-\frac14
\{v^{-\dot\alpha}\cs[{\cal A}\cs v^-_{\dot\alpha}] \}\,.\label{eq12_}
\end{eqnarray}
The expressions (\ref{eq10__})--(\ref{eq12_}) are presented in a
superfield form since their exact component structure is of no importance for out further
consideration.

Now we use the expression (\ref{A}) to find
the component structure of the classical action
(\ref{ab-SYM1}):
\begin{eqnarray}
S_{SYM}&=&S_\phi+S_{\Psi}+S_A,\label{s1.6}\\
S_\phi&=&-\frac12\int d^4x\square\bar\phi\left[\phi+\frac{4IA_mA_m}{1+4I\bar\phi}
+\frac{16I^3\partial_m\bar\phi\partial_m\bar\phi}{1+4I\bar\phi}\right],\label{s1.7}\\
S_\Psi&=&{\rm i}\int d^4x\left(
\Psi^{i\alpha}+\frac{4IA_m\sigma_m{}^{\alpha}{}_{\dot\alpha}\bar\Psi^{i\dot\alpha}}{
1+4I\bar\phi}
\right)(\sigma_n)_{\alpha\dot\beta}\partial_n
\left(\frac{\bar\Psi_i^{\dot\beta}}{1+4I\bar\phi} \right)
\nonumber\\
&&+\frac14\int d^4x
\frac1{(1+4I\bar\phi)^2}
\left(\frac{8I\bar\Psi^i_{\dot\alpha}\bar\Psi^{j\dot\alpha}}{1+4I\bar\phi}
+{\cal D}^{ij}\right)\left(\frac{8I\bar\Psi_{i\dot\alpha}\bar\Psi_j^{\dot\alpha}}{1+4I\bar\phi}
+{\cal D}_{ij}\right),\label{s1.8}\\
S_A&=&\int d^4x\left[-\frac12 A_n\square A_n
-\frac12\partial_m A_m \partial_n A_n
+\frac12 A_nA_n\square\ln(1+4I\bar\phi)\right.\nonumber\\&&
-\varepsilon_{mnrs}\partial_rA_sA_n\partial_m\ln(1+4I\bar\phi)
+\frac12
A_nA_n\partial_m\ln(1+4I\bar\phi)\partial_m\ln(1+4I\bar\phi)\nonumber\\&&\left.
-\frac12 A_mA_n\partial_m\ln(1+4I\bar\phi)\partial_n\ln(1+4I\bar\phi)
+\partial_n A_m A_n\partial_m\ln(1+4I\bar\phi)\right].
\label{s1.9}
\end{eqnarray}

Let us make two comments on the symmetries of the action (\ref{s1.6}). First
of all, it is invariant under the gauge
transformations (\ref{resid-comp}). Secondly, it respects the
residual $\cN{=}(1,0)$ supersymmetry,
\begin{eqnarray}
\delta_\epsilon\phi&=&2(\epsilon^k\Psi_k),\qquad
\delta_\epsilon\bar\phi=0,\nonumber\\
\delta_\epsilon A_m&=&(\epsilon^k\sigma_m\bar\Psi_k),\nonumber\\
\delta_\epsilon\Psi^k_\alpha&=&-\epsilon_{\alpha l}{\cal D}^{kl}
+\frac12(1+4I\bar\phi)(\sigma_{mn}\epsilon^k)_\alpha F_{mn}
-4{\rm i} I\epsilon^k_\alpha A_m\partial_m\bar\phi,\nonumber\\
\delta_\epsilon\bar\Psi&=&-{\rm i}(1+4I\bar\phi)(\epsilon^k\sigma_m)_{\dot\alpha}
\partial_m\bar\phi,\nonumber\\
\delta_\epsilon{\cal D}^{kl}&=&{\rm i}\partial_m
[(\epsilon^k\sigma_m\bar\Psi^l+\epsilon^k\sigma_m\bar\Psi^l)(1+4I\bar\phi)].
\label{susy1}
\end{eqnarray}
We observe that both gauge transformations (\ref{resid-comp}) and the
supersymmetry (\ref{susy1}) are deformed by the non-anticommutativity
parameter $I$.

It is well known that the classical action of the undeformed
supergauge theory can be equivalently written in either chiral
or antichiral superspace, because of the equality
\footnote{Note that the analogous relation for the $\cN{=}1$
supersymmetric theories reads $\int d^4x d^2\theta\, W^\alpha
W_\alpha=\int d^4x d^2\bar\theta\, \bar W_{\dot\alpha}\bar
W^{\dot\alpha}$, where $W_\alpha$, $\bar W_{\dot\alpha}$ are the
$\cN{=}1$ gauge superfield strengths. As shown in \cite{Seiberg},
this relation also holds in the corresponding non-anticommutative
gauge theory with $\cN{=}(1/2,0)$ supersymmetry.}
\be
\frac14\tr\int d^4x_L d^4\theta\,
W^2= \frac14\tr\int d^4x_R d^4\bar\theta\, \bar W^2.
\label{ab-SYM-}
\ee

Surprisingly, the relation (\ref{ab-SYM-})
fails to be valid in the $\cN{=}(1,0)$ non-anticommutative supergauge model.
Moreover, it is inconsistent to treat the expression $\frac14\tr\int d^4x_R d^4\bar\theta\,
\bar W^2$ as any action since it bears the explicit dependence on
Grassmann and harmonic variables.

This statement can be most
easily proved in the abelian case. To this end, we
consider the covariantly antichiral superfield
strength in the antichiral coordinates (\ref{xR}),
\be
\bar{W}=-\frac{1}{4}(D^+)^2V^{--}=\bar{\cal
A}+\theta^{+\alpha}\bar{\tau}^-_\alpha+ (\theta^+)^2\bar\tau^{-2},
\ee
where $\bar{\cal A}$, $\bar{\tau}^-_\alpha$ and
$\bar\tau^{-2}$ are purely antichiral superfields defined on the
coordinate set  $x_R^m, \bar\theta^\pm_{\dot\alpha}, u^\pm\,$. This
superfield strength, as well as the prepotential $V^{--}$, depend
on the parameter of non-anticommutativity $I$. Let us expand $\bar W$
in powers of $I$
\be
\bar W=\sum_{n=0}^\infty I^n\bar W_n,
\ee
where the coefficients $\bar W_n$ are some superfields. Clearly, the first term
$\bar W_0$ in this series is a purely antichiral superfield
which has the same component structure as the undeformed superfield
strength,
\begin{eqnarray}
\bar{W}_0=\bar{\cal A}_0&=&\bar\phi-2\bar\theta^+_{\dot\alpha} \bar\Psi^{-\dot\alpha}
+2\bar\theta^-_{\dot\alpha} \bar\Psi^{+\dot\alpha}+{\rm i}\bar\theta^{-\dot\alpha}
\bar\theta^{+\dot\beta}(\partial^\alpha_{\dot\alpha} A_{\alpha\dot\beta}+
\partial^\alpha_{\dot\alpha} A_{\alpha\dot\beta})\nonumber\\
&&+(\bar\theta^+)^2{\cal D}^{--}+(\bar\theta^-)^2{\cal D}^{++}
-2(\bar\theta^+\bar\theta^-){\cal D}^{+-}
\nonumber\\
&&-2{\rm i}(\bar\theta^-)^2\bar\theta^{+\dot\alpha} u^+_k\partial_{\alpha\dot\alpha}
\Psi^{\alpha k}-2{\rm i}(\bar\theta^+)^2\bar\theta^{-\dot\alpha} u^-_k
\partial_{\alpha\dot\alpha}\Psi^{\alpha k}-(\bar\theta^+)^2(\bar\theta^-)^2\Box\phi.
\label{xxx1}
\end{eqnarray}
Note that $\bar W_0$ is harmonic-independent, $D^{\pm\pm}W_0=0$,
whereas the next term $\bar W_1$ bears such a dependence,
\be
ID^{++}\bar{W}_1=-[V^{++}\cs \bar{W}_0]\ne0.
\ee

Now we are going to prove that the expression
\be
A=\int d^4x_R d^4\bar\theta\, \bar W^2_\star
\ee
depends on Grassmann variables and harmonics,
\be
D^{\pm\pm}A\ne0,
\qquad
D^-_\alpha A\ne0.
\label{q2}
\ee
For this purpose we expand it in powers of $I$,
\be
A=\sum_{n=0}^\infty I^n A_n\,,
\ee
and check the unequalities (\ref{q2}) in the first order in $I$.
Up to terms of the second order in $I$ we have
\be
A_0+IA_1=\int d^4x_R d^4\bar\theta\,
(\bar{W}^2_0+2I\bar{W}_0\bar{W}_1).
\label{xxx}
\ee
Clearly, the term $A_0=\int d^4x_R d^4\bar\theta\,
\bar{W}^2_0$ in (\ref{xxx}) does not depend on
harmonics since $D^{\pm\pm}W_0=0$.
Therefore we have to consider only the harmonic derivative of
$A_1$ which is given by
\begin{eqnarray}
D^{++}A_1=4{\rm i}\int d^4x_R d^4\bar\theta\,\partial^\alpha_+V^{++}\bar\theta^{+\dot\alpha}
\partial_{\alpha\dot\alpha}(\bar{W}^2_0).
\end{eqnarray}
It is a technical exercise to derive the component structure
of $A_1$, given the component expansions (\ref{Vwz}), (\ref{xxx1}) of the superfields
$V^{++}$ and $\bar{W}_0$. It is sufficient to consider only two terms
in the expression $\partial_+^\alpha V^{++} \bar\theta^{+\dot\alpha}=
(\bar\theta^+)^2  A^{\alpha\dot\alpha}+2\theta^{+\alpha}\bar\theta^{
+\dot\alpha}\bar\phi+\ldots$
to come to the conclusion that
\be
D^{++}A_1=16{\rm i}\int d^4x_R\,[A_m\partial_m(\bar\phi{\cal D}^{++})+\theta^{
+\alpha}\bar\phi\partial_{\alpha\dot\beta}(\bar\phi
\partial^{\beta\dot\beta}\Psi^+_\beta) ]+\ldots\ne 0.
\label{xxx2}
\ee
The terms written down in (\ref{xxx2}) cannot be cancelled by any other ones (which are omitted here).
The manifest dependence on harmonics implied by (\ref{xxx2})
entails also the dependence on $\theta^+$ variables owing to the
commutation relation $[D^{--},D^+_\alpha]=D^-_\alpha$.
Therefore, (\ref{xxx2}) proves the unequalities (\ref{q2}) which
show that $\tr\int d^4x_R d^4\bar\theta\, \bar
W^2$ cannot be treated as a superfield action.

It is easy to argue that the more general expression $\int
d^4x_R d^4\bar\theta \bar{\cal F}_\star(\bar W)$ also involves a manifest dependence  on
the Grassmann variables and harmonics. This implies that among the candidate
contributions to the effective action of the supersymmetric gauge model there are no such ones which are
given by  integrals of some functions of the superfield strength $\bar W$ over
the antichiral superspace. In Sect.\ 6 we will demonstrate that the contributions to the effective action in the
non-anticommutative case are naturally written as integrals over the full $\cN{=}(1,1)$ superspace.

\subsection{Seiberg-Witten transform in the abelian supergauge
model}\label{SW-gauge}
The equations (\ref{resid-comp}),
(\ref{susy1}) show that both gauge and supersymmetry
transformations depend on the parameter of the chiral singlet
deformation $I$. A natural question is whether there exist any
change of the variables in the functional integral which would bring these transformations
to the undeformed form. For example, for the gauge models with the
bosonic noncommutative deformation such a transformation was found in
\cite{SW1}, and it is known as the Seiberg-Witten map.
Remarkably, for the chiral singlet deformation such a field redefinition
also exists. It was found in \cite{FILSZ,Araki}:
\begin{eqnarray}
\phi&\to&\varphi=\frac1{(1+4I\bar\phi)^2}\left[
\phi+\frac{4I(A_m A_m+4I^2\partial_m\bar\phi\partial_m\bar\phi)}{1+4I\bar\phi}
\right],\nonumber\\
A_m&\to&a_m=\frac{A_m}{1+4I\bar\phi},\qquad
\bar\Psi^k_{\dot\alpha}\to\bar\psi^k_{\dot\alpha}=
\frac{\bar\Psi^k_{\dot\alpha}}{1+4I\bar\phi},\nonumber\\
\Psi^k_\alpha&\to&\psi^k_\alpha=\frac1{(1+4I\bar\phi)^2}
\left[\Psi^k_\alpha+\frac{4I A_{\alpha\dot\alpha}\bar\Psi^{\dot\alpha k}}{1+4I\bar\phi}
\right],\nonumber\\
{\cal D}^{kl}&\to&d^{kl}=\frac1{(1+4I\bar\phi)^2}\left[
{\cal D}^{kl}+\frac{8I\bar\Psi^k_{\dot\alpha}\bar\Psi^{\dot\alpha l}}{1+4I\bar\phi}
\right].
\label{SW}
\end{eqnarray}
It is easy to check that the supertranslations (\ref{susy1}), being
rewritten in terms of the fields (\ref{SW}), read
\begin{eqnarray}
\delta_\epsilon\varphi&=&2(\epsilon^k\psi_k),\qquad
\delta_\epsilon\bar\phi=0,\nonumber\\
\delta_\epsilon a_m&=&(\epsilon^k\sigma_m\bar\psi_k),\nonumber\\
\delta_\epsilon\psi^k_\alpha&=&-\epsilon_{\alpha l}d^{kl}
+\frac12(\sigma_{mn}\epsilon^k)_\alpha f_{mn},\nonumber \\
\delta_\epsilon\bar\psi^k_{\dot\alpha}&=&-{\rm i}(\epsilon^k\sigma_m)_{\dot\alpha}
\partial_m\bar\phi,\nonumber\\
\delta_\epsilon d^{kl}&=&{\rm i}\partial_m(\epsilon^k\sigma_m\bar\psi^l
+\epsilon^k\sigma_m\bar\psi^l),
\label{susy2}
\end{eqnarray}
where $f_{mn}=\partial_m a_n-\partial_n a_m$. The gauge
transformations of the fields (\ref{SW}) also coincide with those for
the undeformed fields. Namely, all fields are the gauge group singlets, except for $a_m$ which
transforms as
\be
\delta_r a_m=\partial_m\lambda.
\label{gauge5}
\ee

Surprisingly, the field redefinition (\ref{SW}) drastically simplifies the
structure of the action (\ref{s1.6}). In terms of
fields $\varphi$, $\bar\phi$, $\psi^\alpha_k$,
$\bar\psi^{\dot\alpha k}$, $a_m$, $d^{kl}$ it is given by \be
S_{SYM}= \int d^4x\,{\cal L}=\int d^4x\,(1+4I\bar\phi)^2 {\cal
L}_0, \label{SYM3} \ee where \be {\cal
L}_0=-\frac12\varphi\square\bar\phi+\frac14(f_{mn}f_{mn}+\frac12\varepsilon_{mnrs}
f_{mn}f_{rs})-{\rm i}
\psi_k^\alpha\partial_{\alpha\dot\alpha}\bar\psi^{\dot\alpha k}
+\frac14d^{kl}d_{kl}. \label{L0_} \ee The expression ${\cal L}_0$ is
none other than a Lagrangian of $\cN{=}(1,1)$ supersymmetric ${\rm
U}(1)$ gauge theory. As a result,  the net effect of the chiral
singlet deformation in terms of the new fields is the appearance of the factor $(1+4I\bar\phi)^2$
in front of the undeformed Lagrangian.

It is also worth pointing out that the Seiberg-Witten map is not
unique. Indeed, since the scalar field $\bar\phi$ is a singlet
of both the gauge transformations and $\cN{=}(1,0)$ supersymmetry,
one can rescale the fields as
\be
\hat\varphi=L^2\varphi,\quad
\hat\psi^k_\alpha=L^2\psi^k_\alpha,\quad
\hat{d}^{kl}=Ld^{kl},\label{SWsimp}
\ee
which does not affect
gauge transformations and supersymmetry. When written in terms of the fields (\ref{SWsimp}),
the Lagrangian $\cal L$ takes the most simple form, \be
{\cal
L}=-\frac12\hat\varphi\square\bar\phi+\frac14L^2(f_{mn}f_{mn}+\frac12\varepsilon_{mnrs}
f_{mn}f_{rs})-{\rm i}
\hat\psi_k^\alpha\partial_{\alpha\dot\alpha}\bar\psi^{\dot\alpha
k} +\frac14\hat{d}^{kl}\hat{d}_{kl}.\label{Lsimp} \ee We see
that the only remaining interaction is that between the gauge field strength $f_{mn}$ and the scalar field $\bar\phi\,$.
The lagrangian (\ref{Lsimp}) is bilinear in all other fields, like in the free case.

Let us now discuss the problem of a superfield representation for
the Seiberg-Witten-like map (\ref{SW}). For this purpose we need a
relation between the superfield $\cal A$ given by
(\ref{A}) and the undeformed superfield strength $W_0$
given by the expression
\begin{eqnarray}
W_0(x_L,\theta^+,\theta^-,u)&=&
\varphi+2\theta^+\psi^-
-2\theta^-\psi^++(\theta^-\sigma_{mn}\theta^+)f_{mn}
\nonumber\\&&
+(\theta^+)^2d^{--}
+(\theta^-)^2d^{++}
-2(\theta^+\theta^-)d^{+-}\nonumber\\&&
+2{\rm i}(\theta^-)^2\theta^+\sigma_m\partial_m\bar\psi^+
+2{\rm i}(\theta^+)^2\theta^-\sigma_m\partial_m\bar\psi^-
-(\theta^+)^2(\theta^-)^2\square\bar\phi\,.
\label{add2}
\end{eqnarray}
Here we use the notation $\psi^\pm_\alpha=\psi^i_\alpha u^\pm_i$,
$d^{+-}=u^+_ku^-_l d^{kl}$, etc, as for the original
fields. By definition, the superfield strength (\ref{add2}) is gauge
invariant, \be \delta_\lambda W_0=0\,, \label{add3} \ee and
it transforms under $\cN{=}(1,0)$ supersymmetry in the standard way,
\be \delta_\epsilon W_0=(\epsilon^{-\alpha}\partial_{-\alpha}
+\epsilon^{+\alpha}\partial_{+\alpha})W_0. \label{add4} \ee
There is a simple relation between the expressions
(\ref{A}) and (\ref{add2}) \cite{FILSZ},
\be
{\cal
A}(x_L,\theta^+,\theta^-,u)=(1+4I\bar\phi)^2W_0(x_L,\theta^+,
(1+4I\bar\phi)^{-1}\theta^-,u).
\label{add5}
\ee
The equation (\ref{add5}) plays the role of the superfield
Seiberg-Witten transform. It is essential that, up to an overall scalar factor, it
amounts to rescaling the variable $\theta^-$ by the factor $(1+4I\bar\phi)^{-1}$.

Let us now introduce the following differential operator
\begin{eqnarray}
R_\theta=\exp(L^{-1}\theta^-\partial_-)=L^{-2}
+\partial_{-\alpha}\{1-L^{-1}-\frac14(L^{-1}-1)^2[2\theta^{-\alpha}
-(\theta^-)^2\partial_-^\alpha]\},
\label{add6}
\end{eqnarray}
where $L=1+4I\bar\phi$. Using this operator, the superfield Seiberg-Witten
transform (\ref{add5}) can be rewritten as
\be
{\cal A}=L^2R_\theta W_0.
\label{add7}
\ee
Owing to the simple property
$R_\theta A R_\theta B=R_\theta (A B)$, we have
\be
{\cal A}^2=L^4R_\theta W_0^2.
\label{add8}
\ee
Employing now the relations (\ref{add6}), (\ref{add8}), one easily
constructs the Seiberg-Witten transform of the classical action (\ref{ab-SYM1}),
\be
S_{SYM}=\frac14\int d^4x_Ld^4\theta\, {\cal A}^2=
\frac14\int d^4x_L d^4\theta\,(1+4I\bar\phi)^2W_0^2.
\label{add9}
\ee
This is just the action (\ref{SYM3}) derived before.

The Seiberg-Witten transform found here for the classical action (\ref{add9})
can be readily generalized to the action with an arbitrary chiral
potential,
\be
\int d^4x_Ld^4\theta\,{\cal F}_\star({\cal A}),
\label{add10}
\ee
where ${\cal F}_\star({\cal A})$ is some function given by a series,
\be
{\cal F}_\star({\cal A})=\sum_{n=2}^\infty
c_n{\cal A}_\star^n.
\label{add11}
\ee
The function ${\cal A}_\star^n$ is expressed through the undeformed
superfield strength (\ref{add2}) as follows
\be
{\cal A}_\star^n=L^{2n-2}(W_0)^n_{\hat\star}+\ldots,
\label{add12}
\ee
where we introduced a modified $\star$-product,
\be
\hat\star=\exp(L^{-1}P_s).
\label{add13}
\ee
Dots in (\ref{add12}) stand for terms which involve full spinor derivatives
$\partial_{-\alpha}$ coming from the expansion of the operator
(\ref{add6}). These terms are not essential when they are considered under the integral
over chiral superspace. As a result, the action (\ref{add10}) is
expressed through the undeformed superfield strengths (\ref{add2}),
\be
\int d^4x_Ld^4\theta\,{\cal F}_\star({\cal A})=
\int d^4x_Ld^4\theta\,L^{-2}{\cal F}_{\hat\star}(L^2W_0).
\label{add14}
\ee
Here, the function ${\cal F}_{\hat\star}(L^2W_0)$ is given by the
series (\ref{add11}) with the $\hat\star$-product of superfields.
The relation (\ref{add14}) plays the role of Seiberg-Witten
transform for the chiral effective action.

Let us point out that the choice (\ref{SWsimp}) not only brings the
classical action to the most simple form but also is very useful
for studying the contributions to chiral effective potentials.
In particular, given the expansion of the superfield $\cal A$ in terms of these
fields,
\begin{eqnarray}
{\cal A}&=&\hat\varphi+2\theta^{+\alpha}\hat{\psi}^-_\alpha-2L^{-1}\theta^{-\alpha}
\hat{\psi}^+_\alpha
+ L(\theta^+\sigma_{mn}\theta^-)f_{mn}\nonumber\\
&&+L(\theta^+)^2\hat{d}^{--}-2(\theta^+\theta^-)\hat{d}^{+-}+L^{-1}
(\theta^-)^2\hat{d}^{++}\nonumber\\
&&+2{\rm i}[(\theta^-)^2\theta^{+\alpha}\partial_{\alpha\dot\alpha}
\bar\psi^{+\dot\alpha}+L(\theta^+)^2\theta^{-\alpha}\partial_{\alpha\dot\alpha}
\bar\psi^{-\dot\alpha}]
-(\theta^+)^2(\theta^-)^2\Box\bar\phi\,,
\label{Asimp}
\end{eqnarray}
one can readily find the component structure of cubic and quartic
terms in the effective potential ${\cal F}_\star
({\cal A})$ in the bosonic sector,
\begin{eqnarray}
\int d^4\theta{\cal
A}^3_\star&=&-3\hat\varphi^2\Box\bar\phi+3\hat\varphi
(\hat{d}_{kl})^2+\frac34L^2\hat\varphi (f_{\alpha\beta})^2\nonumber\\
&&-3I^2\Box\bar\phi[L^2
(f_{\alpha\beta})^2-4(\hat{d}_{kl})^2]-16I^4(\Box\bar\phi)^3,
\label{xxx4}
\\
\int d^4\theta {\cal A}^4_\star&=&-4\hat\varphi^3
\Box\bar\phi+6\hat\varphi^2(\hat{d}_{kl})^2
+\frac32L^2\hat\varphi^2 (f_{\alpha\beta})^2\nonumber\\
&&+2I^2[L^2 (f_{\alpha\beta})^2-4(\hat{d}_{kl})^2]
[-6\hat\varphi\Box\bar\phi
+(\hat{d}_{kl})^2+\frac14L^2 (f_{\alpha\beta})^2]
\nonumber\\
&&+8I^4(\Box\bar\phi)^2[3L^2(f_{\alpha\beta})^2+12(\hat{d}_{kl})^2-
8\hat\varphi\Box\bar\phi],
\label{xxx5}
\end{eqnarray}
where $(f_{\alpha\beta})^2=f^{\alpha\beta}f_{\alpha\beta}=(f_{mn})^2+
f_{mn}\tilde{f}_{mn}$ and
$(\hat{d}_{kl})^2=\hat{d}^{kl}\hat{d}_{kl}$.
These expressions are the chiral singlet deformation
of the corresponding terms $\int d^4\theta W^3_0$ and $\int d^4\theta W_0^4$
in the undeformed ${\cal N}{=}(1,1)$ holomorphic effective action. The equations (\ref{xxx4}),
(\ref{xxx5}) show that the chiral singlet deformation manifests
itself not only in the appearance of induced interaction of vector field with scalars
but also in the presence of new terms with the field derivatives.
Another important consequence of the non-anticommutative deformation
of the effective potentials is the appearance of non-linear self-coupling of
the auxiliary fields $I^2(\hat{d}_{kl})^4$.

\subsection{Neutral hypermultiplet model}
The $\cN{=}(1,1)$ hypermultiplet on-shell content is four real
scalar fields $f^{ak}$, $a,k=1,2\,$ and two independent spinors
$\rho^{\alpha a}$, $\chi^{\dot\alpha}_a$. Here we assume that
these fields are neutral with respect to the U$(1)$ group. We are
going to find the interactions of these fields with the vector multiplet
$\phi$, $\bar\phi$, $\Psi^\alpha_k$, $\bar\Psi^{\dot\alpha k}$, $D^{kl}$.

The classical action of the neutral hypermultiplet is given by
(\ref{Sq-pg}). Taking into account the equations (\ref{nabla-ad}) and
(\ref{comm-anal}), we make explicit the $\star$-product in (\ref{Sq-pg}),
\be
S_{ad}[q^+,V^{++}]=\frac12\int d\zeta du\, q^+_a[
D^{++}q^{+a}+4{\rm i} I\bar\theta^{+\dot\alpha}
(\partial^\alpha_+ V^{++}\partial_{\alpha\dot\alpha}q^{+\alpha}
-\partial_{\alpha\dot\alpha}V^{++}\partial_+^\alpha q^{+a})]\,.
\label{Sn}
\ee
The hypermultiplet superfield has the following component filed
expansion
\begin{eqnarray}
q^{+a}&=&f^{+a}+\theta^{+\alpha}\pi^a_\alpha+\bar\theta^+_{\dot\alpha}
\kappa^{\dot\alpha a}+\theta^+\sigma_m\bar\theta^+r_m^{-a}
+(\theta^+)^2 g^{-a}+(\bar\theta^+)^2 h^{-a}\nonumber\\&&
+(\bar\theta^+)^2\theta^{+\alpha}\Sigma_\alpha^{--a}
+(\theta^+)^2\bar\theta^{+\dot\alpha}\bar\Sigma_{\dot\alpha}^{--a}
+(\theta^+)^2(\bar\theta^+)^2\omega^{-3\,a}\,,
\label{q-comp}
\end{eqnarray}
where all the component fields depend only on the variables $x^m_A$ and $u^\pm_i$.
These fields can be further expanded over the harmonic variables,
giving rise to an infinite number of the auxiliary fields. The auxiliary fields
should be eliminated from the action using the classical equation
of motion for the hypermultiplet superfield $q^{+a}\,$. This equation is
easily obtained from the action (\ref{Sn}),
\be
D^{++}q^{+a}+4{\rm i} I\bar\theta^{+\dot\alpha}
(\partial_+^\alpha V^{++}\partial_{\alpha\dot\alpha} q^{+a}
-\partial_{\alpha\dot\alpha}V^{++}\partial_+^\alpha q^{+a})=0\,.
\label{q-eom}
\ee
Substituting (\ref{q-comp}), (\ref{Vwz}) into (\ref{q-eom})
we find the explicit expressions of the hypermultiplet component
fields in terms of the physical scalars $f^{ia}$ and fermions
$\rho^a_\alpha$, $\chi^{\dot\alpha a}$,
\begin{eqnarray}
&&f^{+a}=f^{ak}u^+_k\,,\quad
\pi^a_\alpha=\rho^a_\alpha\,,\quad
\kappa^{\dot\alpha a}=\chi^{\dot\alpha a},\quad
r^{-a}_m=r_m^{ak}u^-_k,\quad g^{-a}=0\,,\nonumber\\&&
h^{-a}=h^{ak}u^-_k\,,\quad
\Sigma_\alpha^{--a}=\Sigma_\alpha^{kl\, a}u^-_k u^-_l,\quad
\bar\Sigma_{\dot\alpha}^{--a}=0\,,\quad
\omega^{-3\,a}=0\,,\nonumber\\&&
r_m^{ak}=2{\rm i}(1+4I\bar\phi)\partial_m f^{ak}\,,\quad
h^{ak}=-8{\rm i} IA_m\partial_m f^{ak},\nonumber\\&&
\Sigma_\alpha^{kl\, a}=-4{\rm i} I(\bar\Psi^{\dot\alpha k}
\partial_{\alpha\dot\alpha}f^{al}+\bar\Psi^{\dot\alpha l}
\partial_{\alpha\dot\alpha}f^{ak})\,.
\label{comp1}
\end{eqnarray}
Taking into account (\ref{comp1}), we obtain the following action for
the physical components in the neutral hypermultiplet model,
\begin{eqnarray}
S_{ad}&=&\int d^4x[
\frac12(1+4I\bar\phi)^2\partial_m f^{ak}\partial_m f_{ak}
+\frac12{\rm i}(1+4I\bar\phi)\rho^{\alpha a}\partial_{\alpha\dot\alpha}
\chi_a^{\dot\alpha}\nonumber\\&&
+4{\rm i} I\bar\Psi_k^{\dot\alpha}\rho_a^\alpha\partial_{\alpha\dot\alpha}
f^{ak}+2{\rm i} I\rho^{\alpha a}A_m\partial_m\rho_{\alpha a}
+{\rm i} I\rho^{\beta
a}\rho_a^\alpha\partial_{(\alpha\dot\alpha}A^{\dot\alpha}_{\beta )}
]\,.
\label{Sn-comp}
\end{eqnarray}
Note that only the fields  $\bar\phi$,
$A_m$, $\bar\Psi _k^{\dot\alpha}$ from the vector multiplet
interact with the hypermultiplet fields in (\ref{Sn-comp}).

Let us study the symmetries of the action (\ref{Sn-comp}). This action is invariant
in the evident way under the gauge
transformations (\ref{gauge-tr2}). Using the relation (\ref{comm-anal}),
these gauge transformations can be cast in the following form
\be
\delta_\Lambda q^{+a}=4{\rm i}\bar\theta^+_{\dot\alpha}
(\partial^{\alpha\dot\alpha}\Lambda\partial_{+\alpha}q^{+a}
-\partial_{+\alpha}\Lambda\partial^{\alpha\dot\alpha}q^{+a})\,.
\label{gauge-tr3}
\ee
Recall that the component structure of the vector multiplet
(\ref{Vwz}) is given in the Wess-Zumino gauge. Therefore
it makes sense to discuss here only the residual gauge
transformations with the parameter $\Lambda_r={\rm i}\lambda(x_A)$.
With such a choice of the gauge
parameter the hypermultiplet gauge transformations (\ref{gauge-tr3})
are reduced to
\be
\delta_r q^{+a}=-4I\bar\theta^+_{\dot\alpha}\partial^{\alpha\dot\alpha}
\lambda(x_A)\partial_{+\alpha}q^{+a}\,.
\label{gauge-tr4}
\ee
Equation (\ref{gauge-tr4}) leads to the following gauge transformations of
the hypermultiplet component fields
\be
\delta_r f^{ak}=0\,,\quad
\delta_r \rho^a_\alpha=0\,,\quad
\delta_r\chi^{\dot\alpha a}=-4I\partial^{\alpha\dot\alpha}
\lambda \rho^a_\alpha\,.
\label{gauge-tr5}
\ee

Let us also consider the $\cN{=}(1,0)$ supersymmetry
transformations for the hypermultiplet \be \delta_\epsilon
q^{+a}=(\epsilon^{-\alpha}Q^+_\alpha
-\epsilon^{+\alpha}Q^-_\alpha)q^{+ a}
=(\epsilon^{+\alpha}\partial_{+\alpha}
-2{\rm i}\epsilon^{-\alpha}\bar\theta^{+\dot\alpha}\partial_{\alpha\dot\alpha})
q^{+ a}\,. \label{susy3} \ee In the Wess-Zumino gauge (\ref{susy3})
leads to the following
component field transformations \be \delta_\epsilon
f^{ak}=\epsilon^{ak}\rho^a_\alpha,\quad \delta_\epsilon
\rho^a_\alpha=0,\quad
\delta_\epsilon\chi^a_{\dot\alpha}=2{\rm i}\epsilon^{\alpha k}
(1+4I\bar\phi)\partial_{\alpha\dot\alpha}f^a_k\,. \label{susy4}
\ee It is an easy exercise to check that the action
(\ref{Sn-comp}) is invariant with respect to the hypermultiplet
gauge transformations (\ref{gauge-tr5}) combined with
(\ref{resid-comp}), as well as with respect to the $\cN{=}(1,0)$ supersymmetry
(\ref{susy4}) combined with (\ref{susy1}).

Both supersymmetry transformations (\ref{susy3})
and the gauge transformations (\ref{gauge-tr5}) are deformed due to the
explicit presence of the parameter $I\,$.
Let us consider the following transform of the hypermultiplet
fields
\begin{eqnarray}
f^{ak}&\to&f_0^{ak}=(1+4I\bar\phi)f^{ak},\nonumber\\
\rho^{\alpha a}&\to&\rho_0^{\alpha a}=(1+4I\bar\phi)\rho^{\alpha
a}\,,\nonumber\\
\chi^{\dot\alpha a}&\to&\chi_0^{\dot\alpha a}=\chi^{\dot\alpha a}
+\frac{4IA^{\alpha\dot\alpha}\rho^a_\alpha}{1+4I\bar\phi}
-\frac{8I\bar\Psi^{\dot\alpha k} f^a_k}{1+4I\bar\phi}\,.
\label{SW-hyper}
\end{eqnarray}
One can easily check that the new fields $f_0^{ak}$,
$\rho_0^{\alpha a}$, $\chi_0^{\dot\alpha a}$ transform under ${\rm
U}(1)$ gauge group and $\cN{=}(1,0)$ supersymmetry in the standard
way, i.e. as in the undeformed case with $I=0\,$.
Therefore we can refer to the transformation (\ref{SW-hyper}) as a
Seiberg-Witten map for the neutral hypermultiplet model. In terms of the fields
 $f_0^{ak}$, $\rho_0^{\alpha a}$,
$\chi_0^{\dot\alpha a}$ the
action (\ref{Sn-comp}) is rewritten as
\begin{eqnarray}
S_{ad}&=&\int d^4x[\frac12\partial_m f_0^{ak}\partial_m f_{0\, ak}
+\frac{\rm i} 2\rho_0^{\alpha a}\partial_{\alpha\dot\alpha}\chi^{\dot\alpha}_{0\, a}
+\frac{2{\rm i} I\rho_0^{\beta a}\rho^\alpha_{0\, a}\partial_{(\alpha\dot\alpha}
a^{\dot\alpha}_{\beta)}}{1+4I\bar\phi}\nonumber\\&&
+\,\frac{2If_0^{ak}f_{0\, ak}\square\bar\phi}{1+4I\bar\phi}
+\frac{4{\rm i} I\rho_0^{\alpha a}f_{0\, ak}\partial_{\alpha\dot\alpha}\bar\psi^{\dot\alpha k}}{
1+4I\bar\phi}
]\,.
\label{Sn-comp1}
\end{eqnarray}

Let us now turn to the full $\cN{=}(1,0)$ supersymmetric gauge model which
is defined by the sum of the classical actions (\ref{SYM3}) and
(\ref{Sn-comp}).
In this model one can perform the further change of fields
of the vector multiplet in order to bring the total action
$S_{SYM}+S_{ad}$ to the simplest form,
\begin{eqnarray}
\varphi&\to&\hat\varphi=(1+4I\bar\phi)^2\varphi
-\frac{4I(f_0^{ak}f_{0\, ak})}{1+4I\bar\phi}\,,\nonumber\\
\psi_k^\alpha&\to&\hat\psi_k^\alpha=
(1+4I\bar\phi)^2\psi_k^\alpha-\frac{4I\rho_0^{ak}f_{0\,
ak}}{1+4I\bar\phi},\nonumber\\
d_{kl}&\to&\hat d_{kl}=(1+4I\bar\phi)d_{kl}\,.
\label{SW-sym+hyper}
\end{eqnarray}
In terms of these new fields (\ref{SW-sym+hyper}) the
action $S_{SYM}+S_{ad}$ reads
\begin{eqnarray}
S_{SYM}+S_{ad}&=&\int d^4x(L_0+L_{int}),
\label{SYM+hyper}\\
L_0&=&-\frac12\hat\varphi\square\bar\phi+\frac12\partial_m f_0^{ak}
\partial_m f_{0\, ak}-\frac1{16}f^{\alpha\beta}f_{\alpha\beta}
-{\rm i}\hat\psi_k^\alpha\partial_{\alpha\dot\alpha}\bar\psi^{\dot\alpha k}
\nonumber\\&&
+\frac{\rm i}2\rho_0^{\alpha
a}\partial_{\alpha\dot\alpha}\chi^{\dot\alpha}_{0\, a}+\frac14
\hat d^{kl}\hat d_{kl},
\label{L0}\\
L_{int}&=&-\frac12I\bar\phi(1+2I\bar\phi)f^{\alpha\beta}f_{\alpha\beta}
+\frac{I\rho_0^{\beta a}\rho^\alpha_{0\,
a}f_{\alpha\beta}}{1+4I\bar\phi}\,.
\label{Lint}
\end{eqnarray}
Here
$f_{\alpha\beta}={\rm i}\partial_{\alpha\dot\alpha}a_\beta^{\dot\alpha}
+{\rm i}\partial_{\beta\dot\alpha}a_\alpha^{\dot\alpha}=(\sigma_{mn})_{\alpha\beta}f_{mn}$.
Note that $L_0$ coincides with the  Lagrangian of the free undeformed
$\cN{=}(1,1)$ ${\rm U}(1)$ supergauge theory while $L_{int}$ presents
the interaction of hypermultiplet fields with the
gauge multiplet. We see that the whole interaction  is
proportional to the deformation parameter $I$. Thereby, the
interaction in this model is entirely an effect of chiral
singlet deformation.

\subsection{Charged hypermultiplet model}
Consider now the classical action (\ref{Sqad}) of the charged hypermultiplet
model. Using the representation (\ref{Sq-pg}) and the explicit expression (\ref{nabla-f}), this
action can be written as
\be
S_f[q^+,V^{++}]=\frac12\int d\zeta du\,
q^+_a\left(D^{++} q^{+a}+\frac12[V^{++}\cs q^{+a}]
-\frac12(\tau_3)^a_b\{V^{++}\cs q^{+b}\}
\right).
\label{Se}
\ee
The relevant superfield equation of motion reads
\be
D^{++} q^{+a}+\frac12[V^{++}\cs q^{+a}]
-\frac12(\tau_3)^a_b\{V^{++}\cs q^{+b}\}=0\,.
\label{Se-eom}
\ee

To derive the component structure of the action
(\ref{Se}), we follow the same steps as for the neutral
hypermultiplet model considered in the previous subsection.
We take the component expansions of the hypermultiplet (\ref{q-comp})
and the gauge superfield in the Wess-Zumino gauge (\ref{Vwz}) and substitute
them into (\ref{Se-eom}). As usual, the equations of motion for the auxiliary
fields have the algebraic form and can be easily solved
to eliminate these fields. As a result, we find the following
component structure of the charged hypermultiplet superfield in terms of physical
fields
\begin{eqnarray}
q_e^{+a}&=&u^+_k f^{ak}+\theta^{+\alpha}\rho^a_\alpha
+(\theta^+)^2u^-_kg^{ak}+\bar\theta^+_{\dot\alpha}
[\chi^{\dot\alpha a}+(\theta^+)^2u^-_k u^-_l \sigma^{\dot\alpha a\,kl}]
+\theta^+\sigma_m\bar\theta^+ r_m^{ak}u^-_k\nonumber\\&&
+(\bar\theta^+)^2[u^-_kh^{ak}+\theta^{+\alpha}u^-_k u^-_l
\sigma_\alpha^{a\, kl}+(\theta^+)^2 u^-_k u^-_l u^-_j X^{a\, klj}]\,,
\label{qe-comp}
\end{eqnarray}
where
\begin{eqnarray}
&&g^{ak}=(\tau_3)^a_b\bar\phi f^{bk},\quad
r_m^{ak}=2{\rm i}(1+2I\bar\phi)\partial_m f^{ak}
+2(\tau_3)^a_bA_m f^{bk}\,,\nonumber\\&&
h^{ak}=-4{\rm i} IA_m\partial_m f^{ak}+(\tau_3)^a_b
(\phi f^{bk}+2I^2\bar\phi\square f^{bk})\,,\nonumber\\&&
\sigma^{\dot\alpha a\,
kl}=2(\tau_3)^a_b\bar\Psi^{\dot\alpha(k}f^{bl)},\quad
\sigma_\alpha^{akl}=-4{\rm i} I\bar\psi^{\dot\alpha(k}
 \partial_{\alpha\dot\alpha}f^{al)}
+2(\tau_3)^a_b\Psi_\alpha^{(k}f^{bl)},\nonumber\\&&
X^{a\,klj}=(\tau_3)^a_b{\cal D}^{(kl}f^{bj)}\,.
\label{qe-comp1}
\end{eqnarray}
Now we substitute the expansions (\ref{qe-comp}) and
(\ref{Vwz}) into the action (\ref{Se}) and integrate over
the Grassmann and harmonic variables to obtain the component
form of the action of charged hypermultiplet in terms of the physical fields,
\begin{eqnarray}
S_f&=&\int d^4x[
\frac12(1+4I\bar\phi)\partial_m f_{ak}\partial_m f^{ak}
+{\rm i} (\tau_3)_a^b A_m f_{bk}\partial_m f^{ak}+\frac12 (A_m)^2
(f^{ak})^2\nonumber\\&&
+\frac12\phi\bar\phi(f^{ak})^2+I^2(f^{ak})^2\square(\bar\phi^2)
-\frac12(\tau_3)_b^a f_a^kf^{bl}{\cal D}_kl
+2{\rm i}
I\bar\Psi_k^{\dot\alpha}\rho_a^\alpha\partial_{\alpha\dot\alpha}f^{ak}
\nonumber\\&&
+(\tau_3)_b^a\Psi_k^\alpha\rho_{\alpha a}f^{bk}
+(\tau_3)_b^a f_a^k\bar\psi_{\dot\alpha k}\chi^{\dot\alpha b}
+\frac{\rm i}2(1+2I\bar\phi)\rho^{\alpha a}\partial_{\alpha\dot\alpha}
\chi_a^{\dot\alpha}\nonumber\\&&
-\frac12(\tau_3)^a_b\rho_a^\alpha A_{\alpha\dot\alpha}\chi^{\dot\alpha b}
+\frac14(\tau_3)^a_b(\bar\phi\chi_{\dot\alpha a}\chi^{\dot\alpha b}
+\phi\rho^\alpha_a\rho_\alpha^b)+{\rm i} I\rho^{\alpha a}A_m \partial_m
\rho_{\alpha a}\nonumber\\&&
+\frac{\rm i}2I\rho^{\beta
a}\rho^\alpha_a\partial_{(\alpha\dot\alpha}A^{\dot\alpha}_{\beta)}
+I^2(\tau_3)_b^a\bar\phi\partial_{\alpha\dot\alpha}\rho_{\beta a}
\partial^{\beta\dot\alpha}\rho^{\alpha b}
]\,.
\label{Se-comp}
\end{eqnarray}
Note that in the limit $I\to0$ the action (\ref{Se-comp})
still retains an interaction. It has the standard form of the interaction between the physical fields of $\cN{=}(1,1)$
supersymmetric electrodynamics.

As in the neutral hypermultiplet model, from
(\ref{gauge-tr1}) one can derive the residual gauge transformations
for the physical component fields (in the Wess-Zumino gauge for
the vector multiplet), \be \delta_r f^{ak}={\rm i}\lambda (\tau_3)_b^a
f^{bk},\qquad \delta_r\rho^a_\alpha={\rm i}\lambda
(\tau_3)_b^a\rho^b_\alpha\,, \qquad \delta_r\chi^{\dot\alpha
a}={\rm i}\lambda (\tau_3)_b^a\chi^{\dot\alpha b}
-2I\partial^{\alpha\dot\alpha}\lambda\rho^a_\alpha\,.
\label{gauge-tr6} \ee The $\cN{=}(1,0)$ supersymmetry
transformations for these fields are given by \be \delta_\epsilon
f^{ak}=\epsilon^{\alpha k}\rho^a_\alpha,\quad
\delta_\epsilon\rho^a_\alpha=2\epsilon^k_\alpha(\tau_3)^a_b
\bar\phi f^b_k\,,\quad
\delta_\epsilon\chi^a_{\dot\alpha}=-2\epsilon^\alpha_k
[{\rm i}(1+2I\bar\phi)\partial_{\alpha\dot\alpha}f^{ak} +(\tau_3)^a_b
A_{\alpha\dot\alpha} f^{bk}]\,. \label{susy5} \ee
It is easy to check that the action (\ref{Se-comp}) is invariant
under both the gauge transformations (\ref{gauge-tr6}) and
the supersymmetry ones (\ref{susy5}).

In the charged hypermultiplet model there also exists
a field redefinition (Seiberg-Witten map) which casts the transformations
(\ref{gauge-tr6}), (\ref{susy5}) in the undeformed form,
\begin{eqnarray}
f^{ak}&\to&f_0^{ak}=(1+2I\bar\phi)f^{ak}\,,\nonumber\\
\rho^{\alpha a}&\to&\rho_0^{\alpha a}=(1+2I\bar\phi)\rho^{\alpha
a}\,,\nonumber\\
\chi^a_{\dot\alpha}&\to&\chi^a_{\dot\alpha\, 0}
=\chi^a_{\dot\alpha}-\frac{2IA_{\alpha\dot\alpha}\rho^{\alpha a}}{1+4I\bar\phi}
+\frac{4I\bar\Psi_{\dot\alpha k} f^{ak}}{1+4I\bar\phi}\,.
\label{SWqe}
\end{eqnarray}
However, in contrast to the neutral hypermultiplet model,
the map (\ref{SWqe}) does not lead to the substantial
simplifications of the classical action.
Therefore, here we do not give how the classical action of charged
hypermultiplet looks in terms of the new fields $f_0^{ak}$, $\rho^{a\alpha}_0$,
 $\chi^a_{\alpha\, 0}$.

\section{Renormalizability of the ${\cal N}{=} (1,0)$
non-an\-ti\-com\-mu\-ta\-ti\-ve abelian models}
In this section we explore the quantum aspects of the
non-anticommutative theories defined by the classical actions
 (\ref{s1.6}), (\ref{Sn-comp}) and (\ref{Se}).

From the point of view of physical applications only
renormalizable theories play a fundamental role in quantum field
theory while the non-renormalizable ones are usually treated
as some effective theories. By the term ``renormalizability''
we will mean the multiplicative renormalizability, when all
the divergent quantum corrections in a given theory have the form
of some terms of the classical action and hence can be
taken away by some redefinition of coupling constants or fields in the
classical action. According to the customary lore of quantum field theory,
a model is power-counting non-renormalizable if it involves coupling
constants of the negative mass dimension. The
supersymmetric models with the chiral singlet deformation under
consideration contain the parameter of non-anticommutativity $I$
with the negative mass dimension, $[I]=-1$. If one treats this
parameter as a coupling constant, the considered models are
formally non-renormalizable. Nevertheless, we will show that in
our case the standard arguments towards non-renormalizability fail
and all the models considered here are renormalizable. A key feature
of the non-anticommutativity is that all such models are formulated
only in the Euclidean superspace and the deformation is present only in
the chiral sector of superspace, while the antichiral one remains intact.
Therefore, the interaction terms in the actions appear
in a non-symmetric way, still preserving the reality with respect to
the conjugation (\ref{C2}) in the Euclidean space.
These interactions lead to the quantum divergences of a special
form which do not violate the renormalizability. As a result, the
non-anticommutative theories with $\cN{=}(1,0)$ supersymmetry are
renormalizable and so they can bear certain interest for the further study
in the framework of quantum field theory.

Here we will prove the renormalizability of the models with the
classical actions (\ref{s1.6}), (\ref{Sn-comp}), (\ref{Se}).
For this purpose we will calculate the divergent parts of the
effective actions of these models. By definition, the effective action in quantum
field theory is a generating functional of all connected one-particle
irreducible Green functions. It encodes the
full information about the quantum dynamics of the given field theory and, in
particular, allows one to find the structure of quantum
divergences. To obtain the divergent parts of the effective
actions we employ here the standard methods of quantum field theory based
on the Feynman diagram techniques.

\subsection{Gauge superfield model}\label{SYM-quant}
Consider the non-anticommutative model of abelian gauge superfield
in its component formulation with the classical action
(\ref{s1.6}). As a first step, we eliminate the auxiliary field
${\cal D}^{ij}$ by its equation of motion,
\be
{\cal
D}^{ij}=-\frac{8I\bar\Psi^i_{\dot\alpha}\bar\Psi^{j\dot\alpha}}{
1+4I\bar\phi}.
\label{aux}
\ee
Upon substituting (\ref{aux}) into (\ref{s1.8}),
the action for the spinor fields takes the form
\be
S_\Psi={\rm i}\int d^4x\left(
\Psi^{i\alpha}+\frac{4IA_m\sigma_m{}^{\alpha}{}_{\dot\alpha}\bar\Psi^{i\dot\alpha}}{
1+4I\bar\phi}
\right)(\sigma_n)_{\alpha\dot\beta}\partial_n
\left(\frac{\bar\Psi_i^{\dot\beta}}{1+4I\bar\phi} \right).
\label{s1.10}
\ee
In what follows we will consider the quantization of the model (\ref{s1.6})
with the action $S_\Psi$ given by (\ref{s1.10}).

Since the action (\ref{s1.6}) is invariant under the gauge
transformations (\ref{resid-comp}) one needs to fix the gauge
to quantize the theory. It is convenient to choose the
following gauge-fixing condition
\be
\partial_m\frac{A_m}{1+4I\bar\phi}=0.
\label{s1.12}
\ee
Note that (\ref{s1.12}) is none other than the Lorentz gauge
condition $\partial_ma_m=0$ for the gauge field
$a_m=A_m/(1+4I\bar\phi)$ which transforms in a standard way
under the ${\rm U}(1)$ gauge group, $\delta a_m=\partial_m\lambda\,$.

Further we follow the routine of Faddeev-Popov procedure to fix the
gauge freedom in the functional integral.
Let us introduce the corresponding gauge-fixing function
\be
\chi=\partial_m\frac{A_m}{1+4I\bar\phi}=
\frac{\partial_m A_m-A_m G_m}{1+4I\bar\phi},
\label{s1.13}
\ee
where
\be
G_m(x)=\partial_m \,\ln[1+4I\bar\phi(x)]\,.
\label{s1.14}
\ee
The function $\chi$ transforms under gauge transformations
(\ref{resid-comp}) as follows
\be
\delta\chi=\partial_m\frac{\delta
A_m}{1+4I\bar\phi}=\square\lambda.
\label{s1.15}
\ee
The relation (\ref{s1.15}) shows that the action for the
ghost fields is just the action of free scalars
\be
S_{FP}=\int d^4x\, b\square c\,.
\label{s1.16}
\ee
The generating functional for the Green's functions is now defined
as
\be
Z[J]=\int{\cal D}(\phi,\bar\phi,\Psi,\bar\Psi,A_m,b,c)
\delta(\chi-\tfrac{\partial_m A_m-A_m G_m}{1+4I\bar\phi})
e^{-\frac12( S_{SYM}+S_{FP}+S_J)}\,,
\label{s1.17}
\ee
where
\be
S_J=\int d^4x[\phi J_\phi+\bar\phi J_{\bar\phi}
+\Psi^i_\alpha (J_{\Psi})_i^\alpha+
\bar\Psi_{i\dot\alpha} (J_{\bar\Psi})^{i\dot\alpha}+ A_m (J_A)_m]
\label{s1.18}
\ee
and $J_\phi$, $J_{\bar\phi}$,
$(J_\Psi)_i^\alpha$, $(J_{\bar\Psi})^{i\dot\alpha}$, $(J_A)_m$
are sources for the fields
$\phi$, $\bar\phi$, $\Psi^i_\alpha$, $\bar\Psi_{i\dot\alpha}$,
$A_m$. We have inserted into (\ref{s1.17})
the functional delta-function that fixes the gauge degrees of freedom in
the functional integral over the gauge fields. This delta-function
can be easily written in the Gaussian form by
averaging (\ref{s1.17}) with the factor
\be
1=\int {\cal D}\chi e^{-\frac\alpha2\int d^4x  \chi^2(1+4I\bar\phi)^2}
={\rm Det}^{-1/2}[\delta^4(x-x')(1+4I\bar\phi)^2]\,.
\label{s1.19}
\ee
The functional integral (\ref{s1.19}) produces the following
gauge-fixing action
\begin{eqnarray}
S_{gf} &=&\frac\alpha2\int d^4x(\partial_m A_m-A_m G_m)^2
\nonumber\\
&=&\frac\alpha2\int d^4x[(\partial_m A_m)^2-2\partial_m A_m A_n G_n
+A_mA_n G_mG_n]\,.
\label{s1.20}
\end{eqnarray}
Here $\alpha$ is an arbitrary parameter. For simplicity, in the sequel we set
$\alpha=1\,$. As a result, the generating functional
(\ref{s1.17}) can be represented in the following form
\be
Z[J]=\int{\cal D}(\phi,\bar\phi,\Psi,\bar\Psi,A_m,b,c)
e^{-\frac12(S_{\rm tot}+S_{FP}+S_J)}\,,
\label{s1.21}
\ee
where
\begin{eqnarray}
S_{\rm tot}&=&S_{SYM}+S_{gf}\nonumber\\
&=&-\frac12\int d^4x\square\bar\phi(\phi
+4I^2\partial_m\bar\phi G_m)\nonumber\\
&&+\,{\rm i}\int d^4x\left(
\Psi^{i\alpha}+\frac{4IA_m\sigma_m{}^{\alpha}{}_{\dot\alpha}\bar\Psi^{i\dot\alpha}}{
1+4I\bar\phi}
\right)(\sigma_n)_{\alpha\dot\beta}\partial_n
\left(\frac{\bar\Psi_i^{\dot\beta}}{1+4I\bar\phi} \right)\nonumber\\
&&
-\int d^4x\left[\frac12 A_n\square A_n- A_n G_m\partial_n A_m+
A_n G_n\partial_m A_m+\varepsilon_{mnrs}G_mA_n\partial_r
A_s\right].
\label{s1.22}
\end{eqnarray}

The functional integral (\ref{s1.21}) with the action (\ref{s1.22})
requires several comments.

\begin{enumerate}
\item The ghost fields $b,\ c$ enter the action only through
their kinetic term. Hence, they fully decouple and can be integrated
out.
\item The fermionic fields $\Psi^i_\alpha,\
\bar\Psi^i_{\dot\alpha}$ do not contribute to the effective action.
Indeed, the action $S_\Psi$ (\ref{s1.10}) can be brought to the
form of free action by the following change of fields
$\psi^{i\alpha}=\Psi^{i\alpha}+\frac{4IA_m\sigma_m{}^{\alpha}{}_{\dot\alpha}\bar\Psi^{i\dot\alpha}}{
1+4I\bar\phi}$, $\bar\psi_i^{\dot\beta}=\frac{\bar\Psi_i^{\dot\beta}}{1+4I\bar\phi}
$. One can also check this observation by the direct computations
of the corresponding Feynman diagrams.
\item The contribution to the effective action from the scalar
field $\phi$ is also trivial since it appears in the action
(\ref{s1.22}) without interaction with other fields.
\item A non-trivial contribution to the effective action in this
model comes only from the terms in the last line of (\ref{s1.22}).
These terms are quadratic in the vector field $A_m$ and linear
with respect to $G_m$. Hence, the field $G_m$ appears only on the
external lines while $A_m$ works only inside the Feynman
diagrams. Moreover, there are only one-loop diagrams since there
are no self-interaction of $A_m$. Since the field $G_m$
is expressed only through $\bar\phi$ as in (\ref{s1.14}),
we conclude that the effective action is a functional of $\bar\phi$
only. The dimensional considerations allow one to construct only
the following three terms in the effective action
\be
\Gamma=\int d^4x[
f_1(I\bar\phi)I^2\square\bar\phi\square\bar\phi
+f_2(I\bar\phi)I^3\square\bar\phi
\partial_m\bar\phi\partial_m\bar\phi
+f_3(I\bar\phi)I^4(\partial_m\bar\phi\partial_m\bar\phi)^2]\,,
\label{s1.23}
\ee
where $f_1,\ f_2,\ f_3$ are some functions. The Feynman graph
computations should specify these functions.
\end{enumerate}

Taking into account these comments, we conclude that the
effective action in this model can be represented by the following
formal expression
\footnote{Note that the one-loop effective action
in the Euclidean space is given by $\Gamma=\frac 12\Tr\ln S''[\Phi]$ rather
than the Minkowski space expression $\Gamma=\frac{\rm i}2\Tr\ln S''[\Phi]\,$.
Here $S''[\Phi]$ is the second functional derivative of the classical action.}
\be
\Gamma^{SYM}=\frac12\Tr\ln\frac{\delta^2 \tilde S}{\delta A_p(x)\delta
A_q(x')}\,,
\label{s1.23_}
\ee
where $\tilde S$ is the last line in (\ref{s1.22}),
\be
\tilde S=\int d^4x\left[-\frac12 A_n\square A_n+ A_n G_m\partial_n A_m-
A_n G_n\partial_m A_m-\varepsilon_{mnrs}G_mA_n\partial_r
A_s\right].
\label{s1.24}
\ee
The second functional derivative of the action (\ref{s1.24}) can be easily
calculated,
\be
\frac{\delta^2\tilde S}{\delta A_p(x)\delta A_q(x')}=
-\delta_{pq}\square\delta^4(x-x')
+4G_{[q}\partial_{p]}\delta^4(x-x')
+2\varepsilon_{pqmn} G^m\partial^n\delta^4(x-x')\,.
\label{s1.25}
\ee
Substituting (\ref{s1.25}) into (\ref{s1.23_}) we have
\begin{eqnarray}
\Gamma^{SYM}&=&\frac 12 \Tr\ln[
\delta_{pq}\delta^4(x-x')
+4G_{[p}\partial_{q]}\frac{1}\square\delta^4(x-x')
-2\varepsilon_{pqmn} G_m\frac{\partial_n}\square\delta^4(x-x')]
\nonumber\\&=&
\frac 12 \Tr\sum_{j=1}^\infty \frac{(-1)^{j+1}}j
[4G_{[p}\partial_{q]}\frac{1}\square\delta^4(x-x')
-2\varepsilon_{pqmn} G_m\frac{\partial_n}\square\delta^4(x-x')]^j\,.
\label{s1.26}
\end{eqnarray}
The expression (\ref{s1.26}) provides us with the perturbative
expansion of the effective action in a form of Feynman diagram
series with the external lines $G_m$\,.

The propagators in (\ref{s1.26}) appear in the combination
$\partial_m{\square}^{-1}\delta^4(x-x')\,$. On the dimensionality
grounds, only the expressions like
\be
\left[\frac{\partial_m}\square\delta^4(x-x')\right]^2,\quad
\left[\frac{\partial_m}\square\delta^4(x-x')\right]^3,\quad
\left[\frac{\partial_m}\square\delta^4(x-x')\right]^4
\label{s1.26_}
\ee
are divergent and all higher powers of $\partial_m{\square}^{-1}\delta^4(x-x')$ produce
finite contributions to the effective action. Therefore, only two-
three- and four-point diagrams lead to the divergent contributions
in the effective action (note that the external line is that of the field
$G_m$). We are interested solely in the divergent contributions to the
effective action, and consider the calculations of two-,
three- and four-point functions separately.

Let us consider only the terms in the series (\ref{s1.26})
which are responsible for the two-, three- and four-point
diagram contributions,
\begin{eqnarray}
\Gamma_2^{SYM}&=&- \int d^4x_1 d^4x_2
[2G_{[q}(x_1)\partial_{p]}\frac1\square\delta^4(x_1-x_2)
+\varepsilon_{pqmn}G_m(x_1)\partial_n
\frac1\square\delta^4(x_1-x_2)]\nonumber\\&&
\qquad\times[
2G_{[p}(x_2)\partial_{q]}\frac1\square\delta^4(x_2-x_1)+
\varepsilon_{qprs}G_r(x_2)\partial_s\frac1\square\delta^4(x_2-x_1)]\,,
\label{s1.27}\\
\Gamma_3^{SYM}&=&\frac{4}3\int d^4x_1 d^4x_2 d^4x_3
[(G_t(x_1)\partial_u-G_u(x_1)\partial_t-\varepsilon_{tumn}G_m(x_1)\partial_n)
\frac1\square\delta^4(x_1-x_2)]\nonumber\\
&&\qquad\times
[(G_u(x_2)\partial_w-G_w(x_2)\partial_u-\varepsilon_{uwrs}G_r(x_2)\partial_s)
\frac1\square\delta^4(x_2-x_3)]\nonumber\\
&&\qquad\times
[(G_w(x_3)\partial_t-G_t(x_3)\partial_w-\varepsilon_{wtpq}G_p(x_3)\partial_q)
\frac1\square\delta^4(x_3-x_1)]\,,
\label{s1.32}\\
\Gamma_4^{SYM}&=&-2\int d^4x_1d^4x_2d^4x_3d^4x_4
[(G_p\partial_q-G_q\partial_p-\varepsilon_{p'qq'p} G_{p'}\partial_{q'})\frac1\square
\delta^4(x_1-x_2)]\nonumber\\&&\times
[(G_q\partial_m-G_m\partial_q-\varepsilon_{m'mn'q} G_{m'}\partial_{n'})\frac1\square
\delta^4(x_2-x_3)]\nonumber\\&&\times
[(G_m\partial_n-G_n\partial_m-\varepsilon_{r'ns'm} G_{r'}\partial_{s'})\frac1\square
\delta^4(x_3-x_4)]\nonumber\\&&\times
[(G_n\partial_p-G_p\partial_n-\varepsilon_{tpun} G_t\partial_u)\frac1\square
\delta^4(x_4-x_1)]\,.
\label{s1.34}
\end{eqnarray}
To proceed, one has to perform the integrations over $x_2$, $x_3$,
$x_4$ in the expressions (\ref{s1.27}), (\ref{s1.32}),
(\ref{s1.34}) using the Fourier representation for the fields and
delta-functions. The corresponding divergent momentum integrals
should be regularized by the standard methods of quantum field
theory using, e.g., the dimensional regularization. Here we omit
the details of these computations which can be found in \cite{BILSZ}.
As a result, the divergent parts of the functions
(\ref{s1.27}), (\ref{s1.32}), (\ref{s1.34}) are given by
\begin{eqnarray}
\Gamma_{2,div}^{SYM}&=&\frac1{16\pi^2\varepsilon}\int d^4x \ln(1+4I\bar\phi)
\square^2\ln(1+4I\bar\phi)\,,
\label{s1.31}\\
\Gamma_{3,div}^{SYM}&=&
-\frac{1}{4\pi^2\varepsilon}
\int d^4x\partial_m\ln(1+4I\bar\phi)\partial_m\ln(1+4I\bar\phi)
\square\ln(1+4I\bar\phi)\,,
\label{s1.33}\\
\Gamma_{4,div}^{SYM}&=&-\frac{5}{16\pi^2\varepsilon}\int d^4x
[\partial_m\ln(1+4I\bar\phi)\partial_m\ln(1+4I\bar\phi)]^2\,.
\label{s1.35}
\end{eqnarray}
Here $\varepsilon$ is a parameter of dimensional regularization,
$\varepsilon=2-d/2\,$, where $d$ is the dimension of space-time.
The limit $\varepsilon\to0$ takes off the regularization.
The full divergent contribution to the effective action is given
by the sum of (\ref{s1.31}), (\ref{s1.33}) and (\ref{s1.35}).
Using the integration by parts, the divergent contribution to the effective
action can be brought to the form
\be
\Gamma_{div}^{SYM}=\Gamma^{SYM}_{2,div}+
\Gamma^{SYM}_{3,div}+\Gamma^{SYM}_{4,div}
=\frac1{\pi^2\varepsilon}\int d^4x \frac{I^2\square\bar\phi
\square\bar\phi}{(1+4I\bar\phi)^2}
-\frac6{\pi^2\varepsilon}\int d^4x \frac{
4I^3\square\bar\phi\partial_m\bar\phi\partial_m\bar\phi}{(1+4I\bar\phi)^3}\,.
\label{s1.37}
\ee
Note that the action (\ref{s1.37}) matches with
the previously guessed structure (\ref{s1.23}).

At first sight, the non-anticommutative supergauge model looks non-renormalizable, since the quantum
computations produce the expressions (\ref{s1.37}) which are absent
in the classical action (\ref{s1.6}). However, it is easy to see
that the divergent terms (\ref{s1.37}) being added to the
classical action (\ref{s1.6}) can be completely compensated by the
following shift of scalar field $\phi$
\be
\phi\longrightarrow\phi-\frac2{\pi^2\varepsilon}
\frac{I^2\square\bar\phi}{(1+4I\bar\phi)^2}
+\frac{12}{\pi^2\varepsilon}
\frac{4I^3\partial_m\bar\phi\partial_m\bar\phi}{(1+4I\bar\phi)^3}\,.
\label{s1.39}
\ee
Therefore, the $\cN{=}(1,0)$ gauge model is renormalizable in the sense that all
divergences can be removed by the redefinition of the scalar field $\phi\,$.
One can treat (\ref{s1.39}) as a change of fields in the
functional integral (\ref{s1.21}). Since the Jacobian of such a
change of functional variables is equal to unity, the terms (\ref{s1.37}), being
added to the classical action (\ref{s1.6}), do not contribute to
the effective action. Moreover, this model
is finite since the shift (\ref{s1.39})
allows one to completely eliminate the divergences from the effective
action.

This situation is analogous to the $\cN{=}(1/2,0)$ SYM model
considered in \cite{Jack}, where it was demonstrated that the
quantum computations in this model generate the divergent terms
which are not present in the classical action of the model, but
these extra divergences can be removed by a simple shift of the
gaugino field (the lowest component in $\cN{=}(1/2,1/2)$ gauge
multiplet). In our case the divergences can also be removed by
the shift of lowest component of $\cN{=}(1,1)$ gauge multiplet
(scalar field).

It should also be noted that the divergent expression
(\ref{s1.37}) vanishes on the classical equation of motion for the
scalar field $\bar\phi$ given by $\square\bar\phi=0$.
Therefore the $S$-matrix in this model is free of divergences and in
this sense one can say that the model under consideration is finite.

\subsection{Neutral hypermultiplet model}
\label{ren-hyp}
Consider the model of neutral hypermultiplet with the classical
action (\ref{Sn-comp}). Clearly, the hypermultiplet fields
$f^{ak}$, $\rho^{\alpha a},\ \chi_a^{\dot\alpha}$ work
only inside the loops of Feynman diagrams while the vector multiplet
fields $\bar\phi$, $\bar\Psi_k^{\dot\alpha}$, $A_m$ appear only on
external lines. Moreover, since the action (\ref{Sn-comp}) is
quadratic with respect to the hypermultiplet fields, the
corresponding effective action is one-loop exact.
It is easy to observe also that the terms in the second line of
the action (\ref{Sn-comp}) correspond to the interaction vertices
which do not couple with the other vertices in one-loop diagrams.
Indeed, to form a loop with these vertices one needs the
propagators $\langle\rho^{\alpha a} f^{bk}\rangle$,
$\langle\rho^{\alpha a}\rho_{\beta b}\rangle$ which are absent in
this model.

Let us analyze also the term $\frac{\rm i}2(1+4I\bar\phi)\rho^{\alpha a}\partial_{\alpha\dot\alpha}
\chi^{\dot\alpha}_a$ in the first line of (\ref{Sn-comp}). It
is easy to see that this term does not contribute to the effective
action,
\begin{eqnarray}
\Gamma_{ferm}&=&-\Tr\ln\frac{\delta^2 S_{ad}}{\delta\rho^{\alpha a}(x)
\delta\chi_b^{\dot\alpha}(x')}=
-\Tr\ln\left[\frac{\rm i}2(1+4I\bar\phi)\partial_{\alpha\dot\alpha}\delta^4(x-x')
\delta_a^b\right]\nonumber\\&=&
-2\Tr\ln[\frac{\rm i}2(1+4I\bar\phi)\delta^4(x-x')]
-2\Tr\ln[\partial_{\alpha\dot\alpha}\delta^4(x-x')]
\simeq0\,.
\label{s2.5}
\end{eqnarray}

As a result, the non-trivial contribution to the effective action
comes only from the loops with the internal lines given by the scalar
fields $f^{ak}$ and with the field $\bar\phi$ on external lines.
This contribution is given by the following formal expression
\be
\Gamma^{hyp}=\frac12\Tr\ln\frac{\delta^2}{\delta f^{ak}(x)\delta f_{a'k'}(x')}
\left[\frac12\int d^4x (1+4I\bar\phi)^2\partial_m
f^{ak}\partial_m f_{ak} \right].
\label{s2.6}
\ee
Calculating the variational derivative in (\ref{s2.6}), we
obtain
\be
\Gamma^{hyp}=
2\Tr\ln\left[(1+4I\bar\phi)^2\square\delta^4(x-x') \right]
+2\Tr\ln\left[\delta^4(x-x')+2\frac1\square G_m(x)\partial_m\delta^4(x-x')
 \right].
\label{s2.7}
\ee
The first term in the r.h.s. of (\ref{s2.7}) is trivial since it
is proportional to $\delta^4(0)$ that is zero in the sense of
dimensional regularization. The second term in the r.h.s. of
(\ref{s2.7}) provides us with the following perturbative
representation for the effective action
\be
\Gamma^{hyp}=2\Tr\sum\limits_{n=1}^\infty\frac{(-1)^{n+1}}{n}
\left[\frac2\square G_m(x)\partial_m\delta^4(x-x')\right]^n.
\label{s2.8}
\ee
Note that the fields $G_m$ in the series (\ref{s2.8}) play the
role of external lines of corresponding Feynman diagrams.
Taking into account the dimensions of field $G_m$
and propagators $\frac1\square\partial_m\delta^4(x-x')$
we conclude that only the diagrams with two, three and four
external lines are divergent. Let us consider these divergent
terms in the series (\ref{s2.8}), which corresponds to $n=2,3,4$:
\begin{eqnarray}
\Gamma^{hyp}_2&=&-4\int d^4x_1d^4x_2 G_m(x_1) G_n(x_2)
\frac{\partial_m}{\square}\delta^4(x_1-x_2)
\frac{\partial_n}{\square}\delta^4(x_2-x_1)\,,
\label{s2.9}\\
\Gamma_3^{hyp}&=&\frac{16}3\int d^4x_1d^4x_2d^4x_3 G_m(x_1)G_n(x_2)G_p(x_3)
\nonumber\\&&\times
\frac{\partial_m}{\square}\delta^4(x_1-x_2)
\frac{\partial_n}{\square}\delta^4(x_2-x_3)
\frac{\partial_p}{\square}\delta^4(x_3-x_1)\,,
\label{s2.13}\\
\Gamma_4^{hyp}&=&-8\int d^4x_1 d^4x_2d^4x_3d^4x_4
G_m(x_1)G_n(x_2)G_p(x_3)G_r(x_4)\nonumber\\&&\times
\frac{\partial_m}\square\delta^4(x_1-x_2)
\frac{\partial_n}\square\delta^4(x_2-x_3)
\frac{\partial_p}\square\delta^4(x_3-x_4)
\frac{\partial_r}\square\delta^4(x_4-x_1).
\label{s2.17}
\end{eqnarray}
The expressions (\ref{s2.9}), (\ref{s2.13}), (\ref{s2.17}) can be
calculated by the standard methods of quantum field theory, see
\cite{BILSZ} for details. Here we give only the results,
\begin{eqnarray}
\Gamma_{2,div}^{hyp}&=&-\frac1{16\pi^2\varepsilon}\int d^4x\ln(1+4I\bar\phi(x))\square^2
\ln(1+4I\bar\phi(x))\,,
\label{s2.12}\\
\Gamma_{3,div}^{hyp}&=&-\frac1{8\pi^2\varepsilon}
\int d^4x\square\ln(1+4I\bar\phi)
\partial_n\ln(1+4I\bar\phi)\partial_n\ln(1+4I\bar\phi)\,,
\label{s2.16}\\
\Gamma_{4,div}^{hyp}&=&-\frac{1}{16\pi^2\varepsilon}\int d^4x
[\partial_m\ln(1+4I\bar\phi)
\partial_m\ln(1+4I\bar\phi)]^2\,.
\label{s2.19}
\end{eqnarray}
Summarizing (\ref{s2.12}), (\ref{s2.16}), (\ref{s2.19}) we
obtain the full divergent contribution to the hypermultiplet
effective action,
\be
\Gamma_{div}^{hyp}=-\frac{1}{\pi^2\varepsilon}\int d^4x\frac{I^2
\square\bar\phi\square\bar\phi}{(1+4I\bar\phi)^2}\,.
\label{s2.21}
\ee
Note that (\ref{s2.21}) agrees with the general
structure of the effective action (\ref{s1.23}) guessed before.

As in the deformed gauge model, there are no terms in the classical
action (\ref{s1.6}) having the field structure similar to
(\ref{s2.21}). Therefore, at first sight the multiplicative
renormalizability of the model can be spoiled by the divergent
contribution (\ref{s2.21}). However, it is easy to observe that
the term (\ref{s2.21}), being added to the classical action
(\ref{s1.6}), can be completely compensated by the following shift
of scalar field
\be
\phi\longrightarrow
\phi+\frac2{\pi^2\varepsilon}\frac{I^2\square\bar\phi}{(1+4I\bar\phi)^2}\,,
\label{s2.23}
\ee
while the other fields remain intact.
Since the Jacobian of the change (\ref{s2.23}) is equal
to unity, we conclude that the term (\ref{s2.21}) does not spoil the
renormalizability and finiteness of the model.
Moreover, the divergent term (\ref{s2.21}) vanishes on the
classical equation of motion for the scalar field $\bar\phi$
given by $\square\bar\phi=0$. Therefore the finiteness of
$S$-matrix is also evident.

Let us consider finally the general abelian $\cN{=}(1,0)$
non-anticommutative model of gauge superfield interacting with the
hypermultiplet matter. It is described by the classical action
\be
S=S_{SYM}+S_{ad},
\label{s2.24}
\ee
where $S_{SYM}$ and $S_{hyp}$
are given by (\ref{s1.6}), (\ref{Sn-comp}), respectively. It is easy to
see that the divergent part of the effective action in this model
is given by the sum of the expressions (\ref{s1.37}) and
(\ref{s2.21}),
\be
\Gamma_{div}=\Gamma_{div}^{SYM}+\Gamma_{div}^{hyp}=
-\frac6{\pi^2\varepsilon}\int d^4x \frac{
4I^3\square\bar\phi\partial_m\bar\phi\partial_m\bar\phi}{(1+4I\bar\phi)^3}\,.
\label{s2.25}
\ee
The divergent expression (\ref{s2.25}) can also
be completely compensated by the following shift of the scalar
$\phi$, \be \phi\longrightarrow\phi
+\frac{12}{\pi^2\varepsilon}\frac{4I^3\partial_m\bar\phi\partial_m\bar\phi}{(1+4I\bar\phi)^3}\,.
\label{s2.26} \ee As a result, the general $\cN{=}(1,0)$
supergauge theory is also renormalizable and finite.

\subsection{Charged hypermultiplet model}
\label{hyp}
The model of charged hypermultiplet is described by the superfield
action (\ref{Sqf}) or the corresponding component field action
(\ref{Se-comp}). Note that the component action (\ref{Se-comp})
is much more complicated than the actions of neutral
hypermultiplet and gauge superfield considered above. Therefore,
to prove the renormalizability of the charged hypermultiplet model
we prefer to use the superfield description (\ref{Sqf}).

In superfields, the effective action of charged hypermultiplet is
given by the following formal expression
\be
\Gamma=\Tr\ln\frac{\delta^2S_{f}}{\delta\tilde q^+(1)\delta q^+(2)}
=\Tr\ln(D^{++}+V^{++}\star).
\label{eq1}
\ee
The free Green function of hypermultiplet superfield has the
standard form \cite{Book},
\be
G_0^{(1,1)}(1|2)=-\frac1\square(D^+_1)^4(D^+_2)^4
\frac{\delta^{12}(z_1-z_2)}{(u^+_1u^+_2)^3}.
\label{eq2}
\ee
It solves free equation of motion with the delta-source,
$D^{++}G_0^{(1,1)}(1|2)=\delta_A^{(1,3)}(1|2)$,
where $\delta_A^{(1,3)}(1|2)$ is an analytic delta-function
(see \cite{Book} for the details of the harmonic superspace approach).
The effective action (\ref{eq1}) can be formally expressed
through the free Green function (\ref{eq2}) as
\be
\Gamma=\Tr\ln[\delta_A^{(3,1)}(1|2)+V^{++}(1)\star
G_0^{(1,1)}(1|2)].
\label{eq4}
\ee
Expanding the logarithm in (\ref{eq4}) in a series, we obtain
a perturbative representation for the effective action,
\begin{eqnarray}
\Gamma&=&\sum_{n=2}^{\infty}\Gamma_n,\label{eq5_}\\
\Gamma_n&=&\frac{(-1)^{n+1}}{n}\int d\zeta_1 du_1\ldots d\zeta_n du_n
V^{++}(1)\star G_0^{(1,1)}(1|2)\ldots V^{++}(n)\star G_0^{(1,1)}(n|1).
\label{eq5}
\end{eqnarray}
We calculate further the divergent parts of the $n$-point
functions $\Gamma_n$.

As the first step we restore in (\ref{eq5}) the full
$\cN{=}(1,1)$ superspace integration measure with the help of
$(D^+)^4$ factors of the propagator (\ref{eq2}). For this purpose
we apply the following standard identity \be d^{12}z=d^4x
d^8\theta=d\zeta(D^+)^4. \label{full_measure} \ee As a result, the
$n$-point Green function reads
\begin{eqnarray}
\Gamma_n&=&-\frac{1}{n}\int d^{12}z_1 du_1 \ldots d^{12}z_n du_n
V^{++}(1)\star \frac1\square (D^+_1)^4\frac{\delta^{12}(z_1-z_2)}{
(u^+_1u^+_2)^3}\nonumber\\&&\times
V^{++}(2)\star \frac1\square (D^+_2)^4\frac{\delta^{12}(z_2-z_3)}{
(u^+_2u^+_3)^3}\ldots
V^{++}(n)\star \frac1\square (D^+_n)^4\frac{\delta^{12}(z_n-z_1)}{
(u^+_nu^+_1)^3}.
\label{e3.1}
\end{eqnarray}
Further we integrate by parts and take off the integration over Grassmann
variables $\theta_2,\ldots,\theta_{n-1}$ using the corresponding
delta-functions. As a consequence, for the expression (\ref{e3.1})
we have
\begin{eqnarray}
\Gamma_n&=&-\frac{1}{n}\int d^{12}z_1 d^{12}z_n d^4x_2\ldots
d^4x_{n-1} d^nU X_n[U]\delta^8(\theta_1-\theta_n)
\nonumber\\&&\times
V^{++}(n)\star(D^+_{n-1})^4[(D^+_{n-2})^4[\ldots
\nonumber\\&&
(D^+_2)^4[ (D^+_1)^4(D^+_n)^4\frac1\square \delta^{12}(z_n-z_1)
\star V^{++}(1)\frac1\square\delta^4(x_1-x_2)]
\nonumber\\&&
\star V^{++}(x_2,\theta_1,u_2)\frac1\square\delta^4(x_2-x_3)]
\ldots\star V^{++}(x_{n-2},\theta_1,u_{n-2})\frac1\square
\delta^4(x_{n-2}-x_{n-1})]\nonumber\\&&
\star V^{++}(x_{n-1},\theta_1,u_{n-1})\frac1\square\delta^4(x_{n-1}-x_n).
\label{e3.2}
\end{eqnarray}
Here we have introduced the denotations
 $d^nU= du_1\ldots du_n$, $X_n[U]=1/[(u^+_1u^+_2)^3\ldots (u^+_nu^+_1)^3]$.
Note that the covariant derivatives in (\ref{e3.2}) commute
with the $\star$-product operators and hit only the corresponding $V^{++}$
superfields and delta-functions.

To integrate over the remaining Grassmann variables in (\ref{e3.2})
we make the Fourier transform for superfields and delta-functions,
\begin{eqnarray}
V^{++}(1)&=&\int \frac{d^4p_1}{(2\pi)^4} d^4\rho_1
e^{{\rm i} p_1 x_1}e^{\rho_1\theta_1}\tilde V^{++}(p_1,\rho_1,\bar\theta_1,u_1),
\label{eq8}\\
\delta^{12}(z_1-z_2)&=&\int \frac{d^4k_1}{(2\pi)^4}d^4\pi_1
e^{{\rm i} k_1(x_1-x_2)}e^{\pi_1(\theta_1-\theta_2)}\delta^4(\bar\theta_1-\bar\theta_2),
\label{eq10}
\end{eqnarray}
where we denote $\pi\theta=\pi_i^\alpha\theta^i_\alpha=
-\theta_i^\alpha\pi^i_\alpha=-\theta\pi$. The $\star$-product of
Fourier transforms of arbitrary two superfunctions $f$, $g$
is given by
\be
f\star g =\int \frac{d^4p d^4k}{(2\pi)^8}d^4\rho d^4\pi
\tilde f(p,\rho)\tilde g(k,\pi)e^{{\rm i}(p+k)x}e^{(\rho+\pi)\theta}
e^{-IQ^\alpha_i(p,\rho,\bar\theta)Q^i_\alpha(k,\pi,\bar\theta)},
\label{eq14}
\ee
where $Q^i_\alpha(p,\rho,\bar\theta)=(-\rho^i_\alpha+
\bar\theta^{i\dot\alpha}\sigma^m_{\alpha\dot\alpha}p_m)$.
As a result, the equation (\ref{e3.2}) reads
\begin{eqnarray}
\Gamma_n&=&\frac{(-1)^{n+1}}{n}
\int d^4\bar\theta_1 d^4\bar\theta_n \frac{d^4p_1\ldots
d^4p_n}{(2\pi)^{4n}}d^4\rho_1\ldots d^4\rho_n
d^4k_n d^4\pi_n d^nU X_n[U]\nonumber\\&&
\delta^4(\bar\theta_1-\bar\theta_n)
\delta^4(\tsum p_i) \delta^4(\tsum \rho_i)
\frac{\exp[\tsum^n_{i,j=2 (i<j)}-IQ(\rho_i)Q(\rho_j)]
}{k_n^2(k_n-p_1)^2\ldots (k_n-\tsum_{i=1}^{n-1}p_i)^2}
\nonumber\\&&
(D^+_{n-1})^4
[(D^+_{n-2})^4[\ldots
[(D^+_2)^4
[(D^+_1)^4
(D^+_n)^4
\delta^{4}(\bar\theta_n-\bar\theta_1)]
\nonumber\\&&
V^{++}(p_1,\rho_1,\bar\theta_1,u_1)]
V^{++}(p_2,\rho_2,\bar\theta_1,u_2)]
\ldots \nonumber\\&&
V^{++}(p_{n-2},\rho_{n-2},\bar\theta_1,u_{n-2})]
V^{++}(p_{n-1},\rho_{n-1},\bar\theta_1,u_{n-1})
V^{++}(p_n,\rho_n,\bar\theta_n,u_n).
\label{e3.3}
\end{eqnarray}
Note that in this representation the derivatives $\bar D^+_{\dot\alpha}$
differentiate only $\bar\theta$ variables while $D^+_\alpha(\pi,u)$
is nothing but a multiplication operator on $u^+_i\pi ^i_\alpha$.
Hence, we can apply the following identities
\begin{eqnarray}
\frac1{16}\delta^4(\bar\theta_1-\bar\theta_n)(\bar D^+_1)^2(\bar D^+_n)^2\delta^4(\bar\theta_n
-\bar\theta_1)&=&(u^+_1u^+_n)^2,
\label{e3.4}\\
\frac1{16}\int d^4\pi (D^+(\pi,u_{n-1}))^2((D^+(\pi,u_{n}))^2&=&(u^+_{n-1}u^+_n)^2
\label{e3.5}
\end{eqnarray}
to simplify (\ref{e3.3}). After these manipulations we are left with $n-2$
differential operators in (\ref{e3.3}) which give at most
the momentum $(k^2)^{n-2}$ on condition that these
derivatives do not hit the external lines. Clearly, this
corresponds to the logarithmic divergence of the momentum
integral over $d^4k$. The terms with derivatives on the external
lines $V^{++}$ are always finite since they have the power of
momentum less than $n-2$. Here we are interested only in the
divergent contributions to the effective action and consider therefore only
the terms in (\ref{e3.3}) without derivatives on $V^{++}$ superfield.
As a result, the divergent part of the effective action (\ref{e3.3})
is given by
\begin{eqnarray}
\Gamma_{n,div}&=&\frac{(-1)^{n+1}}{n}\frac1{16^n}\int \frac{d^4p_1}{(2\pi)^4}du_1 d^4\rho_1
\ldots \frac{d^4p_n}{(2\pi)^4}du_n d^4\rho_n d^4\bar\theta_1
d^4\bar\theta_nd^4\pi d^4k
\nonumber\\&&\times
V^{++}(p_1,\rho_1,\bar\theta_1,u_1)\ldots
V^{++}(p_n,\rho_n,\bar\theta_1,u_1)
\exp[-I\tsum_{i,j=2; i<j}^n Q(\rho_i)
Q(\rho_j)]
\nonumber\\&&\times
\delta^4(\bar\theta_1-\bar\theta_n)
(D^+_{n-1})^2(\bar D^+_{n-1})^2\ldots
(D^+_{2})^2(\bar D^+_{2})^2(D^+_{1})^2(\bar D^+_{1})^2
(D^+_{n})^2(\bar D^+_{n})^2\delta^4(\bar\theta_1-\bar\theta_n)\nonumber\\&&\times
\frac{\delta^4(\tsum p_i)\delta^4(\tsum \rho_i)}{
k^2(k-p_2)^2\ldots (k-\sum_{l=2}^np_l)^2}
\frac1{(u^+_1u^+_2)^3(u^+_2u^+_3)^3\ldots (u^+_nu^+_1)^3}.
\label{eq33}
\end{eqnarray}
Note that the factor $e^{-I\Sigma QQ}$ in (\ref{eq33})
allows us to restore the $\star$-product of gauge superfields
$V^{++}(u_1)\star V^{++}(u_2)\star\ldots\star V^{++}(u_n)$ after
the inverse Fourier transform.

Now we simplify the chain of covariant derivatives in
(\ref{eq33}). Consider, e.g., the block
 $(D^+_{2})^2(\bar D^+_{2})^2(D^+_{1})^2(\bar
D^+_{1})^2$. Using the identity $D^+_{1\alpha}=(u^+_1u^-_2)D^+_{2\alpha}
-(u^+_1u^+_2)D^-_{2\alpha}$ and anticommutation relations for the
derivatives, $\{D^+_\alpha,\bar D^-_{\dot\alpha}
\}=2k_{\alpha\dot\alpha}$, this expression simplifies to
\be
(D^+_{2})^2(\bar D^+_{2})^2(D^+_{1})^2(\bar D^+_{1})^2
\longrightarrow
16(u^+_1u^+_2)^2k^2 (D^+_2)^2(\bar D^+_1)^2.
\label{eq35}
\ee
Applying the relation (\ref{eq35}) $n-2$ times, we rewrite the
chain of derivatives in (\ref{eq33}) as
\begin{eqnarray}
\delta^4(\bar\theta_1-\bar\theta_n)
(D^+_{n-1})^2(\bar D^+_{n-1})^2\ldots
(D^+_{2})^2(\bar D^+_{2})^2(D^+_{1})^2(\bar D^+_{1})^2
(D^+_{n})^2(\bar D^+_{n})^2\delta^4(\bar\theta_1-\bar\theta_n)&&
\nonumber\\
=16^{n-2}(k^2)^{n-2}
(u^+_1u^+_2)^2\ldots (u^+_{n-2}u^+_{n-1})^2
(D^+_n)^2(D^+_{n-1})^2\delta^4(\bar\theta_1-\bar\theta_n)
(\bar D^+_1)^2(\bar D^+_n)^2
\delta^4(\bar\theta_1-\bar\theta_n).&&
\label{eq36}
\end{eqnarray}
Finally, integrating over $d^4\bar\theta_n d^4\pi_n$  in (\ref{eq33})
and using the identities (\ref{e3.4}), (\ref{e3.5}), we obtain
\be
\Gamma_{n,div}=\frac1{16\pi^2\varepsilon}\frac{(-1)^n}{n}
\int d^{12}z du_1\ldots du_n\frac{V^{++}(u_1)\star V^{++}(u_2)
\star\ldots\star
V^{++}(u_n)}{(u^+_1u^+_2)(u^+_2u^+_3)\ldots(u^+_nu^+_1)}.
\label{eq37}
\ee
The factor $\frac1{16\pi^2\varepsilon}$ appears here due to the
dimensional regularization of logarithmicaly divergent momentum
integral.

The full divergence in the model of charged hypermultiplet is
obtained now by summarizing the divergent parts of $n$-point
functions given by (\ref{eq37}),
\be
\Gamma_{div}=\frac1{16\pi^2\varepsilon}
\sum_{n=2}^\infty\frac{(-1)^n}{n} \int d^{12}z du_1\ldots
du_n\frac{V^{++}(u_1)\star V^{++}(u_2) \star\ldots\star
V^{++}(u_n)}{(u^+_1u^+_2)(u^+_2u^+_3)\ldots(u^+_nu^+_1)}.
\label{eq38}
\ee
As a result, we  conclude that the divergent part
of the effective action in the charged hypermultiplet model
coincides, up to a divergent factor, with the classical action
(\ref{SYM1}) in the model of gauge superfield. This proves the
renormalizability of the $\cN{=}(1,0)$ model of gauge superfield
interacting with the hypermultiplet.

\subsection{Seiberg-Witten transform and renormalizability}
In this section we have proven the renormalizability of abelian
models of gauge multiplet and hypermultiplets by direct
computations of divergent contribution to the effective actions of
these models. Recall that there is a change of classical
fields in these actions (\ref{SW}), (\ref{SW-hyper}), (\ref{SWqe}) which not
only brings the supersymmetry and gauge transformations to the
undeformed form but also essentially simplifies  the structures of the classical
action. We refer to these transformations as the Seiberg-Witten maps.
The Seiberg-Witten maps prove also very useful for the proof of
renormalizability and finiteness of neutral hypermultiplet and gauge multiplet
models. The quantum computations in terms of the transformed fields
explicitly demonstrate  the exact cancellations of divergent terms in the
corresponding effective actions.

The Seiberg-Witten map in the abelain $\cN{=}(1,0)$ supergauge
model (\ref{s1.6}) was derived in Sect.\ \ref{SW-gauge}. Here we
use these  transformations in the form (\ref{SWsimp}),
\begin{eqnarray}
\phi&\rightarrow&\hat\varphi=\phi+\frac{4I}{1+4I\bar\phi}[
A_mA_m+4I^2(\partial_m\bar\phi)^2],\nonumber\\
A_m&\rightarrow&a_m=\frac{A_m}{1+4I\bar\phi},\qquad
\bar\Psi^k_{\dot\alpha}\rightarrow\bar\psi^k_{\dot\alpha}=\frac{\bar\Psi^k_{\dot\alpha}}{1+4I\bar\phi},
\nonumber\\
\Psi^k_\alpha&\rightarrow&\hat\psi^k_\alpha=\Psi^k_\alpha+
\frac{4IA_{\alpha\dot\alpha}\bar\Psi^{\dot\alpha k}}{1+4I\bar\phi},
\nonumber\\
{\cal D}^{kl}&\rightarrow&
 \hat{d}^{kl}=\frac1{1+4I\bar\phi}\left[{\cal D}^{kl}+\frac{8I
\bar\Psi^k_{\dot\alpha}\bar\Psi^{\dot\alpha
l}}{1+4I\bar\phi}\right].
\label{s3.1}
\end{eqnarray}
The action (\ref{s1.6}) in terms of these new fields is rewritten as
\begin{eqnarray}
S_{SYM}&=&\int d^4x ( -\frac12\hat\varphi\square
\bar\phi-{\rm i}\hat\psi_k^\alpha\partial_{\alpha\dot\alpha}
\bar\psi^{\dot\alpha k}+\frac14\hat{d}^{kl}\hat{d}_{kl}) \nonumber\\&&+
\frac14\int d^4x(1+4I\bar\phi)^2
(f_{mn}f_{mn}+\frac12\varepsilon_{mnrs}f_{mn}f_{rs})\,,
\label{s3.2}
\end{eqnarray}
where $f_{mn}=\partial_m a_n-\partial_n a_m$. The form
(\ref{s3.2}) is more preferable for further quantum computations as compared to
the one given by (\ref{SYM3}), since the scalar $\hat{\varphi}$, as well as the spinor and
auxiliary fields, are free in (\ref{s3.2}). The only interaction term is present in the
second line of (\ref{s3.2}). It is an interaction between the vector field strength
and the scalar $\bar\phi$.

The action (\ref{s3.2}) is invariant under the abelian gauge
transformation,
\be
\delta a_m=\partial_m \lambda,
\label{s3.3}
\ee
$\lambda$ being the gauge parameter. We use here the Lorentz gauge
\be
\partial_m a_m=0,
\label{s3.4}
\ee
since the transformation (\ref{s3.3}) has the same form as
in the classical electrodynamics. Further we follow the
the Faddeev-Popov procedure of constructing the functional
integral. Let us introduce the gauge-fixing function
\be
\chi=\partial_m a_m\,,
\label{s3.5}
\ee
which transforms under (\ref{s3.3}) as
$\delta \chi=\square\lambda\,$. Obviously, the ghost fields do
not interact with any other ones and so they completely decouple. The ghost action
is given again by (\ref{s1.16}).
The generating functional for Green's functions is now given by
\footnote{Note that the Jacobian of the change of functional variables
(\ref{s3.1}) is unity since this redefinition of fields is local.}
\be
 Z[J]=\int {\cal D}(\hat\varphi,\bar\phi,\hat\psi,\bar\psi,
a_m,b,c) \delta(\chi-\partial_m a_m)
e^{-\frac12(S_{SYM}+S_{FP}+S_J)}\,,
\label{s3.6}
\ee
where
\be
S_J=\int d^4x[\hat\varphi J_{\hat\varphi}+\bar\phi J_{\bar\phi}
+\hat\psi^i_\alpha (J_{\hat\psi})_i^\alpha+ \bar\psi_{i\dot\alpha}
(J_{\bar\psi})^{i\dot\alpha}+ a_m (J_a)_m]\,.
\label{s3.6.1}
\ee
To represent the delta-function in
the Gaussian form, we average the equation (\ref{s3.6}) with the functional
factor (\ref{s1.19}). As a result, we obtain the gauge-fixing action in the form
\be
 S_{gf}=\frac\alpha2\int
d^4x(1+4I\bar\phi)^2\partial_m a_m
\partial_n a_n\,.
\label{s3.7}
\ee
For simplicity we choose the gauge -fixing parameter $\alpha$ to be equal to unity,
$\alpha=1\,$. Now, the generating functional (\ref{s3.6}) reads
\be
 Z[J]=\int{\cal
D}(\hat\varphi,\bar\phi,\hat\psi,\bar\psi, a_m, b,c) e^{-\frac12(S_{\rm
tot}+S_{FP}+S_J)}\,,
\label{s3.8}
\ee
where
\be
 S_{\rm tot}
=S_{SYM}+S_{gf} =\int d^4x ( -\frac12\hat\varphi\square
\bar\phi-{\rm i}\hat\psi_k^\alpha\partial_{\alpha\dot\alpha}
\bar\psi^{\dot\alpha k}+\frac14\hat{d}^{kl}\hat{d}_{kl}) +S_a\,,
\label{s3.9}
\ee
and
\be
S_a=\frac12\int d^4x(1+4I\bar\phi)^2 (\partial_m a_m
\partial_n a_n+
\partial_m a_n\partial_m a_n-
\partial_m a_n\partial_n a_m
+\varepsilon_{mnrs}\partial_m a_n\partial_r a_s)\,.
\label{s3.10}
\ee

It is evident that the scalar and spinor fields, as well as the ghosts, do not
contribute to the effective action. The only contribution comes from the
part (\ref{s3.10}), namely
\begin{eqnarray}
\Gamma^{SYM}&=&\frac12\Tr\ln\frac{\delta^2 S_a}{\delta
a_p(x)\delta a_q(x')}
\nonumber\\
&=&\frac12\,\Tr\ln[
\delta_{pq}\square\delta^4(x-x')+2\delta_{pq}G_m\partial_m\delta^4(x-x')
\nonumber\\&& +\,4
G_{[p}\partial_{q]}\delta^4(x-x')-2\varepsilon_{pqmn}G_m
\partial_n\delta^4(x-x')
]\,. \label{s3.11}
\end{eqnarray}
The field $G_m(x)$ was defined in (\ref{s1.14}). The
expression (\ref{s3.11}) is the starting point for perturbative
calculations of one-loop effective action in the $\cN{=}(1,0)$
non-an\-ti\-com\-mu\-ta\-ti\-ve SYM model. Note that it resembles
the first line of (\ref{s1.26}), except for the term
$2\delta_{pq}G_m\partial_m\delta^4(x-x')\,$. Therefore the further
computations are very similar to the ones in Sect. 2. As usual,
only two-, three- and four-point diagrams are divergent. The
two-point function is given by
\begin{eqnarray}
\Gamma_2^{SYM}&=&-\int d^4x_1 d^4x_2
[\delta_{pq}G_m(x_1)\partial_m\frac{1}{\square}\delta^4(x_1-x_2)+
2G_{[p}(x_1)\partial_{q]}\frac1\square\delta^4(x_1-x_2)
\nonumber\\&& +\,\varepsilon_{qpmn}G_m(x_1)\partial_n
\frac1\square\delta^4(x_1-x_2)]
[\delta_{qp}G_n(x_2)\partial_n\frac{1}{\square}\delta^4(x_2-x_1)
\nonumber\\&&+
\,2G_{[q}(x_2)\partial_{p]}\frac1\square\delta^4(x_2-x_1)+
\varepsilon_{pqrs}G_r(x_2)\partial_s\frac1\square\delta^4(x_2-x_1)]\,.
\label{s3.12}
\end{eqnarray}
To proceed, we pass to momentum space and
compute the divergent momentum integrals using the standard methods of
quantum field theory. As a result, we find that the two-point function
(\ref{s3.12}) has no divergent contributions,
\be
\Gamma^{SYM}_{2,div}=0\,.
\label{s3.13}
\ee
The absence of divergences here is owing to the term
$2\delta_{pq}G_m\partial_m\delta^4(x-x')$ in (\ref{s3.11}) and (\ref{s3.12}).
It gives the contribution which
exactly cancels the expression (\ref{s1.31}) obtained by similar
calculations without this term.

The three- and four-point functions are defined by the following formal
expressions
\begin{eqnarray}
\Gamma_3^{SYM}&=&\frac43\Tr[(\delta_{pq}G_m(x)\partial_m
+2G_{[p}(x)\partial_{q]}-\varepsilon_{pqmn}G_m(x)\partial_n)
\frac1\square\delta^4(x-x')]^3,
\label{s3.14}\\
\Gamma_4^{SYM}&=&-2\Tr[(\delta_{pq}G_m(x)\partial_m
+2G_{[p}(x)\partial_{q]}-\varepsilon_{pqmn}G_m(x)\partial_n)
\frac1\square\delta^4(x-x')]^4. \label{s3.15}
\end{eqnarray}
The further computations are very similar to those in Sect.\
\ref{SYM-quant}, but with taking into account the term
$2\delta_{pq}G_m\partial_m\delta^4(x-x')\,$. After carefully
tracking the coefficients during the computations, we find that
the three- and four-point functions also have no divergences, \be
\Gamma^{SYM}_{3,div}=0\,,\qquad \Gamma^{SYM}_{4,div}=0\,.
\label{s3.16} \ee As a result, we conclude that the abelian
$\cN{=}(1,0)$ non-anticommutative gauge model (\ref{s3.2}) is
completely finite, \be \Gamma^{SYM}_{div}=0, \label{s3.17} \ee
without the necessity to perform any field redefinition such as
(\ref{s1.39}).

One more important comment to be added is as follows. The abelian
$\cN{=}(1,0)$ non-anticommutative  gauge model is described by the
classical actions (\ref{s1.6}) or (\ref{s3.2}) which are related
to each other by the Seiberg-Witten map (\ref{s3.1}). It is
obvious that the Jacobian of such a change of functional variables
(\ref{s3.1}) is unity (in the sense of dimensional
regularization). Therefore the effective actions in these two
models should also be related by the Seiberg-Witten map. As for
the divergent part, we observe that it is trivial for both models
(\ref{s1.6}) and (\ref{s3.2}), since it can be removed by the
shift (\ref{s1.39}) of the scalar field $\phi\,$. Note that this
explains the appearance of only two out of three possible
divergent terms (\ref{s1.23}). Indeed, if the third term
proportional to $I^4\int d^4x
f_3(I\bar\phi)(\partial_m\bar\phi\partial_m\bar\phi)^2$ appeared
in the divergent part of the effective action, it could not be
removed by any shift of the scalar field $\phi\,$, that would
mean the presence of a nontrivial divergence in the model.
However, we have seen in this Section that the effective action in
$\cN{=}(1,0)$ non-anticommutative gauge theory is finite.

Let us also consider the general model of an abelian $\cN{=}(1,0)$
non-anticommutative gauge superfield interacting with a neutral
hypermultiplet. It is described by the sum of the classical
actions (\ref{s1.6}) and (\ref{Sn-comp}). Upon performing the Seiberg-Witten
map (\ref{SW-sym+hyper}), this action turns into (\ref{SYM+hyper}).
We see that the non-trivial interaction terms in (\ref{Lint}) are
the ones given by \be \int
d^4x\frac14(1+4I\bar\phi)^2(f_{mn}f_{mn}+\frac12\varepsilon_{mnrs}
f_{mn}f_{rs})\,. \label{s3.22.1} \ee This expression just
coincides with the one present in the gauge theory action
(\ref{s3.2}). Thus the quantum computations tell us once again
that the general abelian $\cN{=}(1,0)$
non-an\-ti\-com\-mu\-ta\-ti\-ve model is finite \be
\Gamma_{div}=0\,. \label{s3.23} \ee This result agrees with the
one of Sect.\ \ref{ren-hyp}, modulo some divergent redefinition
(\ref{s2.26}) of the scalar field $\phi\,$.

To summarize, the use of the Seiberg-Witten map in the models under
consideration makes it possible to avoid the divergent
expressions in the effective action from the very beginning. Otherwise,
such expressions appear but they are removable by some divergent
redefinition of the scalar field $\phi\,$.

\section{Holomorphic potential in the
non-anticommutative \protect\\ abelian charged hypermultiplet model}
The previous section was devoted to the calculation of divergent
parts of the effective actions in the models of gauge superfield
and hypermultiplets. In this section we study the structure of
finite parts of these effective actions. It is
clear that the chiral singlet deformation modifies the effective
action of the original undeformed theory in some way. Since we restrict
our consideration only to the abelian case, it makes sense to study only
the issue of non-anticommutative corrections to the effective
action in the charged hypermultiplet model. Indeed, in the limit
$I\to0$ the undeformed abelian models of neutral hypermultiplet
and gauge superfield become free and exhibit no any quantum dynamics, while
the charged hypermultiplet model turns into the $\cN{=}(1,1)$
supersymmetric electrodynamics which is non-trivial at the quantum
level.

It is well known \cite{Buch,non-hol} that the low-energy effective
action in the undeformed charged hypermultiplet model has the
following structure in the sector of gauge superfields
\be
\Gamma=\int d^4x d^4\theta\, {\cal F}(W)+\int d^4x d^4\bar\theta\,
\bar{\cal F}(\bar W)+
\int d^4x d^8\theta\, {\cal H}(W,\bar W),
\label{e3}
\ee
where ${\cal F}$ is a holomorphic potential, $\bar{\cal F}$ is
an antiholomorphic potential and ${\cal H}$ is a non-holomorphic
potential. The superfield strengths $W$, $\bar W$ are expressed
through the prepotential $V^{--}$ as in (\ref{W}).
In the abelian case these superfields obey the (anti)chirality
conditions,
\be
D^\pm_\alpha \bar W=0,\qquad \bar D^{\pm}_{\dot\alpha}W=0.
\label{e7}
\ee
The holomorphic and antiholomorphic parts of the effective action
(\ref{e3}) result in the following effective equations of motion
\be
(D^+)^2{\cal F}^\prime(W)+(\bar D^+)^2\bar{\cal F}^\prime(\bar{W})=0.
\label{effequ}
\ee

By now, the perturbative contributions to the effective action in the
undeformed charged hypermultiplet model have been thoroughly
studied (see, e.g., \cite{Buch}-\cite{Dragon}). In particular, the
holomorphic potential in this model is given by the following
simple formula
\be
{\cal F}(W)=-\frac{1}{32\pi^2}W^2\ln \frac W\mu,
\label{e8}
\ee
where $\mu$ is some constant of mass dimension $+1$.

The superfield strengths $W$, $\bar W$ have the scalar fields
$\phi$, $\bar\phi$ and the Maxwell field strength $F_{mn}=\partial_m A_n-\partial_n
A_m$ as their bosonic component fields,
\be
W=\phi+(\theta^+\sigma_{mn}\theta^-)F_{mn}+\ldots,
\qquad
\bar W= \bar\phi+(\bar\theta^+\tilde
\sigma_{mn}\bar\theta^-)F_{mn}+\ldots.
\label{e9}
\ee
Here we do not consider the dependence of these superfields on
the spinors $\Psi^i_\alpha$, $\bar\Psi_{i\dot\alpha}$
and auxiliary fields ${\cal D}^{kl}$. Substituting the field strength
(\ref{e9}) into (\ref{e8}), one readily obtains the component
structure of the holomorphic effective action in the bosonic
sector,
\be
\Gamma_{hol}=\int d^4x d^4\theta\, {\cal F}(W)=
-\frac1{32\pi^2}\int d^4x (F_{mn}F_{mn}+F_{mn}\tilde F_{mn})\left(
\ln\frac\phi\mu +\frac32
\right)+\ldots.
\label{e11}
\ee
Here $\tilde F_{mn}=\frac12\varepsilon_{mnrs}F_{rs} $ and dots
stand for the terms with derivatives of fields and spinor fields.
We stress that the expression (\ref{e11}) corresponds to the
bosonic terms in the effective action which are leading in the
following approximation
\be
\begin{array}{c}
\phi=const,\quad \bar\phi=const,\quad F_{mn}=const,\\
\Psi^i_\alpha=\bar\Psi_{i\dot\alpha}={\cal D}^{kl}=0.
\end{array}
\label{e10}
\ee
Note that the constant $3/2$ in (\ref{e11}) can be removed by
a shift of the parameter $\mu$, however it will be important
when we will consider the non-anticommutative deformation of (anti)holomorphic
effective action. The antiholomorphic
part of the effective action can be obtained by the complex conjugation of
the action (\ref{e11}).

\subsection{General structure of the effective action}
Let us discuss the general structure of the effective action in
the charged hypermultiplet model. Since the classical action (\ref{Sqf}) is
a simple $\star$-product generalization of
the corresponding classical action of undeformed theory, one can
assume that the chiral part of the effective potentials in (\ref{e3})
is also given by the $\star$-deformation of the holomorphic
potential,
\be
{\cal F}(W)\longrightarrow {\cal F}_\star(W).
\label{e23}
\ee
However, the antiholomorphic contributions to the effective action
cannot be accounted by such a naive considerations. As was shown
in Sect.\ \ref{SYMsect}, one cannot construct any action in the
antichiral superspace having the form
$\bar{W}^n_\star$. We will show that the
corresponding contributions to the effective action are naturally given
by the full superspace integrals.

For the further consideration it will be more convenient to study
the variation of effective action $\delta\Gamma\,$, rather than
$\Gamma$ itself. In particular, given the holomorphic effective
action \be \Gamma_{hol}=\int d^4x d^4\theta\, {\cal F}_\star(W)\,,
\label{e29} \ee one can write its variation either in the analytic
superspace, \be \delta\Gamma_{hol}=\int d\zeta du\, \delta
V^{++}\star[-\frac 14 D^{+\alpha}D^+_\alpha{\cal F}'_\star(W)],
\label{e30} \ee or in full superspace, \be \delta\Gamma_{hol}=\int
d^{12}z du\, \delta V^{++}\star V^{--}\star\frac1W\star {\cal
F}'_\star(W). \label{e31} \ee To derive the expressions
(\ref{e30}), (\ref{e31}) one should follow the same steps as in
\cite{Dragon} for the non-abelian $\cN{=}2$ superymmetric gauge model
without deformations.

We assume that the antiholomorphic contributions to the effective
action can be accounted by the following variation \be
\delta\Gamma_{antihol}=\int d^{12}z du\, \delta V^{++}\star
V^{--}\star\frac{1}{\bar{W}}\star \bar{\cal F}'_\star(\bar{W}),
\label{varanti} \ee which is written in full $\cN{=}(1,1)$
superspace rather than in the antichiral one. In particular, it
reproduces the correct effective equations of motion of the form
(\ref{effequ}). Clearly, the full superspace action
$\delta\Gamma_{antihol}$ can always be reduced to the antichiral
superspace by integrating over $d^4\theta$, but the result cannot
be written as $\int d^4xd^4\bar\theta\, \bar{\cal F}_\star(\bar
W)$.

\subsection{One-loop effective action}
In this subsection we explicitly calculate the leading
contributions to the charged hypermultiplet effective action in
harmonic superspace. Consider the full propagator $G^{(1,1)}(1|2)$
of the charged hypermultiplet defined as a solution of the equation
\be
\nabla^{++}\star
G^{(1,1)}(1|2)=\delta_A^{(3,1)}(1|2).
\label{e19}
\ee
In contrast
to the free propagator $G_0^{(1,1)}(1|2)$ given by
(\ref{eq2}), $G^{(1,1)}(1|2)$ describes the dynamics of the
charged hypermultiplet interacting with the background gauge
superfield $V^{++}$. It is straightforward to check that the
solution of (\ref{e19}) can be written as
\be
G^{(1,1)}(1|2)=-\frac1{\hat\square_\star}\star
(D^+_1)^4(D^+_2)^4\left\{ e_\star^{\Omega(1)}\star
e_\star^{-\Omega(2)}
\star\frac{\delta^{12}(z_1-z_2)}{(u^+_1u^+_2)^3} \right\},
\label{e20}
\ee
where $\hat\square_\star$ is the covariant box
operator,
\be
\hat\square_\star=-\frac12(D^+)^4\nabla^{--}\star\nabla^{--},
\label{e21}
\ee
and $\Omega(z,u)$ is a ``bridge'' superfield in
the full $\cN{=}(1,1)$ superspace defined by the relations
\be
\nabla^{++}=e^{\Omega}_\star\star D^{++}e^{-\Omega}_\star,\qquad
\nabla^{--}=e^{\Omega}_\star\star D^{--}e^{-\Omega}_\star.
\label{e17}
\ee
The bridge superfield was originally introduced in
\cite{HSS} for the undeformed $\cN{=}2$ supergauge theory as an
operator relating $\cN{=}2$ superfields in the so-called
$\lambda$- and $\tau$-frames. Using the superfield $\Omega(z,u)$
one can write the relation (\ref{V--}) between the prepotentials
$V^{--}$ and $V^{++}$ in the following simple forms,
\be
V^{--}(z,u)=\int du'\frac{e_\star^{\Omega(z,u)}\star
e_\star^{-\Omega(z,u')}\star V^{++}(z,u')}{(u^+u'^+)^2} =\int
du'\frac{V^{++}(z,u')\star e_\star^{\Omega(z,u')}\star
e_\star^{-\Omega(z,u)}}{(u^+u'^+)^2}.
\label{e18}
\ee
The relations (\ref{e18}) can be directly checked using
(\ref{e17}) and the properties of harmonic distributions.

Note that the operator $\hat\square_\star$ takes any analytic
superfield into an analytic one. This operator,
while acting on analytic superfields, can be represented in the
following form
\begin{eqnarray}
\hat\square_\star&=&\nabla^m\star\nabla_m
-\frac 12(\nabla^{+\alpha}\star W)\star\nabla^-_\alpha
-\frac 12(\bar\nabla^+_{\dot\alpha}\star \bar W)\star\bar\nabla^{-\dot\alpha}
+\frac 14(\nabla^{+\alpha}\star\nabla^+_\alpha\star W)\star\nabla^{--}
\nonumber\\&&
-\frac 18[\nabla^{+\alpha}\cs\nabla^-_\alpha]\star W
-\frac12\{W\cs\bar W \},
\label{e22}
\end{eqnarray}
where $\nabla^\pm_\alpha=D^\pm_\alpha+V^\pm_\alpha$,
$\bar\nabla^\pm_{\dot\alpha}=\bar D^\pm_{\dot\alpha}+\bar V^\pm_{\dot\alpha}$
are covariant spinor derivatives. The expression (\ref{e22}) has
the same form as in the undeformed non-abelian gauge theory \cite{backgr}, with the
$\star$-product playing the role of the matrix commutator. This
result is not surprising since (\ref{e22}) is derived using only the algebra
of covariant derivatives which has the same form as in the undeformed case.

Clearly, the charged hypermultiplet effective action is one-loop
exact since the classical action (\ref{Sqf}) is quadratic in
the hypermultiplet superfields. It can also be expressed
through the propagator $G^{(1,1)}(1|2)$,
\be
\Gamma=\Tr\ln\frac{\delta^2S}{\delta\tilde q^+(1)\delta q^+(2)}
=\Tr\ln(\nabla^{++}\star)=-\Tr\ln\,G^{(1,1)}(1|2).
\label{e32}
\ee
The variation of this effective action reads
\be
\delta\Gamma=\Tr[
\delta V^{++}\star G^{(1,1)}
]=\int d\zeta du\, \delta V^{++}(1)\star
G^{(1,1)}(1|2)|_{(1)=(2)}.
\label{e33}
\ee
Using the definition (\ref{e19}), one can derive the following
relation between the free and full hypermultiplet propagators,
\be
G^{(1,1)}(1|3)=G_0^{(1,1)}(1|3)-\int d\zeta_2 du_2\,
G_0^{(1,1)}(1|2)\star V^{++}(2)\star G^{(1,1)}(2|3).
\label{e34}
\ee
Substituting (\ref{e34}) into the variation (\ref{e33}), we
find
\be
\delta\Gamma=-\int d\zeta_1 du_1 d\zeta_2 du_2\,
\delta V^{++}(1)\star G_0^{(1,1)}(1|2)\star V^{++}(2)\star
G^{(1,1)}(2|1).
\label{e35}
\ee
Taking into account the explicit forms of the propagators
(\ref{eq2}), (\ref{e20}), we rewrite (\ref{e35}) as follows
\begin{eqnarray}
\delta\Gamma&=&-\int d\zeta_1 d\zeta_2 du_1 du_2\,
\delta V^{++}(1)\star\frac1\square(D^+_1)^4(D^+_2)^4
\frac{\delta^{12}(z_1-z_2)}{(u^+_1u^+_2)^3}
\nonumber\\&&\times
V^{++}(2)\star\frac1{\hat\square_{\star(2)}}\star
(D^+_1)^4(D^+_2)^4\left\{
e^{\Omega(2)}_\star\star e^{-\Omega(1)}_\star
\frac{\delta^{12}(z_2-z_1)}{(u^+_2u^+_1)^3}
\right\}.
\label{e36}
\end{eqnarray}
Now we take off the spinor derivatives from the first
delta-function to restore full $\cN{=}(1,1)$ superspace measure
with the help of (\ref{full_measure}),
\begin{eqnarray}
\delta\Gamma&=&-\int d^{12}z_1 d^{12}z_2 du_1 du_2\,
\delta V^{++}(1)\star\frac1\square
\frac{\delta^{12}(z_1-z_2)}{(u^+_1u^+_2)^3}
\nonumber\\&&\times
V^{++}(2)\star\frac1{\hat\square_{\star(2)}}\star
(D^+_1)^4(D^+_2)^4\left\{
e^{\Omega(2)}_\star\star e^{-\Omega(1)}_\star\star
\frac{\delta^{12}(z_2-z_1)}{(u^+_2u^+_1)^3}
\right\}.
\label{e37}
\end{eqnarray}
We did not impose any restrictions on the background gauge
superfields so far, therefore (\ref{e37}) is the exact representation for
the hypermultiplet effective action. It should be considered as a starting
point for further calculations of different contributions to the
effective action.

\subsection{Divergent part of the effective action}
The effective action, as a functional of superfield strengths, can
be expanded in a series with respect to these superfields and
their covariant derivatives. This series can be obtained from the
representation (\ref{e37}) for the effective action as a result of the
decomposition of the operator $1/\hat\square_\star$ in this
expression.

Let us omit all superfields in the operator
(\ref{e22}),
\be
\frac1{\hat\square_\star}\approx \frac1\square.
\label{eq77}
\ee
Such an approximation, being applied to (\ref{e37}),
corresponds exactly to the divergent part of the effective action
since the other terms in the covariant box operator
produce higher powers of momenta in the denominator which lead to
the finite contributions. Under the condition (\ref{eq77}), the
variation of the effective action (\ref{e37}) is essentially simplified
\begin{eqnarray}
\delta\Gamma_{div}&=&\int d^{12}z_1 d^{12}z_2 \frac{du_1 du_2}{(u^+_1u^+_2)^6}
\delta V^{++}(1)\star\frac1\square
\delta^{12}(z_1-z_2)
\nonumber\\&&\times
V^{++}(2)\star\frac1\square
(D^+_1)^4(D^+_2)^4\left\{
e^{\Omega(2)}_\star\star e^{-\Omega(1)}_\star\star
\delta^{12}(z_2-z_1)\right\}.
\label{eq78}
\end{eqnarray}
Next, we apply the identity
\be
\delta^8(\theta_1-\theta_2)(D_1^+)^4(D^+_2)^4\delta^{12}(z_1-z_2)=
(u^+_1u^+_2)^4\delta^{12}(z_1-z_2)
\label{eq79}
\ee
to shrink the integration over the Grassmann variables to a point. As a result,
after regularization of the divergent momentum integral, (\ref{eq78})
becomes
\be
\delta\Gamma_{div}=\frac1{16\pi^2\varepsilon}\int d^{12}z du_1\,
\delta V^{++}(z,u_1)\star\int du_2\frac{V^{++}(z,u_2)\star
e^{\Omega(z,u_2)}_\star\star
e^{-\Omega(z,u_1)}_\star}{(u^+_1u^+_2)^2}.
\label{eq80}
\ee
Here $\varepsilon$ is a parameter of dimensional regularization.
Applying now (\ref{e18}), we obtain the following expression
for the variation of the effective action
\be
\delta\Gamma_{div}=\frac1{16\pi^2\varepsilon} \int d^{12}z du\,
\delta V^{++}\star V^{--}.
\label{eq81}
\ee
The variation
(\ref{eq81}) can be easily integrated with the help of
(\ref{e29}), (\ref{e31}),
\be
\Gamma_{div}=\frac1{32\pi^2\varepsilon}\int d^4x d^4\theta\, W^2.
\label{eq81_}
\ee
As a result, we see that the divergent part of
the effective action is proportional to the classical action of
the $\cN{=}(1,0)$ supesymmetric rgauge theory. This result has already been
obtained in Sect.\ \ref{hyp} by a different method.

\subsection{Holomorphic and non-holomorphic contributions}
In this subsection we will study the finite contributions to the
effective actin in the charged hypermultiplet model.
The leading terms in the low-energy effective action are composed of the
superfield strengths without derivatives. Such an approximation is
effectively accounted by considering the background superfield
strengths obeying the following constraints
\be
\nabla^{\pm\alpha}\star W=0,\qquad
\bar\nabla^\pm_{\dot\alpha}\star\bar W=0\,.
\label{eq83}
\ee
Under the constraints (\ref{eq83}) all superfields with
derivatives in the operator $\hat\square_\star$ given by (\ref{e37})
can be neglected \footnote{In (\ref{eq82}) we discard
also the connections covariantizing the vector derivatives $\partial_m$
which are present in the first term of the operator
(\ref{e22}). These connections are always proportional to the
derivatives of superfield strengths and therefore are not
essential for studies of the holomorphic effective action.},
\be
\frac1{\hat\square_\star}\approx\frac1{\square-\frac12\{W\cs\bar W
\}}\,.
\label{eq82}
\ee
As a result, the variation of the effective action is given by
\begin{eqnarray}
\delta\Gamma&=&\int d^{12}z_1 d^{12}z_2 \frac{du_1 du_2}{(u^+_1u^+_2)^6}
\delta V^{++}(1)\star\frac1\square
\delta^{12}(z_1-z_2)
\nonumber\\&&\times
V^{++}(2)\star\frac1{\square-\frac12\{W\cs\bar W \}}
(D^+_1)^4(D^+_2)^4\left\{
e^{\Omega(2)}_\star\star e^{-\Omega(1)}_\star\star
\delta^{12}(z_2-z_1)\right\}.
\label{eq84}
\end{eqnarray}
Next, we apply the identity (\ref{eq79}) and integrate over $d^8\theta_2$
using the corresponding delta-function
\begin{eqnarray}
\delta\Gamma&=&\int d^{12}z_1 d^4x_2 \frac{du_1 du_2}{(u^+_1u^+_2)^2}
\delta V^{++}(x_1,\theta,u_1)\star V^{++}(x_2,\theta,u_2)
\star e^{\Omega(2)}_\star\star e^{-\Omega(1)}_\star
\nonumber\\&&
\star\frac1\square
\delta^4(x_1-x_2)\frac1{\square-\frac12\{W\cs\bar W \}}
\delta^4(x_2-x_1).
\label{eq85}
\end{eqnarray}
The bosonic delta-functions in (\ref{eq85}) result in the
following momentum integral
\be
\int \frac{d^4k}{(2\pi)^4}\frac1{k^2}\frac1{k^2+\frac12\{W\cs\bar W\}}
=-\frac{1}{16\pi^2}\ln_\star\left[\frac{\{W\cs\bar W\}}{2\mu^2}
\right]+(\mbox{divergent terms}),
\label{eq86}
\ee
where $\mu$ is an arbitrary constant of dimension $+1$.
The function $\ln_\star$ here is understood in a sense of the
corresponding Taylor series with the $\star$-product of superfields,
e.g., $\ln_\star(1+X)=X-\frac12 X\star X+\frac13 X\star X\star
X+\ldots$. Since the divergent part of the effective action was
studied in the previous subsection, here we concentrate only on the
finite part. Applying the identity
(\ref{e18}), we conclude,
\be
\delta\Gamma=-\frac1{16\pi^2}\int d^{12}z du\,
\delta V^{++}\star  V^{--}\star \ln_\star\frac{\{W\cs\bar W\}}{
2\mu^2}.
\label{eq88}
\ee
The expression (\ref{eq88}) is responsible for all contributions
to the effective action with the superfield strengths without
derivatives.

Note that in the limit $I\to0$ the $\star$-product becomes the
usual multiplication and (\ref{eq88}) is given by
\be
\delta\Gamma_{(I=0)}=-\frac1{16\pi^2}\int d^{12}z du\,
\delta V^{++} V^{--} \left(\ln\frac W\mu +\ln \frac{\bar W}\mu\right).
\label{eq89}
\ee
The variation (\ref{eq89}) corresponds precisely to the holomorphic
and antiholomorphic parts of the effective action (\ref{e3})
with the holomorphic potential (\ref{e8}) of the undeformed
theory. However, if $I\ne0$, the logarithm in (\ref{eq88})
cannot be written as a sum of two logarithms since there are
mixed terms. Therefore in the non-anticommutative case the
expression (\ref{eq88}) is responsible for both holomorphic,
antiholomorphic and non-holomorphic contributions to the effective
action.

We have to extract the holomorphic and antiholomorphic parts from
the effective action (\ref{eq88}). For this purpose we apply the
following identity for the logarithm in (\ref{eq88})
\be
\ln_\star\frac{\{W\cs\bar W
\}}{2\mu^2}=\ln_\star\frac{W}{\mu}+
\ln_\star\frac{\bar{W}}{\mu}
+\frac{1}{12\mu^3}[W\cs[W\cs\bar W]]
+\frac{1}{12\mu^3}[\bar{W}\cs[\bar{W}\cs W]]+\ldots,
\label{eq89-1}
\ee
where dots stand for terms of fourth and higher orders in superfields
which come with various commutators. Note that the expression (\ref{eq89-1})
is valid without any restrictions on the superfields and is
obtained only with the help of formal manipulations with
$\star$-(anti)commutators of superfields. The terms with
commutators in (\ref{eq89-1}) correspond to the
non-holomorphic contributions to the effective action since they
involve both $W$ and $\bar W$. Keeping only the first two terms in the
r.h.s. of (\ref{eq89-1}), we obtain the following expression
for the variation (\ref{eq88}):
\be
\delta\Gamma=-\frac1{16\pi^2}\int d^{12}z du\,
\delta V^{++}\star  V^{--}\star\left[ \ln_\star\frac W\mu
+\ln_\star\frac{\bar W}{\mu}\right]+\ldots.
\label{eq89-2}
\ee
Here dots stand for the non-holomorphic contributions.
According to the equation (\ref{e31}), the holomorphic part of the
variation (\ref{eq89-2}) can be easily integrated,
\be
\Gamma_{hol}=-\frac1{32\pi^2}\int d^4x d^4\theta\, W\star W\star
\ln_\star \frac{W}{\mu}.
\label{eq93.2}
\ee
As a result, we proved that the holomorphic part of the effective
action in the non-an\-ti\-com\-mu\-ta\-ti\-ve charged hypermultiplet model
is nothing but a $\star$-product generalization of a standard
holomorphic potential (\ref{e8}).

Note that the terms with commutators in (\ref{eq89-1}) can be
eliminated by imposing the further constraints on the background
gauge superfields. Consider, e.g., the following constraints
\be
\partial_m\bar W=0,\qquad
\frac\partial{\partial \theta^i_\alpha}\bar W=0.
\label{eq90}
\ee
One of the consequences of (\ref{eq90}) is the
relation $Q^i_\alpha \bar W=0$ which reduces the
$\star$-product of the superfield strengths to the usual
product. Note that the constraints (\ref{eq90}) are not covariant
and could be too strong. However, they keep the dependence of the
superfield $\bar W$ on the $\bar\theta^{i\dot\alpha}$ variables
and are consistent with the approximation (\ref{e10}) in which we study
the corrections to the holomorphic potential. Moreover, the
constraints (\ref{eq90}) do not violate the covariance of the
effective action in the holomorphic sector.

The constraint (\ref{eq90}) simplifies the antiholomorphic part of
the effective action since it allows one to omit the $\star$-product,
\be
\delta\Gamma_{antihol}=\frac1{16\pi^2}\int d^{12}z du\,
\delta V^{++}  V^{--} \ln\frac{\bar W}\mu.
\label{eq93.3}
\ee
The variation (\ref{eq93.3}) reproduces the standard
antiholomorphic potential,
\be
\Gamma_{antihol}=-\frac1{32\pi^2}\int d^4x d^4\bar\theta
\,\bar W^2\ln\frac{\bar W}{\mu}.
\label{eq93.4}
\ee
Despite the absence of $\star$-product in (\ref{eq93.4}),
this expression implicitly depends on the parameter of chiral
singlet deformation $I$ through the superfield $\bar W$
which involves this parameter by definition.

\subsection{Component structure of the effective action}
The leading contributions to the effective action in the
undeformed hypermultiplet model are given by (\ref{e11}).
Here we study the corrections to these terms due to the
non-anticommutative deformations of supersymmetry. For this purpose
we find the component structure of the actions (\ref{eq93.2}),
(\ref{eq93.3}) in the bosonic sector in the approximation
(\ref{e10}). Here we follow the same steps as in the Sect.\ %
\ref{SYMsect} where the component structure of the classical
action of $\cN{=}(1,0)$ supergauge theory was studied.

In the component expansion of the prepotential (\ref{Vwz})
we keep only the bosonic fields,
\be
V^{++}_{WZ}=(\theta^+)^2\bar\phi+(\bar\theta^+)\phi
+2(\theta^+\sigma_m\bar\theta^+)A_m
-2{\rm i}(\bar\theta^+)^2(\theta^+\theta^-)\partial_m A_m
-(\bar\theta^+)^2(\theta^-\sigma_{mn}\theta^+)F_{mn}.
\label{eqq100}
\ee
Note that both the strength $F_{mn}$ and gauge potential $A_m$ enter
the prepotential (\ref{eqq100}). Therefore the
expression (\ref{eqq100}) depends on the spatial coordinates $x^m$
through the potential $A_m$. Without loss of generality, we choose
the vector potential to be linear in $x^m$, $A_m=\frac12 F_{nm}x^n$,
$F_{mn}=const$. In particular, $\partial_m A_n-\partial_n A_m= F_{mn}$,
$\partial_m A_m=0$.

Analogously to the expression (\ref{V--1}), we look for the prepotential
$V^{--}$ in the form
\begin{eqnarray}
V^{--}&=&v^{--}+\bar\theta^-_{\dot\alpha}v^{-\dot\alpha}+
(\bar\theta^-)^2 A
+(\bar\theta^+\bar\theta^-)\varphi^{--}\nonumber\\
&&+(\bar\theta^+\bar\sigma_{mn}\bar\theta^-)\varphi^{--}_{mn}
+(\bar\theta^-)^2\bar\theta^+_{\dot\alpha}\tau^{-\dot\alpha}
+(\bar\theta^+)^2(\bar\theta^-)^2\tau^{--},
\label{eqq107}
\end{eqnarray}
as a solution of the zero-curvature equation (\ref{zero-curv1}). All the component fields
in the r.h.s. of (\ref{eqq107}) depend only on
the variables $\theta^+_\alpha,\ \theta^-_\alpha$.
Substituting (\ref{eqq100}), (\ref{eqq107}) into (\ref{zero-curv1}), we find
\begin{eqnarray}
v^{--}&=&(\theta^-)^2\frac{\bar\phi}{1+4I\bar\phi},
\label{eqq108}\\
v^{-\dot\alpha}&=&\frac{2(\theta^-\sigma_m)^{\dot\alpha}A_m}{1+4I\bar\phi},
\label{eqq109}\\
{\cal A}&=&\phi+\frac{4IA_mA_m}{1+4I\bar\phi}
+(\theta^+\sigma_{mn}\theta^-)F_{mn},
\label{eqq111}\\
\tau^{-\dot\alpha}&=&-\frac{4I(\theta^-\sigma_{mn})^\alpha F_{mn}\sigma_{
r\alpha}{}^{\dot\alpha}A_r}{1+4I\bar\phi},
\label{eqq112}\\
\tau^{--}&=&\varphi^{--}=\varphi^{--}_{mn}=0.
\label{eqq113}
\end{eqnarray}
Using the definitions (\ref{W}), we find the component structure
of the superfield strengths
\begin{eqnarray}
W&=&\phi+\frac{4IA_m
A_m}{1+4I\bar\phi}+(\theta^+\sigma_{mn}\theta^-)F_{mn}
-4I(\theta^-\sigma_{mn}^\alpha)(\sigma_r\bar\theta^+)_\alpha
\frac{A_r F_{mn}}{1+4I\bar\phi},
\label{eqq114}\\
\bar W&=&\frac{\bar\phi}{1+4I\bar\phi}+
(\bar\theta^+\bar\sigma_{mn}\bar\theta^-)\frac{F_{mn}}{1+4I\bar\phi}
+(\bar\theta^+)^2(\bar\theta^-)^2
\frac{2IF_{mn}F_{mn}+4IF_{mn}\tilde F_{mn}}{1+4I\bar\phi}.
\label{eqq115}
\end{eqnarray}
Note that the superfields $W$ and $\bar W$ are deformed
differently. Moreover, the superfield  $\bar W$ given by
(\ref{eqq115}) does not depend on the $x^m$ and $\theta^i_\alpha$
variables, which agrees with the constraint (\ref{eq90}).

Introducing the notations
\be
\Phi=\phi+\frac{4IA_m A_m}{1+4I\bar\phi},
\label{eqq116}
\ee
we bring the superfield strength (\ref{eqq114}) to the standard form,
\be
W=\Phi+(\theta^+\sigma_{mn}\theta^-)F_{mn}+\ldots,
\label{eqq117}
\ee
where dots correspond to the last term in (\ref{eqq114}) which does
not contribute to the holomorphic effective action.

Now we substitute the superfield strength (\ref{eqq117}) into the
holomorphic potential (\ref{eq93.2}), compute the
$\star$-products and integrate over Grassmann variables.
As a result, we arrive at the following component expression for
the holomorphic effective action
\be
\Gamma_{hol}=-\frac1{32\pi^2}\int d^4x (F^2+F\tilde F)
\left[\ln\frac{\Phi}{\mu}+\Delta(X(\Phi,F_{mn})) \right],
\label{eqq124}
\ee
where
\begin{eqnarray}
\Delta(X)&=&\frac12(1-X)^2
\ln(X-1)+\frac12(1+ X)^2\ln(1+ X)-(1+X^2)\ln X,
\label{eqq126}\\
X(\Phi,F_{mn})&=&\frac\Phi{2I\sqrt{2(F^2+F\tilde
F)}}.
\label{eqq124.1}
\end{eqnarray}
The function $\Delta(X)$ in (\ref{eqq124}) is responsible
for the non-anticommutative corrections to the standard terms in
the holomorphic potential. In the limit $I\to0$ we have
\be
\lim_{I\to0}\Delta(X)=\frac32.
\label{eqq127}
\ee
We see here that (\ref{eqq127}) reproduces the constant $3/2$
in (\ref{e11}). This constant was not essential in the undeformed case, but
now it is replaced by the function $\Delta(X)$.

Let us finally study the non-anticommutative corrections in the
antiholomorphic sector. We substitute the superfield strength
(\ref{eqq115}) into the antiholomorphic potential (\ref{eq93.4})
and integrate there over the Grassmann variables. As a result, we find
the component structure of the antiholomorphic effective
action,
\begin{eqnarray}
\Gamma_{antihol}&=&-\frac1{32\pi^2}\int d^4x \frac{(F^2+F\tilde F)}{
(1+4I\bar\phi)^2}\left(
 \ln\frac{\bar\phi}{\mu(1+4I\bar\phi)}+\frac32\right)\nonumber\\
&&-\frac1{32\pi^2}\int d^4x\frac{F^2+
2F\tilde F}{(1+4I\bar\phi)^2} 2I\bar\phi\left(
1+2\ln\frac{\bar\phi}{\mu(1+4I\bar\phi)}
\right).
\label{eqq135}
\end{eqnarray}
In the limit $I\to0$
the expression (\ref{eqq135}) reproduces the standard antiholomorphic
potential in the undeformed charged hypermultiplet theory.

\section{Conclusions}
In this review we considered ${\cal N}{=}(1,0)$ non-anticommutative
theories with a chiral singlet $Q$-deformation of ${\cal
N}{=}(1,1)$ supersymmetry in harmonic superspace. In particular, we
studied abelian models of the gauge superfield and
hypermultiplets, both classical and quantum. Let us
give a brief summary of the basic results of the review.

In the superfield approach the non-anticommutative deformation of
${\cal N}{=}(1,1)$ supersymmetry is taken into account by
introducing a $\star$-product in ${\cal N}{=}(1,1)$ superspace.
The chiral singlet deformation of ${\cal N}{=}(1,1)$ harmonic
superspace is a particular case. Owing to the fact that the operation of
$\star$-multiplication is compatible with the harmonic and
Grassmann harmonic analyticities, classical actions of the
gauge superfield and hypermultiplet models can be obtained simply
by substituting the $\star$-product for the ordinary local
product in the undeformed superfield actions. At the component
level, the $\star$-products induce a modification of the
actions by new terms proportional to the deformation parameter. We
have presented the component structure of the
deformed classical actions for the abelian models of the neutral
and charged hypermultiplets, as well as for the gauge
supermultiplet.

The quantum aspects of these non-anticommutative models
are remarkable. The deformation
parameter has negative mass dimension, so counterterms are expected to
destroy the renormalizability. Nevertheless,
we proved  the renormalizability of the deformed models of the
abelian gauge superfield and neutral hypermultiplet.
It turned out that the divergent
contributions to the effective action can be eliminated altogether by
an appropriate shift of the scalar field $\phi$, one of two
independent scalar fields present in the Euclidean ${\cal
N}{=}(1,1)$ vector gauge supermultiplet. This field
redefinition has no impact on the quantum
dynamics of the theory, and the theories under
considerations are actually
finite. The renormalizability of the abelian non-anticommutative
model of the charged hypermultiplet was proved using
perturbative quantum calculations in ${\cal N}{=}(1,1)$ harmonic
superspace. It turns out that the divergent part of the effective
action is proportional to the superfield action of the ${\cal
N}{=}(1,0)$ model of the gauge superfield. In this sense the charged
hypermultiplet model is also renormalizable.
Moreover, in this model the holomorphic potential
was calculated and found to follow
from its undeformed analog just by employing $\star$-product universally.
At the level of component
fields this leads to new terms in the
effective action which, at least in the bosonic sector, can be
accommodated in the single function $\Delta(X)$ defined  in
(\ref{eqq126}).

Summarizing, we point out that all
theories with chiral deformations of ${\cal N}{=}(1/2,1/2)$ and
${\cal N}{=}(1,1)$ supersymmetry studied so far are
renormalizable. One naturally conjecture that any deformation of
this type preserves the renormalizability properties. Surely, this
hypothesis requires a rigorous proof and to be supported by
further examples. In this connection, it would be interesting to
prove the renormalizability of non-abelian ${\cal N}{=}(1,0)$
non-anticommutative gauge theories with and without
hypermultiplets, as well as to attack the problem of constructing
the low-energy effective action in these theories.

An important problem for further study is the question of
renormalizability and finiteness of non-anticommutative ${\cal
N}{=}4$ (actually, ${\cal N}{=}(2,2)$ in Euclidean space)
supersymmetric gauge theory corresponding to the chiral singlet
deformation (\ref{sstar}) in harmonic superspace \cite{S04,ILZ,IR06,I07}.
If the chiral deformations of supersymmetry preserve the finiteness
of this model, the quantum aspects of such a deformed  ${\cal
N}{=}4$ supergauge theory shall be very special.

Another possible direction of future investigation concerns the
quantum study of non-singlet deformations of ${\cal N}{=}(1,1)$
supersymmetry. In particular, it is of interest to consider those
deformations which
reduce to the known deformations of ${\cal N}{=}(1/2,1/2)$
supersymmetry upon the appropriate reduction of the Grassmann
sector of ${\cal N}{=}(1,1)$ superspace. In this way, one
might compare the results of calculations in the ${\cal
N}{=}1$ and ${\cal N}{=}2$ superfield approaches.
Non-anticommutative models with non-singlet
deformations of ${\cal N}{=}(1,1)$ supersymmetry were considered at
the classical level in \cite{CILQ}.

To conclude, theories with non-anticommutative deformations of
supersymmetry represent a prospective area for further
studies. Only a small part of this new ``continent'' of
applications of supersymmetry has been developed until today.

\vspace{3mm}\noindent
{\bf Acknowledgements}\nopagebreak

\noindent
The present work is supported particularly by the following grants:
RFBR project No 06-02-04012;
DFG project No 436 RUS 113/669/0-3;
INTAS project No 05-1000008-7928.
E.A.I. and B.M.Z. acknowledge a partial support from
RFBR grant, project No 06-02-16684, and
a grant of the Heisenberg-Landau program.
I.L.B. and I.B.S. acknowledge a partial support from RFBR grant,
project No 06-02-16346, and the grant of the President
of Russian Federation for leading Russian
scientific schools, project No 4489.2006.2.
The work of I.B.S. is also supported by the grant of the
President of Russian Federation for young scientists, project No
7110.2006.2 and INTAS grant, project No 06-1000016-6108.

\def\theequation{A.\arabic{equation}}
\def\thesection{}
\def\thesubsection{\arabic{subsection}}
\section{Appendices}
\subsection{Euclidean $\cN{=}(1,1)$ superspace}
The Euclidean $\cN{=}(1,1)$ superspace is defined as a superspace
parametrized by the coordinates \be
z=\{x^m,\theta_i^\alpha,\bar\theta^{\dot\alpha i} \}, \quad
\alpha,\dot\alpha=1,2,\quad i=1,2. \label{A0} \ee Here
$\theta_i^\alpha$, $\bar\theta^{\dot\alpha i}$ are analytic
Grassmann variables, $x^m=(x^1,x^2,x^3,x^4)$ are coordinates of
the Euclidean space $\mathbb{R}^4$ with the metrics
$g_{mn}=\delta_{mn}$. Since the metrics is given by the unit matrix
$\delta_{mn}$, the objects with upper and lower indices are
equivalent. Therefore, throughout this work we use only the
vectors and tensors with lower indices, except for $x^m$, and the
contraction over repeated indices is assumed.

The spinor ${\rm SU}(2)$ indices $\alpha,\dot\alpha$ are raised
and lowered with the antisymmetric $\varepsilon$-tensor,
\begin{eqnarray}
&&\psi_\alpha=\varepsilon_{\alpha\beta}\psi^\beta,\qquad
\bar\psi_{\dot\alpha}=\varepsilon_{\dot\alpha\dot\beta}\bar\psi^{\dot\beta},
\nonumber\\
&& \varepsilon_{12}=-\varepsilon^{12}=\varepsilon_{\dot1\dot2}=
-\varepsilon^{\dot1\dot2}=1,\quad
\varepsilon^{\alpha\beta}\varepsilon_{\beta\gamma}=\delta^\alpha_\gamma,\quad
\varepsilon^{\dot\alpha\dot\beta}\varepsilon_{\dot\beta\dot\gamma}=
\delta^{\dot\alpha}_{\dot\gamma}.
\label{A1}
\end{eqnarray}

We use the following conventions for the Euclidean sigma-matrices
\begin{eqnarray}
&&(\sigma_m)_{\alpha\dot\alpha}=({\rm i}\vec\sigma,{\bf
1})_{\alpha\dot\alpha},\qquad
(\bar\sigma_m)^{\dot\alpha\alpha}=\varepsilon^{\alpha\beta}
\varepsilon^{\dot\alpha\dot\beta}(\sigma_m)_{\beta\dot\beta},
\nonumber\\&&
\sigma_m\bar\sigma_n+\sigma_n\bar\sigma_m=2\delta_{mn},\qquad
\sigma_{mn}=\frac{\rm i}2(\sigma_m\bar\sigma_n-\sigma_n\bar\sigma_m),
\nonumber\\&&
\tr\,\sigma_n\bar\sigma_m=2\delta_{mn},\quad
\tr(\sigma_n\bar\sigma_m\sigma_p\bar\sigma_r)
=2\delta_{mn}\delta_{pr}-2\delta_{np}\delta_{mr}
+\delta_{nr}\delta_{pm}-2\varepsilon_{nmpr},
\label{A2}
\end{eqnarray}
where $\vec\sigma$ are the Pauli matrices.

Supercharges and covariant spinor derivatives in $\cN{=}(1,1)$
superspace are given by
\begin{eqnarray}
&&Q^i_\alpha=\partial_\alpha^i-{\rm i}\bar\theta^{\dot\alpha i}
(\sigma_m)_{\alpha\dot\alpha}\frac\partial{\partial x^m},\qquad
\bar Q_{\dot\alpha i}=-\bar\partial_{\dot\alpha
i}+{\rm i}\theta^\alpha_i(\sigma_m)_{\alpha\dot\alpha}\frac\partial{\partial
x^m},\nonumber\\
&&D^i_\alpha=\partial_\alpha^i+{\rm i}\bar\theta^{\dot\alpha i}
(\sigma_m)_{\alpha\dot\alpha}\frac\partial{\partial x^m},\qquad
\bar D_{\dot\alpha i}=-\bar\partial_{\dot\alpha
i}-{\rm i}\theta^\alpha_i(\sigma_m)_{\alpha\dot\alpha}\frac\partial{\partial
x^m},
\label{A3}
\end{eqnarray}
where the anticommuting derivatives $\partial^i_\alpha=\frac\partial{\partial\theta_i^\alpha}$,
$\bar\partial_{\dot\alpha i}=\frac\partial{\partial\bar\theta^{\dot\alpha i}
}$ act on the Grassmann variables by the rules
\be
\partial^i_\alpha
\theta^\beta_j=\delta^i_j\delta_\alpha^\beta,\qquad
\bar\partial_{\dot\alpha i}\bar\theta^{\dot\beta j}=
\delta_i^j\delta_{\dot\alpha}^{\dot\beta}.
\label{A4}
\ee
The non-vanishing anticommutation relations between the
operators (\ref{A3}) are as follows
\be
\{D^i_\alpha,\bar D_{\dot\alpha j} \}=
-\{Q^i_\alpha,\bar Q_{\dot\alpha j} \}=
-2{\rm i}\delta^i_j(\sigma_m)_{\alpha\dot\alpha}\frac\partial{\partial
x^m}.
\label{A5}
\ee

\subsection{Euclidean harmonic superspace}
The harmonic variables $u^\pm_i$, $i=1,2$, are defined as the
coordinates parametrizing the coset ${\rm SU}(2)/{\rm U}(1)$
and obeying the following basic relations
\be
u^{\pm k}=\varepsilon^{kj}u^\pm_j,\qquad u^{+k}u^-_k=1.
\label{harm}
\ee
The harmonic variables (\ref{harm}) allow ones to convert the
internal symmetry group indices into the U$(1)$ indices $\pm$, e.g.,
\begin{eqnarray}
&&\theta^{\pm\alpha}=\theta^{\alpha k}u_k^\pm,\qquad
\bar\theta^{\pm\dot\alpha}=\bar\theta^{\dot\alpha k}u^\pm_k,
\nonumber\\
&&Q^\pm_\alpha=Q^k_\alpha u^\pm_k,\quad
\bar Q^\pm_{\dot\alpha}=\bar Q^k_{\dot\alpha}u^\pm_k,\quad
D^\pm_\alpha=D^k_\alpha u^\pm_k,\quad
\bar D^\pm_{\dot\alpha}=\bar D^k_{\dot\alpha}u^\pm_k.
\label{harm1}
\end{eqnarray}

A key feature of the superspace with the coordinates
$(x^m,\theta^{\pm\alpha},\bar\theta^{\pm \dot\alpha}, u^{\pm i})$
is the presence of the so called analytic subspace $(\zeta,
u)=(x^m_A,\theta^+, \bar\theta^+, u^{\pm i})\,$, where
\be
x^m_A=x^m-{\rm i}
(\theta^{+\alpha}(\sigma^m)_{\alpha\dot\alpha}\bar\theta^{-\dot\alpha}
+\theta^{-\alpha}(\sigma^m)_{\alpha\dot\alpha}\bar\theta^{+\dot\alpha}).
\label{anal1}
\ee
The analytic subspace is closed under
supersymmetry,
\be
\delta_\epsilon x^m_A=-2{\rm i}
(\sigma^m)_{\alpha\dot\alpha}(
\epsilon^{-\alpha}\bar\theta^{+\dot\alpha}+\theta^{+\alpha}\bar\epsilon^{-\dot\alpha}),
\qquad
\delta_\epsilon\theta^{\alpha\pm}=\epsilon^{\pm\alpha},\qquad
\delta_\epsilon\bar\theta^{\pm\dot\alpha}=\bar\epsilon^{\pm\dot\alpha}.
\label{anal-susy}
\ee
Here $\epsilon^{\pm\alpha}=\epsilon^{\alpha
k}u^\pm_k$, $\bar\epsilon^{\pm\dot\alpha}
=\bar\epsilon^{\dot\alpha k}u^\pm_k$ and $\epsilon^{\alpha k}$,
$\bar\epsilon^{\dot\alpha k}$ are anticommuting parameters of
supertranslations. Therefore there exist the so called analytic
superfields which ``live'' on the subset of analytic coordinates,
$\Phi_A=\Phi_A(\zeta,u)$. Such superfields are singled out from
the general superfields on $\cN{=}(1,1)$ superspace by the following
covariant analyticity (Grassmann Cauchy-Riemann) conditions \be D^+_\alpha\Phi_A =0,\qquad \bar
D^+_{\dot\alpha}\Phi_A=0, \label{anal2} \ee where the covariant
spinor derivatives $D^\pm_{\alpha}$, $\bar D^\pm_{\dot\alpha}$ in
the analytic basis are given by
\begin{eqnarray}
&&D^+_\alpha=\frac\partial{\partial \theta^{-\alpha}},\qquad
D^-_\alpha=-\frac\partial{\partial \theta^{+\alpha}}+2{\rm i}(\sigma_m)_{\alpha\dot\alpha}
\bar\theta^{-\dot\alpha}\frac\partial{\partial x^m_A},\nonumber\\&&
\bar D^+_{\dot\alpha}=\frac\partial{\partial\bar\theta^{-\dot\alpha}},\qquad
\bar D^-_{\dot\alpha}=-\frac\partial{\partial\bar\theta^{+\dot\alpha}}-2{\rm i}(\sigma_m)_{\alpha\dot\alpha}
\theta^{-\alpha}\frac\partial{\partial x^m_A}.
\label{deriv}
\end{eqnarray}

In the harmonic superspace approach the harmonic variables
$u^\pm_i$ are considered on equal footing with the Grassmann
and space-time ones. In particular, there are covariant
harmonic derivatives, which in the analytic coordinates
($x_{A}^m,\theta^\pm_\alpha,\bar\theta^\pm_{\dot\alpha},u$)
are given by
\begin{eqnarray}
D^{++}&=&u^+_i\dfrac\partial{\partial u^-_i}
 -2{\rm i}\theta^{+\alpha}(\sigma_m)_{\alpha\dot\alpha}\bar\theta^{+\dot\alpha}
 \dfrac\partial{\partial x^m_A}
 +\theta^+_\alpha\dfrac\partial{\partial\theta^-_\alpha}
 +\bar\theta^+_{\dot\alpha}\dfrac\partial{\partial\bar\theta^-_{\dot\alpha}},\nonumber\\
D^{--}&=&u^-_i\dfrac\partial{\partial u^+_i}
 -2{\rm i}\theta^{-\alpha}(\sigma_m)_{\alpha\dot\alpha}\bar\theta^{-\dot\alpha}
 \dfrac\partial{\partial x^m_A}
 +\theta^-_\alpha\dfrac\partial{\partial\theta^+_\alpha}
 +\bar\theta^-_{\dot\alpha}\dfrac\partial{\partial\bar\theta^+_{\dot\alpha}},\nonumber\\
D^0&=&[D^{++},D^{--}]=u^+_i\dfrac\partial{\partial u^+_i}-
 u^-_i\dfrac\partial{\partial u^-_i}
 +\theta^+_\alpha\dfrac\partial{\partial\theta^+_\alpha}
 +\bar\theta^+_{\dot\alpha}\dfrac\partial{\partial\bar\theta^+_{\dot\alpha}}
 -\theta^-_\alpha\dfrac\partial{\partial\theta^-_\alpha}
 -\bar\theta^-_{\dot\alpha}\dfrac\partial{\partial\bar\theta^-_{\dot\alpha}}.
\label{harm-deriv}
\end{eqnarray}
The derivatives (\ref{harm-deriv}) obey the commutation relations
of the algebra ${\rm su}(2)$.

The integration over the Grassmann and harmonic variables is
defined by the rules
\begin{eqnarray}
&&\int d^4\theta (\theta^+)^2(\theta^-)^2=1,\quad
\int d^4\theta^-(\theta^+)^2(\bar\theta^+)^2=1, \quad \int d^8\theta
(\theta^+)^2(\bar\theta^+)^2(\theta^-)^2(\bar\theta^-)^2=1\,,\nonumber\\
&&
\int du\,1 =1\,, \quad \int du\,u^{+ (i_1}\ldots u^{+ i_n}u^{-j_1}\ldots u^{-j_m)} = 0\,.
\label{int-def}
\end{eqnarray}
Here we use the following notation
\be
(\theta^+)^2=\theta^{+\alpha}\theta^+_\alpha,\quad
(\bar\theta^+)^2=\bar\theta^+_{\dot\alpha}\bar\theta^{+\dot\alpha},\quad
\theta^-\sigma_{mn}\theta^+=\theta^{-\alpha}(\sigma_{mn})_\alpha^\beta\theta^+_\beta
=-\theta^+\sigma_{mn}\theta^-.
\label{TT}
\ee

We use also the chiral-analytic coordinates
$Z_C=(z_C,\bar\theta^{\pm\dot\alpha})$, where
\be
z_C=(x_L^m,\theta^{\pm\alpha} ).
\label{Ccoor}
\ee
The covariant
spinor and harmonic derivatives, as well as the $\cN{=}(1,0)$
supercharges, in these coordinates read
\begin{eqnarray}
D^+_\alpha&=&\partial_{-\alpha}+ 2{\rm i}\bar\theta^{+\dot\alpha}\partial_{\alpha\dot\alpha}
\,,\qquad
D^-_\alpha=-\partial_{+\alpha} +2{\rm i}\bar\theta^{-\dot\alpha}\partial_{\alpha\dot\alpha}\,,
\nonumber\\
\bar{D}^+_{\dot\alpha}&=&\bar\partial_{-\dot\alpha}\,,\qquad
\bar{D}^-_{\dot\alpha}=-\bar\partial_{+\dot\alpha}\,\\
D^{++}_C&=&\partial^{++} +\theta^{+\alpha} \partial_{-\alpha} +
 \bar\theta^{+\dot\alpha}\bar\partial_{-\dot\alpha}\,,
\nonumber\\
D^{--}_C&=&\partial^{--} +\theta^{-\alpha} \partial_{+\alpha} +
\bar\theta^{-\dot\alpha}\bar\partial_{+\dot\alpha}\,,\\
Q^+_\alpha&=&\partial_{-\alpha}\,,
\qquad Q^-_\alpha=-\partial_{+\alpha}\,,
\label{QCgen}
\end{eqnarray}
where $\partial^{++}=u^+_i \frac\partial{\partial u^-_i}$,
$\partial^{--}=u^-_i \frac\partial{\partial u^+_i}$.
An analytic superfield $\Lambda(\zeta,u)$ can be represented in
the chiral-analytic coordinates as
\be
\Lambda(\zeta,u)=\Lambda(\zeta_C,u)-2{\rm i}(\theta^-\sigma_m\bar\theta^+)
\partial_m\Lambda(\zeta_C,u)
-(\theta^-)^2(\bar\theta^+)^2\Box\Lambda(\zeta_C,u),
\label{leftanal}
\ee
where $\zeta_C=(x^m_L,\theta^+,\bar\theta^+)$
and the component fields in the $\theta$ and harmonic expansion of $\Lambda(\zeta_C,u)$
depend on the coordinates $x^m_L$.

For the antichiral superfields we use also the
antichiral coordinates,
\be
x_R^m=x_A^m +2{\rm i}\theta^+\sigma^m\bar\theta^-=x_L^m +2{\rm i}\theta^+\sigma^m\bar\theta^-
 -2{\rm i}\theta^-\sigma^m\bar\theta^+.
\label{xR} \ee The $\cN{=}(1,0)$ supercharges and the covariant
spinor derivatives in these coordinates are given by the expressions
\begin{eqnarray}
Q^+_\alpha&=&\partial_{-\alpha}-2{\rm i}\bar\theta^{+\dot\alpha}
\partial_{\alpha\dot\alpha}\,,\qquad  Q^-_\alpha=
-\partial_{+\alpha}-2{\rm i}\bar\theta^{-\dot\alpha}
\partial_{\alpha\dot\alpha},\nonumber\\
D^+_\alpha&=&\partial_{-\alpha}\,,\qquad
D^-_\alpha=-\partial_{+\alpha}.\label{QR}
\end{eqnarray}

\end{document}